\documentclass[fleqn,usenatbib]{mnras}

\usepackage{newtxtext,newtxmath}

\usepackage[T1]{fontenc}
\usepackage{ae,aecompl}

\usepackage{graphicx}   
\usepackage{amsmath}    
\usepackage{amssymb}    

\newcommand{\km}{\rm\thinspace km}
\newcommand{\cm}{\rm\thinspace cm}
\newcommand{\erg}{\rm\thinspace erg}
\newcommand{\s}{\rm\thinspace s}
\newcommand{\Mpc}{\rm\thinspace Mpc}
\newcommand{\kpc}{\rm\thinspace kpc}
\newcommand{\kmps}{\hbox{$\km\s^{-1}$}}
\newcommand{\ergps}{\hbox{$\erg\s^{-1}\,$}}
\newcommand{\ebv}{$E(B-V)$ }
\newcommand{\chisq}{\hbox{$\chi^2$}}
\newcommand{\Msun}{\hbox{$\rm\thinspace M_{\odot}$}}
\newcommand{\kmpspMpc}{\hbox{$\kmps\Mpc^{-1}$}}
\newcommand{\Oiii}{[\ion{O}{iii}] }

\title[{[\ion{O}{iii}]}~Emission in Heavily Reddened Quasars]{[\ion{O}{iii}]~Emission Line Properties in a New Sample of Heavily Reddened Quasars at $z>2$}

\author[M. J. Temple et al.]{Matthew J. Temple,$^{1}$\thanks{E-mail: mtemple@ast.cam.ac.uk}
Manda Banerji,$^{1,2}$
Paul C. Hewett,$^{1}$
Liam Coatman,$^{1}$
\newauthor
Natasha Maddox$^{3}$
and Celine Peroux$^{4}$
\\
$^{1}$Institute of Astronomy, University of Cambridge, Madingley Road, Cambridge CB3 0HA, UK\\
$^{2}$Kavli Institute for Cosmology, University of Cambridge, Madingley Road, Cambridge CB3 0HA, UK\\
$^{3}$Faculty of Physics, Ludwig-Maximilians-Universit\"{a}t, Scheinerstr. 1, 81679 Munich, Germany\\
$^{4}$Aix Marseille Universit\'{e}, CNRS, LAM (Laboratoire d'Astrophysique de Marseille) UMR 7326, 13388, Marseille, France
}

\date{Accepted XXX. Received YYY; in original form ZZZ}

\pubyear{2019}

\begin{document}
\label{firstpage}
\pagerange{\pageref{firstpage}--\pageref{lastpage}}
\maketitle

\begin{abstract}
We present VLT-SINFONI near infra-red spectra of 26 new heavily reddened quasar candidates selected from the UKIDSS-LAS, VISTA VHS and VIKING imaging surveys. This new sample extends our reddened quasar search to both brighter and fainter luminosities. 25 of the 26 candidates are confirmed to be broad line quasars with redshifts $0.7<z<2.6$ and dust extinctions $0.5<E(B-V)<3.0$. Combining with previously identified samples, we study the H\,$\alpha$, H\,$\beta$ and [\ion{O}{iii}] emission line properties in 22 heavily reddened quasars with  $L_{\rm bol}\approx10^{47}$\,erg\,s$^{-1}$  and  $z>2$. We present the first comparison of the [\ion{O}{iii}] line properties in high luminosity reddened quasars to a large sample of 111 unobscured quasars in the same luminosity and redshift range. Broad wings extending to velocities of 2500\,km\,s$^{-1}$ are seen in the [\ion{O}{iii}] emission line profiles of our reddened quasars, suggesting that strong outflows are affecting the ionised gas kinematics. However, we find no significant difference between the kinematics of the [\ion{O}{iii}] emission in reddened and unobscured quasars when the two samples are matched in luminosity and redshift. Our results are consistent with a model where quasar-driven outflows persist for some time after the obscuring dust has been cleared from along the line of sight. Assuming the amount of ionised gas in reddened and unobscured quasars is similar, we use the equivalent width distribution of the [\ion{O}{iii}] emission  to constrain the location of the obscuring dust in our reddened quasars. We find that the dust is most likely to be located on galactic scales, outside the [\ion{O}{iii}] emitting region.

\end{abstract}




\section{Introduction}
\label{sec:intro}

Feedback from active galactic nuclei (AGN) is used in the study of galaxy evolution to explain many observables, such as the shape of the galaxy luminosity function \citep[e.g.][]{Vogelsberger13, Sijacki15} and  correlations between the mass of the central supermassive black hole (SMBH) and  host galaxy properties \citep{Magorrian98, McConnell13}. Evidence for AGN feedback has been found through observations of different gas phases and at different wavelengths \cite[e.g.][]{Fabian12,Harrison17}. 

The most massive galaxies in today's universe must have undergone at least one major merger, and possibly many more, in order to build up their present day masses; thus major mergers are an important component in many models of galaxy evolution. Such merger-driven models normally predict an reddened quasar phase: as the galaxy merger triggers a burst of intense star formation (which in turn produces large amounts of reddening dust), at the same time gas is dynamically shocked and falls onto the central SMBH, triggering highly luminous AGN activity  \citep{Sanders88, DiMatteo05, Hopkins08, Narayanan10}.

This proposed model for quasar activity and galaxy evolution is reliant on a `blow-out' phase, where powerful small scale outflows from the inner regions of the active nucleus couple to gas and dust, transferring energy and momentum and driving large scale outflows which then disrupt the gas in the host galaxy, shutting down star formation and clearing  obscuring material away from the line of sight to reveal a luminous blue quasar. 
Such outflows are likely to be driven by some combination of gas pressure and radiation pressure acting on dust, neutral hydrogen and atomic line transitions, and disk-driven magnetocentrifugal winds {\citep[e.g.][]{Konigl94, Murray95, Ishibashi15, Ishibashi17}}, although the dominant driving mechanism is still a topic of debate.
If this model is correct, one might therefore expect to see evidence for more powerful outflows in observations of luminous reddened  quasars than in unobscured quasars, while the obscuring material is in the process of being cleared from the line of sight.

One  probe of large scale outflows in  quasars is the forbidden \Oiii$\lambda\lambda$4960,5008 doublet transition,  which traces ionised gas in regions of low electron density \citep[$n_e \sim 10^3 \cm^{-3}$;][]{Baskin05}. 
\Oiii is  present as a strong emission line in many active galaxies, and  the shape of the line can be used as a robust estimator of the kinematics of the emitting ionised gas.
Spatially resolved observations of local (i.e.~$z<0.5$, and therefore less luminous) active galaxies have found ionised gas moving with near-escape velocities on scales of $0.5-15\kpc$ in most objects \citep[][]{Greene11, Greene12, Liu13, Harrison14, Tadhunter18}, although different observers use different definitions of spatial extent depending on whether they use spatially resolved kinematic data, narrow-band imaging or other methods \citep{Karouzos16, Baron19}. 

While the classic narrow \Oiii emission line profile can be used to define the systemic redshift, more luminous AGN often have a  broader ($v\sim 1000\kmps$) blueshifted component to the \Oiii line, which is too broad to be tracing gas in dynamical equilibrium with the host galaxy.  If present, broad blue \Oiii emission is therefore usually interpreted as evidence for quasar-driven outflows on kiloparsec scales, driving the emitting gas along the line of sight towards the observer. If these outflows are indeed occurring on such large scales, then they would be capable of transferring significant amounts of kinetic energy from the active nucleus back into the host galaxy \citep[][]{Liu13, Harrison18, Coatman19}.

Using the \Oiii emission line, several authors have recently investigated the ionised gas kinematics in AGN across a range of redshifts, luminosities, and levels of obscuration.
\citet{DiPompeo18} find that at low redshifts ($z<0.4; L_{\rm bol}\approx 10^{46}\erg\s^{-1}$), AGN with redder $u-W3$ colours  have broader, more blueshifted \Oiii on average.
This difference is not seen between spectroscopically classified type 1 and  type 2 quasars in the same sample, suggesting that the difference in ionised gas kinematics is not purely due to orientation, but that evolutionary effects might also be at work.

\citet{Harrison16} find that  the distribution of ionised gas velocities in an X-ray selected ($L_{2-10\,\rm{keV}} > 10^{42}\ergps$) sample of AGN at $1.1<z<1.7$ is consistent with that of a luminosity-matched sample of $z<0.4$ AGN, suggesting that the most powerful ionised gas outflows are driven by AGN activity at all redshifts.
\citet{Brusa15} find that quasars selected to have bright X-ray fluxes and red optical--to--infra-red colours at $z\approx1.5$ and $L_{\rm bol}\approx10^{45.5}\ergps$  have \Oiii lines which are broader than Sloan Digital Sky Survey (SDSS) type 2 AGN at similar \Oiii luminosities.
  
At higher redshifts ($z\approx 2.5$; $L_{\rm bol}\approx 10^{47}\erg\s^{-1}$), \citet{Zakamska16} find broad, strongly blueshifted \Oiii in four of the most extreme quasars in the SDSS. These objects were selected to have strong \ion{C}{iv} emission and red optical--to--infra-red colours. \citeauthor{Zakamska16} argue that these objects may represent a `blow-out' phase, as the emitting gas cannot be confined to the gravitational potential of the galaxy. Such `Extremely Red Quasars' have extinctions \ebv $\approx0.3$ \citep{Hamann17}.
 
The evidence so far therefore suggests that red and/or reddened quasars generally have stronger ionised gas outflows, and has often been interpreted as showing that the red selection (i.e. the selection of objects with more obscuring dust) is responsible for the observed outflow properties. However, there has been no systematic, luminosity-matched comparison of reddened and unobscured quasars at $z>2$, corresponding to the epoch of peak star formation and peak AGN activity \citep[][]{MD14, Kulkarni18}.

Using near infra-red data, \citet{Banerji12, Banerji13, Banerji15} have discovered a population of heavily dust-reddened ($E(B-V)\approx 1$), broad line (i.e., spectroscopic type 1) quasars at redshifts $z\approx 2$. These quasars have been selected to have extinctions similar to those in starburst galaxies, as one would expect following an intense burst of star formation triggered by a major merger. At the same time they show broad emission lines, proving that the obscuration is not solely due to simple orientation effects and also allowing their black hole masses to be estimated from virial methods. Four of these quasars have been followed up with the Atacama Large Millimeter/sub-millimeter Array \citep[ALMA;][]{Banerji17, Banerji18}, finding high levels of both cold dust and molecular gas in their host galaxies, while \citet{Wethers18} exploit the obscuration of the quasar light in the rest frame UV to investigate the host galaxy emission, finding evidence for high rates of ongoing star formation. 

In this work, we present the first direct comparison of the properties of the ionised gas emission in the unobscured and heavily dust-reddened  quasar populations at $L_{\rm bol} \approx 10^{47}\ergps$ and $z>2$. We make use of the catalogue of near infra-red spectroscopic observations from \citet{Coatman19}, which describes the ionised gas kinematics in a large ($>100$) sample of unobscured quasars in this luminosity and redshift range for the first time. 
Quantifying the properties of the ionised gas emission in reddened  and unobscured quasars allows us to compare the feedback mechanisms in the two populations, and simultaneously place constraints on the location of the obscuring material in the reddened quasars. 

The structure of this paper is as follows. In Section \ref{sec:data} we present the photometric selection and spectroscopic follow-up of 25 new heavily reddened broad line quasars. By combining with previous studies, we then construct a sample of 22 heavily reddened quasars at redshift $z>2$ with spectral coverage of the \Oiii$\lambda$5008 emission line, and describe our fitting procedure to derive H\,$\alpha$ and \Oiii emission line properties in Section \ref{sec:fitting}. We compare the ionised gas kinematics in our heavily reddened sample with a sample of unobscured quasars in the same redshift and luminosity range in Section \ref{sec:results} and discuss our results in Section \ref{sec:discuss}. 

We assume a flat $\Lambda$CDM cosmology throughout this work, with $\Omega_m=0.27$, $\Omega_{\Lambda}=0.73$, and $\textrm{H}_0=71 \kmpspMpc$. All emission lines are identified with their wavelengths in vacuum in units of \AA ngstr\"{o}ms. We avoid the inherent uncertainty involved in estimating bolometric luminosities by using the monochromatic continuum luminosity $\lambda L_\lambda$ at $\lambda=5100$\,\AA\ (hereafter $L_{5100}$) as a measure of the strength of the ionising radiation. When estimating Eddington ratios, we assume a bolometric correction 
$L_{\rm bol} = 8 \times L_{5100}$ \citep[e.g.][]{Nemmen10, Runnoe12}.
We correct for dust extinction using the quasar extinction law described in \citet{Wethers18}, which gives wavelength-dependent attenuations 
\[A_\lambda = k(\lambda) \times E(B-V) \]
with $k(\rm{H}\,\beta)=3.57$, $k(5008\,$\AA$)=3.45$,  $k(5100\,$\AA$)=3.38$, $k(V)=3.10$, $k(\rm{H}\,\alpha)=2.53$,  $k(\rm{Pa}\,\gamma)=1.41$, and $k(\rm{Pa}\,\beta)=1.17$.

\section{Data}
\label{sec:data}

\subsection{Photometric sample selection}

Our previous searches for heavily reddened quasars \citep{Banerji12,Banerji13,Banerji15} have made use of near infra-red photometric observations in the $J$, $H$, and $K$-bands from the UKIDSS Large Area Survey (UKIDSS-LAS) and the VISTA Hemisphere Survey (VHS). In these previous studies, the wide-field near infra-red data was used to select extremely red quasar candidates e.g. with $(J-K)_{\rm{Vega}} > 2.5$. Subsequent spectroscopic follow-up then confirmed that many of these candidates are quasars at $2<z<3$, with typical  extinctions of \ebv$\approx 0.8$  towards the quasar continuum. In \citet{Banerji15}, we constructed the first luminosity function for the reddened quasar population, which showed evidence that the reddened quasars outnumber unobscured quasars at the highest quasar luminosities, but that their number counts begin to decline as we approach more typical quasar luminosities around $L^*$. In this work, we extend the luminosity range covered by our reddened quasar sample by specifically searching for these quasars at the bright and faint ends of the luminosity function presented in \citet{Banerji15}.  

\subsubsection{Bright sample}

For the selection of a very luminous sample of extremely red quasars, we make use of near infra-red imaging data over $\approx$6300 deg$^2$ from the UKIDSS-LAS and VHS and mid infra-red imaging from \textit{WISE}. The UKIDSS-LAS search is concentrated in three regions (i) 08h$<$RA$<$16h; Dec$<$20$\degr$ (1920 deg$^2$) (ii) 20h$<$RA$<$00h; Dec$<$20$\degr$ (455 deg$^2$) and (iii) 00h$<$RA$<$04h; Dec$<$20$\degr$ (755 deg$^2$). The VHS search is similarly concentrated in three regions: (i) 10h$<$RA$<$16h; -35$<$Dec$<$0$\degr$ (1510 deg$^2$) (ii) 20h$<$RA$<$00h; -25$<$Dec$<$-2\degr (670 deg$^2$) and 
(iii) 20h$<$RA$<$00h; -65$<$Dec$<$-35\degr (980 deg$^2$). All candidates were selected to be classified as point sources in the $K$-band and have $K_{\rm{Vega}}<17.0$. Where $H$-band data was available, we used a colour cut of $(H-K)_{\rm{Vega}}>1.9$ to select the reddest, most intrinsically luminous quasar candidates. This colour cut corresponds to $E(B-V) \gtrsim 1$ at $z=2$. For some regions in VHS, only $J$ and $K$-band data is available. We therefore used a colour cut of $(J-K)_{\rm{Vega}}>3.4$ in these regions, corresponding to the same extinction criterion,  and also required that sources be fainter in the $Y$-band where available, i.e. $Y>J$. Finally, as in our previous searches for reddened quasars, we applied an additional colour cut based on the \textit{WISE} bands of $(W1-W2)>0.85$. With these selection criteria, a total of 24 candidates were identified over the full 6300 deg$^2$ area. Seven of these candidates were observed and presented in \citet{Banerji12, Banerji15}. A further seven extremely red, bright quasar candidates are spectroscopically followed up and presented in this paper.  

\subsubsection{Faint sample}

To extend the search to fainter intrinsic luminosities than is possible with the UKIDSS-LAS or VHS, we made use of deeper near infra-red data from the VISTA VIKING survey \citep{Edge13}. VIKING is covering 1500 deg$^2$ in southern and equatorial fields and targets were selected over a $\approx$210 deg$^2$ region with 22h$<$RA$<$00h; -40$<$Dec$<$-25\degr. As in our previous work, we required all candidates to have $(J-K)_{\rm{Vega}}>2.5$ and $(W1-W2)>0.85$ but now going down to a fainter magnitude limit of $K_{\rm{Vega}}<18.4$. These initial colour cuts produced a large candidate list of 177 sources, 40 of which are classified as point sources in the $K$-band and are therefore likely to be dominated by the quasar light in the infra-red bands. As discussed extensively in \citet{Banerji12} and \citet{Banerji15}, selecting point sources reduces contamination from low redshift, low luminosity quasars where the light in the $K$-band has significant contributions from the quasar host galaxy. To further reduce the number of candidates, we matched these 40 point sources to new optical data from the VST-KiDS survey and required all candidates to either be undetected in the KiDS $r$ and $i$-bands, or have very red colours of $(i-K)_{\rm AB} >3.5$, reducing our number of quasar candidates to 17. In order to fill our allocation of observing time, we also targeted two of the reddest extended $K$-band sources  with no VST-KiDS detections and $(J-K)_{\rm Vega}>3$, producing a final sample of 19 candidates over the VIKING region. 

\subsection{New spectroscopic data}

All 26 targets were followed up with ESO-SINFONI, a near infra-red Integral Field Unit installed at the Cassegrain focus of VLT-UT4. We use the SINFONI $H$+$K$ grating which provides a spectral resolution of $R\approx 1500$ in the wavelength range 1.45-2.45\,$\mu\textrm{m}$. We use a field of view of 8\arcsec x\,8\arcsec\ which is nodded in four exposures with offsets of $\pm$1.5\arcsec\ in RA and Dec, giving a usable field of view of 5\arcsec x\,5\arcsec\ in the reduced datacubes.
For targets in the redshift range $2.0<z<2.6$, this setup yields coverage of the H\,$\alpha$ emission line in the $K$-band and both H\,$\beta$ and \Oiii in the $H$-band. For redshifts of $1.2<z<2.0$ we have coverage of H\,$\alpha$ in the $H$-band. Lower redshift objects can be identified from their near infra-red spectra using the hydrogen Paschen series.

The data were  reduced using the \textit{easysinf} package, as described in \citet{easysinf}. 
1-dimensional spectra were then extracted from the datacubes using circular apertures with diameters in the range 0.6-1.3\arcsec\ ($\approx5$-10\kpc\ at $z=2$). The size of the aperture was chosen to maximise the signal-to-noise ratio (S/N) of the H\,$\alpha$ line in each spectrum. Choosing instead to maximise the S/N of the continuum or the [\ion{O}{iii}] line does not significantly change the size of the aperture in any of the objects and does not affect the shape of the spectrum, meaning there is no  evidence for spatially extended \Oiii emission on scales $\gtrsim5\kpc$ within the limited S/N  of our observations. 
In the cases where a single object was observed across multiple observing blocks (OBs), spectra were extracted from each datacube before being combined, as different OBs were conducted under different seeing conditions, leading to different optimal aperture sizes. Noise arrays were extracted by considering the spaxel-spaxel variance in the field away from the source at each wavelength pixel. Absolute  flux calibration was carried out for each spectrum using the $K$-band photometry from VISTA and UKIDSS. 

Of the 26 targets which satisfied our selection criteria and  were spectroscopically followed up, 25 are  confirmed to be quasars in the redshift range $0.7<z<2.6$. One target (VHS~J2222-3915) is found to have a red continuum with no emission features. These observations are summarised in Table~\ref{tab:Observations} and spectra are shown in Fig.~\ref{fig:spectra}. 

\begin{table*}
    \centering
\caption{\small Observation log for our 26 targets. Exposure times are given in seconds and the full width at half maximum of the point spread function (i.e., the size of the seeing disc) in arcseconds. Redshifts are taken from the broad component of the fit to H\,$\alpha$, where present, otherwise from the Paschen lines. $JHK$ photometry are taken from UKIDSS-LAS for ULAS sources, and $JHK_s$ photometry from VISTA for VHS and VIK sources. Spectra are shown in Fig.~\ref{fig:spectra}.}
\label{tab:Observations}
    \begin{tabular}{l | l | l | c | c | c | c | c | c | c}
    \hline
Name & RA &  Dec. & $J_{\rm{Vega}}$ & $H_{\rm{Vega}}$ & $K_{\rm{Vega}}$ & Redshift &  Observation date &  Exp. time (s) &  Seeing (\arcsec)  \\
\hline
VHS J1117-1528 & 11:17:03.74   & -15:28:29.6 & 20.75 & 18.44 & 16.94 & 2.428 & 2015-04-06 & 2400 & 1.20\\
VHS J1122-1919 & 11:22:24.43   & -19:19:17.4    & 19.58 & - & 15.96 & 2.464 & 2015-04-06 & 600 & 1.52\\
ULAS J1216-0313 & 12:16:31.78  & -03:13:35.0    & - & 18.17 & 16.20 & 2.574 & 2015-04-06 & 600 & 1.38\\
VHS J1301-1624 & 13:01:31.32   & -16:24:54.0    & 20.12 & - & 16.39 & 2.138 & 2015-04-07 & 1200 & 1.06\\
ULAS J1415+0836 & 14:15:26.74  & +08:36:15.8    & - & 18.67 & 16.64 & 1.120 & 2015-06-05 & 1200 & 1.88\\
VIK J2205-3132 & 22:05:13.68  & -31:32:02.4  & 19.93 & 18.46 & 17.33 & 2.307 & 2015-06-15 & 2400 & 0.95\\
VIK J2214-3100 & 22:14:05.40  & -31:00:38.2  & 20.25 & 18.67 & 16.84 & 1.069 & 2015-06-22 & 2400 & 0.75\\
VHS J2222-3915 & 22:22:10.65  & -39:15:49.3     & 18.77 & - & 14.60 & - & 2015-04-04 & 400 & 0.85\\
VIK J2228-3205 & 22:28:44.98   & -32:05:38.8  & 21.00 & 19.69 & 18.40 & 2.364 & 2015-07-07 & 1200 & 1.22\\
   & &   &  &&      & & 2015-07-02 & 2400 & 0.96\\
VIK J2230-2956 & 22:30:09.36   & -29:56:21.5  & 20.93 & 18.98 & 17.47 & 1.319 & 2015-07-07 & 1200 & 1.36\\
   &&    &   &&     & & 2015-07-02 & 2400 & 0.87\\
VIK J2232-2844 & 22:32:36.96   & -28:44:39.5  & 20.02 & 18.57 & 16.93 & 2.292 & 2015-08-07 & 2400 & 0.99\\
VIK J2238-2836 & 22:38:23.95   & -28:36:40.7  & 19.56 & 18.14 & 17.03 & 1.231 & 2015-07-28 & 2400 & 1.15\\
VIK J2241-3006 & 22:41:12.41   & -30:06:42.5  & 19.56 & 18.27 & 16.84 & 0.720 & 2015-07-21 & 2400 & 0.89\\
VIK J2243-3504 & 22:43:48.96   & -35:04:39.4  & 19.78 & 18.01 & 16.41 & 2.085 & 2015-07-20 & 1200 & 1.12\\
VIK J2245-3516 & 22:45:58.80   & -35:16:58.4  & 20.30 & 18.81 & 17.17 & 1.335 & 2015-08-07 & 2400 & 1.31\\
  &&   &   &&     & & 2015-07-02 & 600 & 1.19\\
VIK J2251-3433 & 22:51:17.90   & -34:33:50.0  & 21.10 & 18.86 & 18.04 & 1.693 & 2015-08-07 & 2400 & 1.31\\
   &&    &   &&     & & 2015-08-11 & 1200 & 1.64\\
VIK J2256-3114 & 22:56:07.97   & -31:14:25.8  & 20.01 & 18.55 & 17.25 & 2.329 & 2015-07-21 & 2400 & 0.80\\
VIK J2258-3219 & 22:58:30.10   & -32:19:11.6  & 18.96 & 17.69 & 16.45 & 0.879 & 2015-05-10 & 1200 & 1.67\\
VIK J2306-3050 & 23:06:50.23   & -30:50:34.1  & 20.72 & 19.25 & 18.07 & 1.060 & 2015-07-07 & 1200 & 1.86\\
  &&    &  &&    & & 2015-07-07 & 2400 & 1.31\\
VIK J2309-3433 & 23:09:49.90   & -34:33:43.6  & 20.77 & 19.50 & 18.26 & 2.159 & 2015-07-19 & 1200 & 0.74\\
  &&    &   &&    &  & 2015-07-19 & 2400 & 0.74\\
ULAS J2312+0454 & 23:12:31.99  & +04:54:20.9  & 20.07 & 18.67 & 16.72 & 0.700 & 2015-08-01 & 1200 & 1.20\\
VIK J2313-2904 & 23:13:51.96   & -29:04:52.7  & 20.22 & 18.00 & 17.60 & 1.851 & 2015-07-10 & 1200 & 0.85\\
  &&    &    &&   &  & 2015-07-02 & 2400 & 1.10\\
VIK J2314-3459 & 23:14:16.87   & -34:59:47.0  & 19.89 & 18.50 & 17.36 & 2.325 & 2015-07-20 & 2400 & 0.94\\
VIK J2323-3222 & 23:23:50.45   & -32:22:16.7  & 20.12 & 18.44 & 16.83 & 2.191 & 2015-05-10 & 1200 & 1.13\\
VIK J2350-3019 & 23:50:45.50   & -30:19:53.4  & 20.12 & 18.81 & 17.58 & 2.324 & 2015-07-19 & 2400 & 0.65\\
  &&    &     &&   & & 2015-06-02 & 1200 & 1.24\\
VIK J2357-3024 & 23:57:45.84   & -30:24:30.6  & 20.33 & 18.75 & 17.58 & 1.129 & 2015-07-02 & 2400 & 0.94\\
  &&    &    &&    & & 2015-07-30 & 1200 & 0.97\\
\hline
\end{tabular}
\end{table*}

\subsection{Existing spectroscopic data}
\label{sec:sample}

13 of the 25 spectroscopically confirmed quasars from the new sample presented above are at $z>2$. We combine these 13 objects with nine quasars from \citet{Banerji12} in the same redshift range and with the same wavelength coverage -- three objects with VLT-SINFONI data and six objects with Gemini-GNIRS data -- to obtain a total of 22 quasars with near infra-red spectra covering  H\,$\alpha$, H\,$\beta$ and [\ion{O}{iii}], which we refer to as the `high redshift sample' and use when reporting results in the rest of this paper. We carry out our own fits to the emission lines in the previously published objects and hence derive parameters in a consistent manner for the whole sample.

\section{Spectral fitting and black hole masses}
\label{sec:fitting}
\subsection{High redshift sample}
\label{sec:High_z_sample_fitting}

For the 22 quasars with $z>2$ in our high redshift sample, we have coverage of [\ion{O}{iii}], which traces ionised gas in regions of low electron density. Measuring the shape and strength of the \Oiii emission therefore allows us to quantify the kinematics of the ionised gas in these objects. 
Using the kinematics of the Balmer emitting gas, we can also estimate black hole masses and Eddington ratios. We model the emission from H\,$\alpha$, H\,$\beta$ and \Oiii  in order to derive more robust properties for these emission lines.

We now describe our modelling procedure for the H\,$\alpha$, H\,$\beta$ and \Oiii emission lines in our high redshift sample, which is very similar to that used by \citet{Shen11}, with the main difference being that we constrain the shape of the H\,$\beta$ line to be the same as that of H\,$\alpha$. Due to the large amount of dust extinction in these reddened quasars, the S/N is considerably higher in H\,$\alpha$ than in H\,$\beta$ and we find that we obtain better fits to the region around H\,$\beta$ by constraining the shape of the emission line in this way.

\begin{figure*}
    \includegraphics[width=\columnwidth]{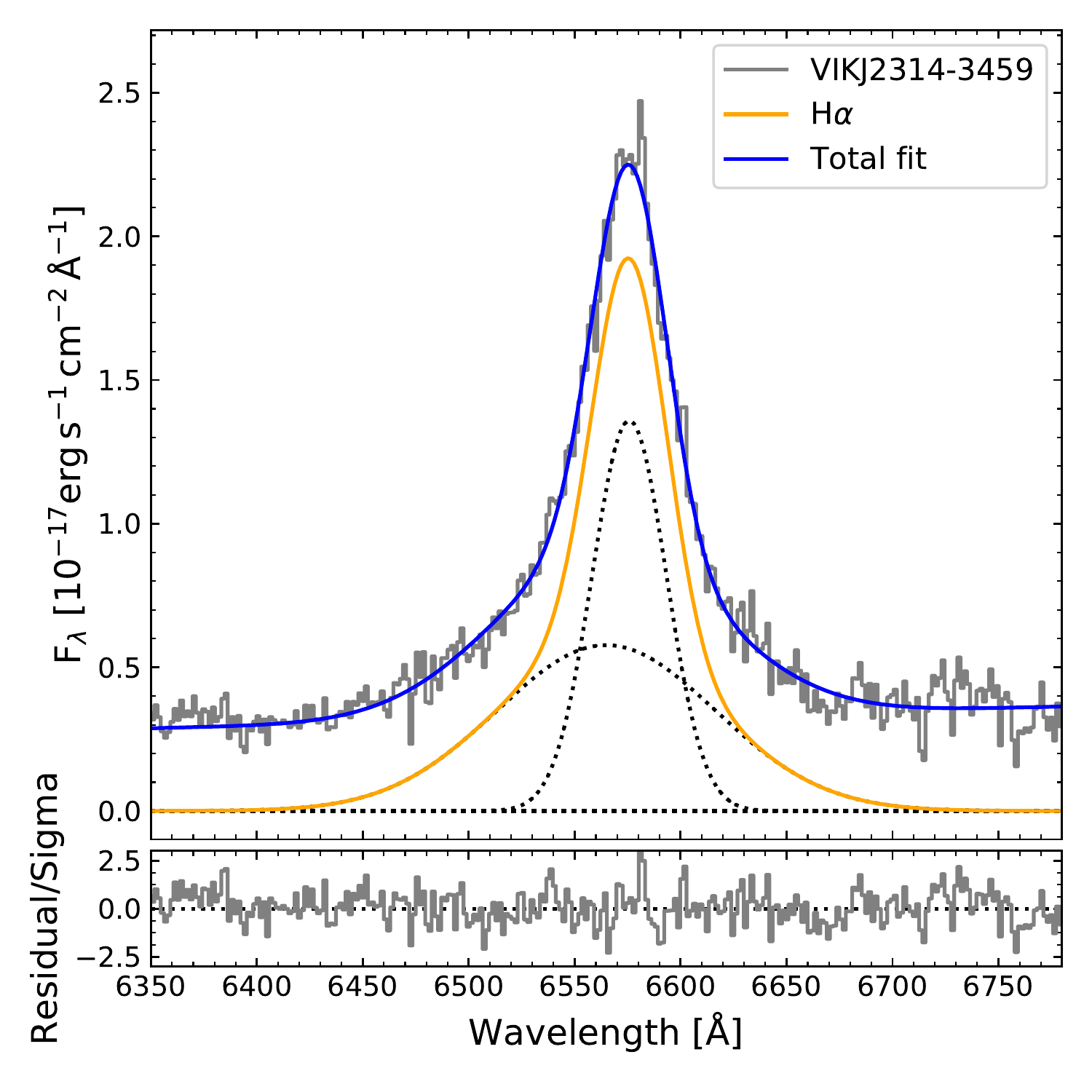}
    \includegraphics[width=\columnwidth]{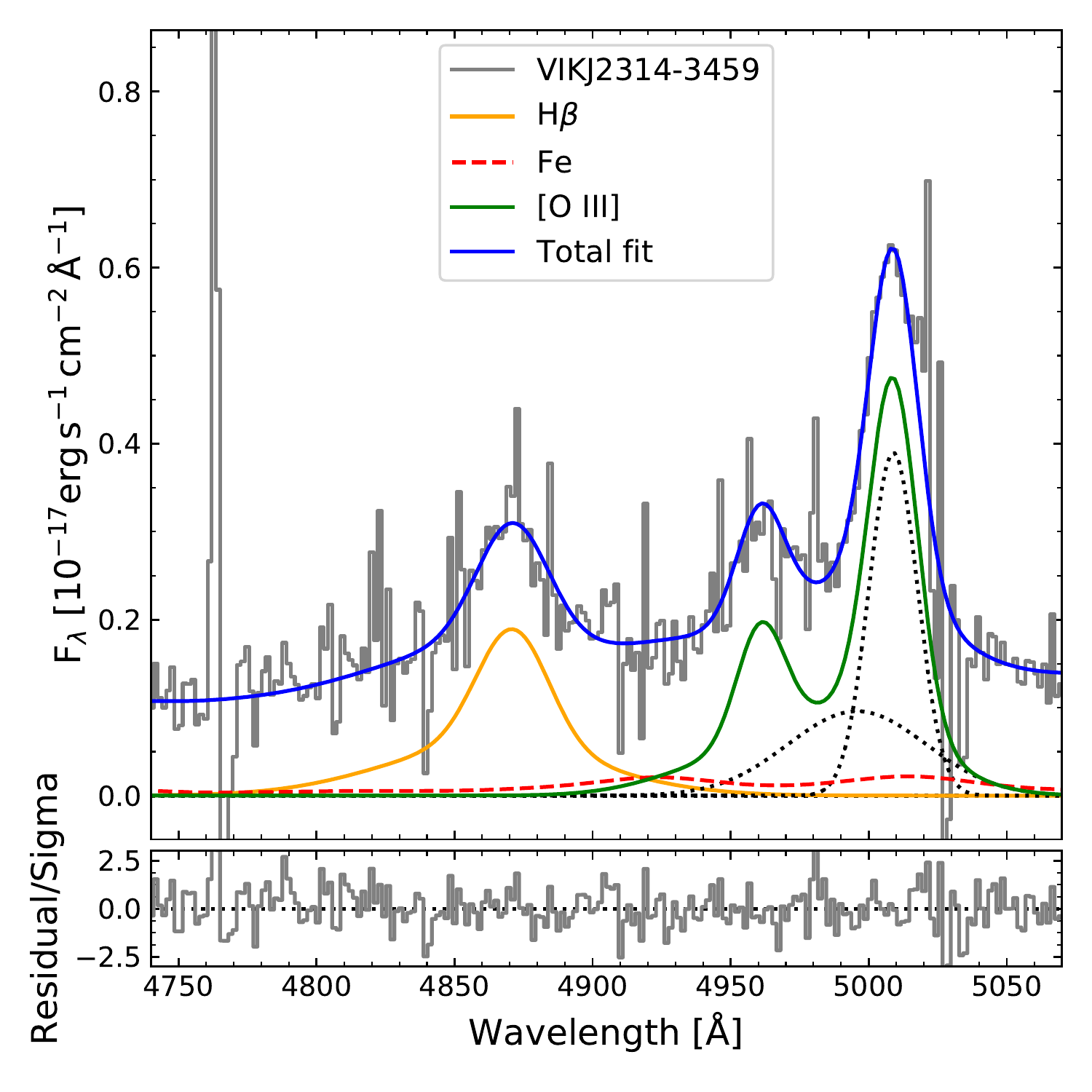}
    \caption{Examples of our fits to the H\,$\alpha$ line (left) and the region around H\,$\beta$ and [\ion{O}{iii}] (right). Individual Gaussian components in the fits to H\,$\alpha$ and [\ion{O}{iii}]~$\lambda$5008  are shown with dashed lines. Residuals are shown in the panels below, scaled by the noise. VIK~J2314-3459 is typical of our high redshift sample in terms of its S/N properties, with the continuum flux being noticeably stronger at 6750\,\AA\ than at 4750\,\AA.}
    \label{fig:fits}
\end{figure*}

\subsubsection{H\,$\alpha$ modelling procedure}

A power law continuum is fit to the data in the wavelength regions 6000-6350 and 6800-7000\,\AA, except for VIK~J2238-2836, where the power law is instead fit to the region 6800-7800\,\AA. This continuum is then subtracted from the spectrum and seven different models are fit to the H\,$\alpha$ line in the window 6350-6800\,\AA.

We fit the continuum-subtracted H\,$\alpha$ line with a model consisting of one, two, or three broad Gaussian components. In the case of two Gaussians, this fit is done twice: once constraining both Gaussians to have the same centroid and once where the two components are unconstrained. For the three models with one or two broad components, a separate fit is also tried with the addition of a narrow line region (NLR) template. The NLR template consists of five Gaussians, each constrained to have the same width and velocity offset: one to fit any narrow component of H\,$\alpha$, and two each to fit the [\ion{N}{ii}]\,$\lambda\lambda$6548,6584 and [\ion{S}{ii}]\,$\lambda\lambda$6717,6731 doublets. The amplitudes of the [\ion{S}{ii}] doublet are constrained to be equal and those of [\ion{N}{ii}] are fixed at the expected ratio of $1:2.96$. All components of all models are constrained to be non-negative. 

Each fit is visually inspected, and models which contain physically unreasonable line widths ($\le 200 \kmps$) are rejected. A model with more free parameters is accepted in favour of a simpler model only if there is a greater than 10 per cent reduction in the $\chisq$ per degree of freedom in the fitting window. These `best' fits are shown in Figs.~\ref{fig:fits}~and~\ref{fig:Ha}, and have a median reduced \chisq~value of 0.978. One object (VIK~J2243-3504) is found to require the inclusion of the NLR template in order to best describe the shape of the H\,$\alpha$ line profile; all the other objects are adequately described by up to three broad Gaussians.

\subsubsection{H\,$\beta$ and \Oiii modelling procedure}
\label{sec:Oiii}

The region outside the observed frame wavelengths 1.45-1.84\,$\mu\textrm{m}$\ is first masked to exclude the portions of the spectrum which are severely affected by atmospheric absorption.
The rest frame wavelength region 4435-5535\,\AA\ is then fit simultaneously with (i) a power law, (ii) an iron template taken from \citet{BG92}, (iii) the profile of the best-fitting H\,$\alpha$ model, with the normalisation free to vary to fit H\,$\beta$, and (iv) one of three models for the \Oiii$\lambda\lambda$4960,5008 doublet. The iron template is first convolved with a Gaussian kernel, the width of which is allowed to vary up to the width of the profile of the best-fitting H\,$\alpha$ model. In all objects, the resulting fits have an iron width consistent with that of the H\,$\alpha$ emission. We show in Appendix~\ref{sec:iron} that our results are robust when comparing the use of different iron templates. 

The \Oiii$\lambda$4960 line is constrained to have the same shape as the $\lambda$5008 line, with the total flux in the lines fixed at the theoretical ratio of $1:2.98$ \citep{Oiii_ratio}. We try fitting zero, one, and two Gaussians to the $\lambda$5008 line, with an extra Gaussian accepted only if it leads to a greater than 10 per cent reduction in the $\chisq$ per degree of freedom in the wavelength range 4700-5100\,\AA. A fourth \Oiii model with a third Gaussian component is found not to lead to a significant reduction in reduced $\chisq$  in any of the objects in this sample. 

The $\lambda$4960 and $\lambda$5008 lines are fit simultaneously, although we only use the $\lambda$5008 line properties when presenting our results. 
Eleven objects are adequately fit with a single Gaussian describing the $\lambda$5008 line. Eight objects require a second Gaussian component to fully reproduce the $\lambda$5008 line profile, providing evidence for an asymmetric line profile in $\approx42$ per cent of the population, similar to the fraction found in unobscured quasars by \citet{Coatman19}.
These fits are shown in Figs.~\ref{fig:fits}~and~\ref{fig:Oiii}, and have a median reduced \chisq~value of 1.113 across the wavelength range 4700-5100\,\AA.

Three objects are found to be consistent with no [\ion{O}{iii}] emission. We do not make any estimate of the velocity width of the [\ion{O}{iii}] lines in these cases.
 In ULAS~J1216-0313 and ULAS~J1234+0907, we estimate an upper limit on the equivalent width of the \Oiii emission line by fitting a template, generated by running our fitting routine over a median composite spectrum constructed from the 19 objects in our sample with reliable [\ion{O}{iii}] measurements.  In one object, ULAS~J0144-0014, we find no \Oiii emission, even when constraining the shape of the line with the template. These three objects are shown in Fig.~\ref{fig:template}.

\subsubsection{Black hole masses}
\label{sec:HighzBHmass}

For intrinsically luminous, heavily reddened quasars at $z>2$, it is known that host galaxy contamination does not significantly affect the $JHK$ colours \citep{Wethers18}. Therefore, for the 22 quasars in our high redshift sample, we derive extinctions using the near infra-red photometric data from VISTA and UKIDSS. Where $J$-band photometry is available, we find the $E(B-V)$ required to best fit the observed $J-K$ colour  when reddening a  quasar template \citep{Maddox08, Wethers18}. If $J$-band photometry is not available, we use the $H-K$ colour instead. 
Rest frame 5100\,\AA\ luminosities are corrected for dust extinction using this $E(B-V)$. We derive black hole masses using the prescription from \citet{vp06}, and use the correction from \citet{Coatman17} to estimate the full width at half maximum (FWHM) of H\,$\beta$ from our fit to H\,$\alpha$:

\[\rm{FWHM}_{\rm{H}\,\beta} = 1.23\times10^3\left(\frac{\rm{FWHM}_{\rm{H}\,\alpha}}{10^3\kmps}\right) ^{0.97} \]
\[\rm{M}_{BH} = 10^{6.91} \left( \frac{L_{5100}}{10^{44} \ergps} \right)^{0.5} \left(\frac{\rm{FWHM}_{\rm{H}\,\beta}}{10^3\kmps}\right) ^{2} \Msun \]

Eddington ratios are estimated assuming $L_{\rm bol} = 8 \times L_{5100}$, and are given in Table~\ref{tab:Oiii_objects}, alongside the derived extinctions, dust-corrected luminosities and black hole masses.

We find extinctions in the range $0.5<E(B-V)<2.0$, as expected from our photometric selection criteria. For our faint sample, we find black hole masses in the range  $10^{8.4-9.4} \Msun$, generally smaller than those presented in \citet{Banerji12,Banerji15}. These quasars are intrinsically fainter than previously known heavily reddened quasars and do not simply appear fainter due to increased extinction. For our bright sample, we find more massive black holes in the range  $10^{9.2-10.5} \Msun$, placing them among the most massive black holes known at these redshifts \citep[e.g.][]{Bischetti17}.

\subsection{Low redshift sample}

Twelve of the 25 spectroscopically confirmed quasars  presented in Section \ref{sec:data} are at $z<2$.
At these lower redshifts, the observed colours may be contaminated by emission from the host galaxy, especially in bluer bands (e.g. $J$) where (by selection) the quasar is faint. To mitigate against this source of uncertainty, we estimate the extinction by comparing the $H-K$ colour from our SINFONI spectra (which are extracted from smaller apertures and will thus be less affected by host galaxy emission)
with the colours predicted from reddening a quasar template \citep{Maddox08, Wethers18}. We find extinctions in the range  $1.3\le E(B-V)\le3.0$, and correct the line luminosities for these extinctions before deriving black hole masses. Results are shown in Table~\ref{tab:lowz}.

For objects with redshifts in the range $1.3<z<2.0$, we fit a single Gaussian to H\,$\alpha$ and estimate the black hole mass using the scaling relation from \citet{GreeneHo}:

\[\rm{M}_{BH} = 2\times10^{6} \left(  \frac{L_{\rm{H}\,\alpha}}{10^{42} \ergps} \right)^{0.55} \left(\frac{\rm{FWHM}_{\rm{H}\,\alpha}}{10^3\kmps}\right) ^{2.06} \Msun \]

We find FWHMs in the range 2000-6500\kmps and black hole masses in the range $10^{8.6-9.6} \Msun$.

For objects with $z<1.3$, we use the scaling relation between the luminosity and width of the Pa\,$\beta$ emission line and the black hole mass given by method 2 of \citet{Kim10}:

\[\rm{M}_{BH} = 10^{7.13} \left(  \frac{L_{\rm{Pa}\,\beta}}{10^{42} \ergps} \right)^{0.48} \left(\frac{\rm{FWHM}_{\rm{Pa}\,\beta}}{10^3\kmps}\right) ^2 \Msun \]

For objects with $z<0.9$, we fit a single Gaussian to the Pa\,$\beta$ line.
For objects with $0.9<z<1.3$ where we have coverage of neither H\,$\alpha$ nor Pa\,$\beta$, we fit kinematically tied Gaussians simultaneously to  Pa\,$\gamma$, Pa\,$\delta$, Pa\,$\epsilon$, \ion{He}{i}\,$\lambda10830$, and \ion{O}{i}\,$\lambda11287$, with the amplitude of each line free to vary. To convert $L_{\rm{Pa}\,\gamma}$ to $L_{\rm{Pa}\,\beta}$, we assume a flux ratio of Pa\,$\gamma:\rm{Pa}\,\beta=1:1.80$  \citep{Storey95, BalmerDecValue}. We find Paschen line FWHMs in the range 1000-4500\kmps and black hole masses in the range $10^{7.5-9.1} \Msun$.

These lower redshift quasars are therefore typically more obscured and have slightly smaller black hole masses than heavily reddened quasars at $z>2$, such as the objects from  \citet{Banerji12} and the high redshift sample in this paper.

\section{Emission line properties}
\label{sec:results}

In this section we explore the [\ion{O}{iii}]\,$\lambda$5008 emission line properties in our high redshift sample (Section~\ref{sec:sample}). We use the best-fitting \Oiii model as described in Section \ref{sec:Oiii} to measure the shape and equivalent width of the line.  For those objects which require multiple Gaussian components to best model the emission line, we do not ascribe any significance to individual components, but instead derive properties for the total line profile. 

\begin{figure}
    \includegraphics[width=\columnwidth]{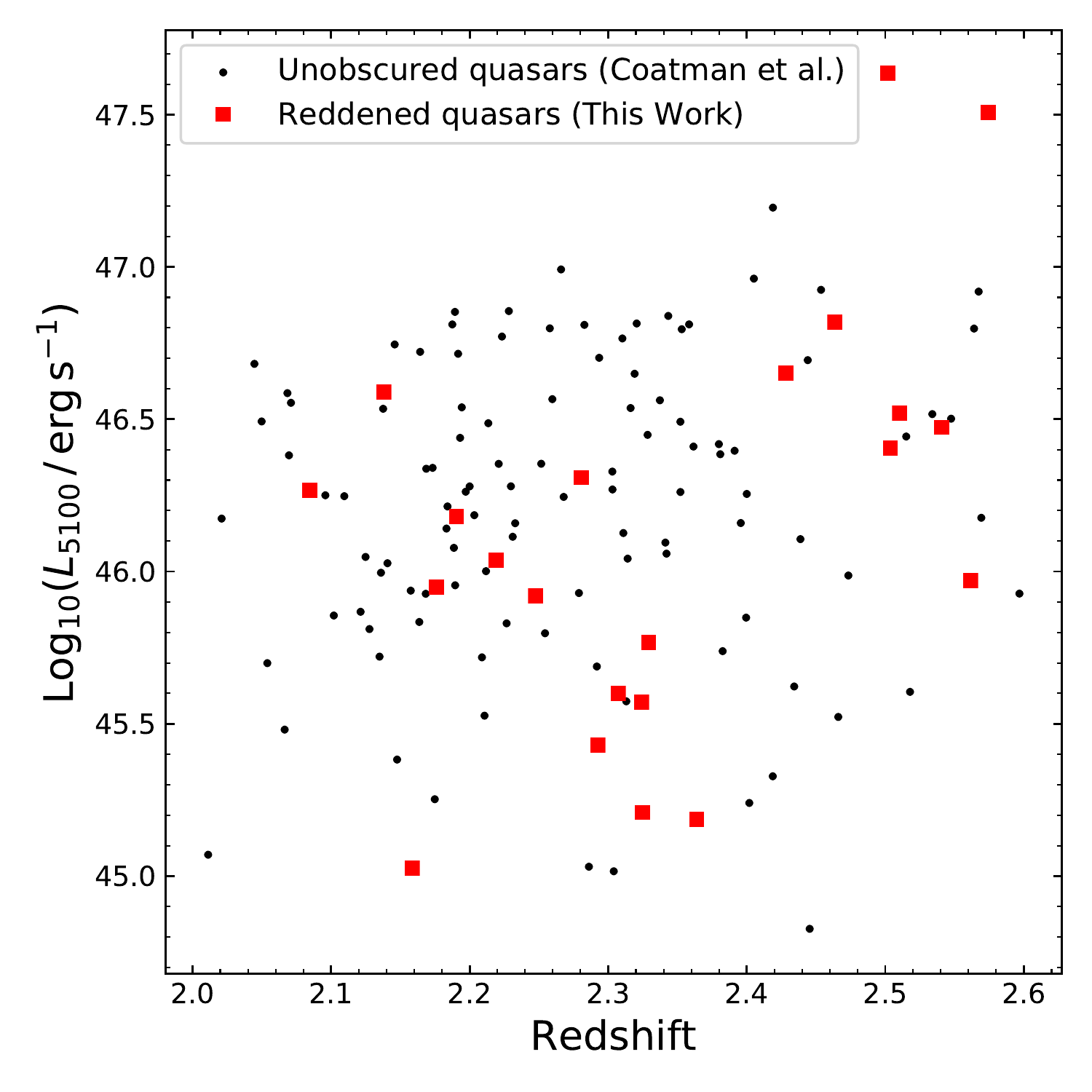}
    \caption{The distribution in redshift-luminosity space of our high redshift reddened quasar sample and our comparison sample of 111 unobscured objects. Luminosities have been corrected for dust extinction, and have a typical uncertainty of 0.2 dex.}
    \label{fig:L5100_z}
\end{figure}
\begin{figure}
    \includegraphics[width=\columnwidth]{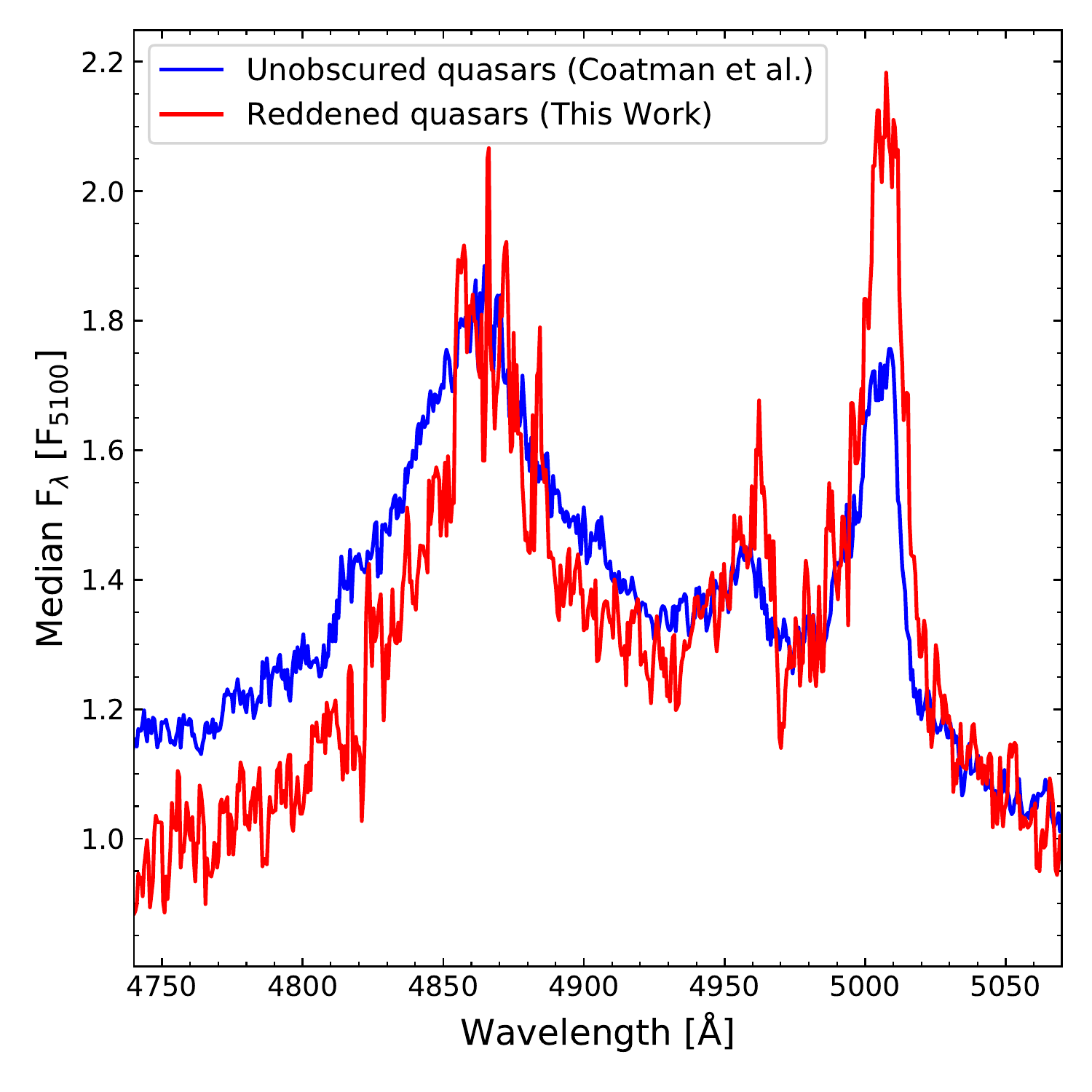}
    \caption{Stacks of 22 highly reddened quasars and a matched sample of 22 unobscured quasars which are the nearest neighbours in $L_{5100}-z$ space. Fluxes have been normalised at 5100\,\AA\ in both samples, with no correction for dust extinction in the reddened sample.}
    \label{fig:Oiii_stack}
\end{figure}

We compare the objects in our high redshift sample to a large sample of unobscured quasars in the same redshift and luminosity range. This comparison sample contains 111 quasars with $2.0<z<2.6$ and near infra-red spectra taken from the catalogue of \citet[][see Fig.~\ref{fig:L5100_z}]{Coatman19}, which are flagged in the catalogue as having robust measurements of the [\ion{O}{iii}] emission line properties.  
The selection of quasars in this catalogue is described in detail by
\citet{Coatman17}. In the context of this work,
the comparison sample essentially provides an unbiased representation of the
optical SEDs of unobscured quasars, matched in redshift and luminosity
to our high redshift reddened quasar sample.
It is worth noting that using a larger comparison sample taken from the full catalogue redshift range of $1.5 <z<4.0$ does not alter any of the results in our subsequent analysis.

To help visualise the quality of data in the two samples, in Fig.~\ref{fig:Oiii_stack} we show stacks of our spectra in the region around the \Oiii emission line, normalised at 5100\,\AA. For the purposes of this stack only, we take a subsample of 22 objects from our comparison sample which are the nearest neighbours to each of our 22 reddened  quasars in the redshift-luminosity space shown in Fig.~\ref{fig:L5100_z}. 
We apply no correction for dust obscuration before stacking, and the slope of the continuum across the narrow wavelength range shown is clearly redder in the reddened sample.
Comparing the two stacks, it can be seen that (a) the reddened objects  generally have lower S/N in this part of the spectrum, (b) the two samples have very similar \Oiii line widths, and (c) the \Oiii emission in the reddened  sample is perhaps slightly stronger relative to the continuum. We will explore the shapes and strengths of the \Oiii lines in our two samples further in Sections~\ref{sec:w80s},~\ref{sec:EWs}~and~\ref{sec:EWs2}.

Uncertainties on derived parameters are estimated using a Monte Carlo approach. 
One hundred realisations of each spectrum are generated by smoothing the observed quasar
spectrum with a 5-pixel inverse-variance weighted top-hat filter, and then adding `noise' drawn from a Gaussian distribution with dispersion equal to the spectrum flux uncertainty at each wavelength. Quoted parameter uncertainties are 1.48 times the median absolute deviation of the parameter distribution in the ensemble of simulations.

We compare the  samples using the two-sample Kolmogorov--Smirnov (KS) test \citep{Peacock83}, which tests the null hypothesis that two samples are drawn from the same underlying probability distribution (i.e., the same population).
We also use the two-sample Anderson--Darling (AD) test \citep{Anderson52, Darling57}, which tests the same null hypothesis, but is more sensitive to differences in the tails of the distributions. 

\begin{figure}
    \includegraphics[width=\columnwidth]{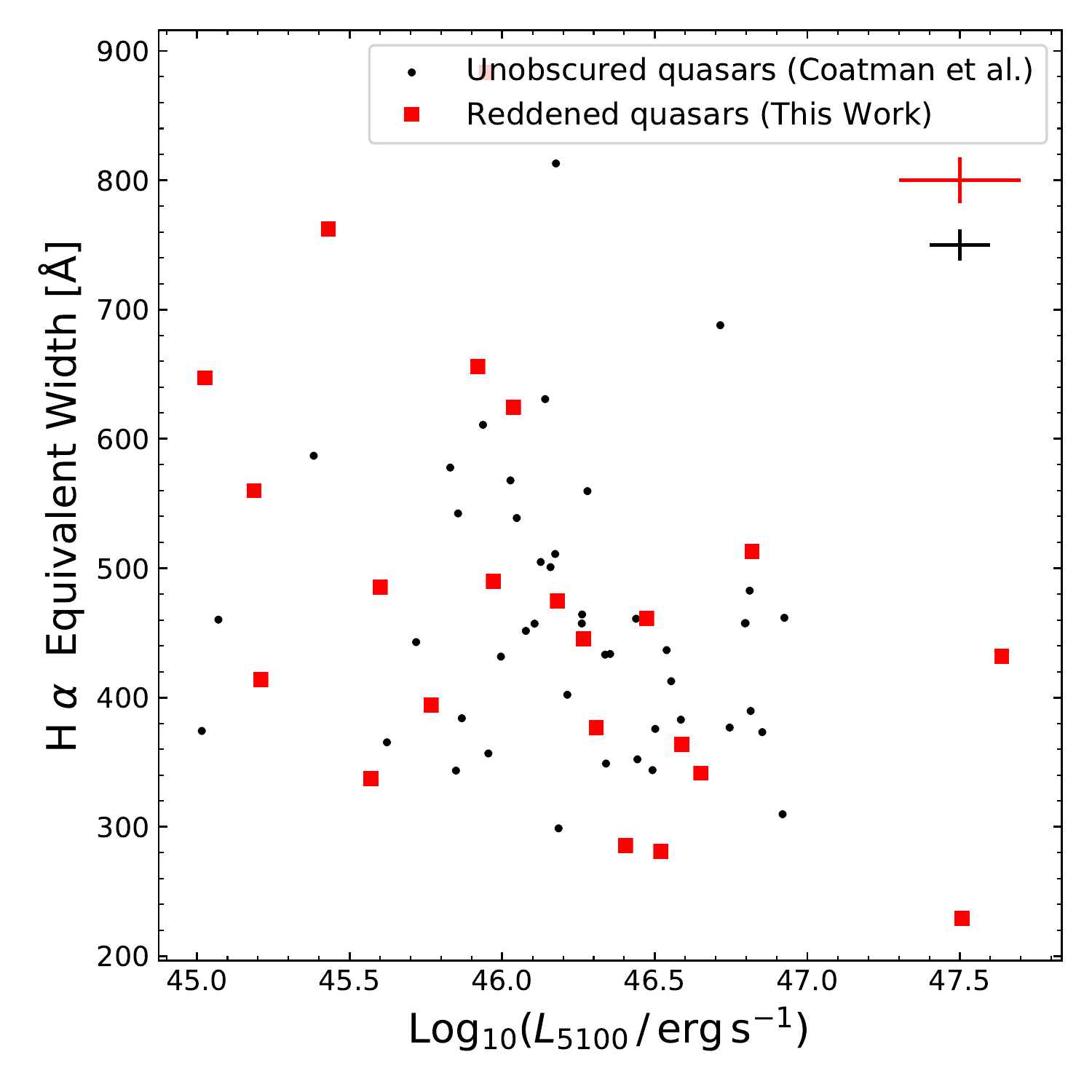}
    \caption{The broad H\,$\alpha$ emission line equivalent widths in our $z>2$ reddened sample and in the 47/111 objects from our comparison sample with spectral coverage of H\,$\alpha$, plotted against the dust-corrected  5100\,\AA\ luminosity. Typical uncertainties are shown in the top right. Using  KS and AD tests, we find no significant difference between the H\,$\alpha$ EWs in the two populations.}
    \label{fig:Ha_EWs}
\end{figure}

We begin by comparing the strengths of the broad H\,$\alpha$ emission in our samples, as measured by the equivalent width. Our heavily reddened quasars were selected by virtue of their red colours, leading to a  selection bias towards objects which are bright in the $K$-band, i.e.~those objects which might have stronger H\,$\alpha$ emission in the redshift range $2.0<z<2.6$. As shown in Fig.~\ref{fig:Ha_EWs}, we find no systematic difference in the equivalent width of the broad line emission between the two samples, suggesting that the amount of ionised gas is not significantly different in the reddened and unobscured samples.

We note that, if one adopts a theoretical value for the unreddened H\,$\alpha$ to H\,$\beta$ ratio, the observed ratio can be used to estimate the extinction in front of the broad line emitting region. However, in addition to the statistical uncertainty of $\sim 0.1$ magnitudes arising from the low S/N in the H\,$\beta$ lines of our heavily reddened quasars, further uncertainty arises due to the unknown temperatures and densities of the gas producing the Balmer emission, i.e. the assumption of a fixed intrinsic flux ratio \citep{Korista04}. 
Adopting an intrinsic ratio of  $\rm{H}\,\alpha : \rm{H}\,\beta = 2.86:1$ \citep{BalmerDecValue}, we find that the extinctions estimated in this way are consistent with the $E(B-V)$s we derive from photometry in Section \ref{sec:High_z_sample_fitting}. However, at a fixed value of the $E(B-V)$ derived from photometry, the scatter in the extinction we measure from the Balmer lines is $\sim 0.5$ magnitudes.
We therefore use  $E(B-V)$s from photometry in the rest of this paper.

\subsection{The shape of the [\ion{O}{iii}] lines}
\label{sec:w80s}

As described in Section \ref{sec:Oiii}, we have information about the shape of the \Oiii lines in 19 of our 22 high redshift reddened quasars. For these 19 objects, we first calculate the 80 per cent velocity width ($w_{80}$) of the $\lambda$5008 line profile. We take the best-fitting [\ion{O}{iii}] model and integrate to find the total flux under the line profile. The velocity width is then defined to be $w_{80} = v_{90}-v_{10}$, where $v_{90}$ and $v_{10}$ are the respective velocities at which 90 and 10 per cent of the cumulative flux in the line profile is found. 
We correct for instrumental broadening of $\sigma = 200 \kmps$, noting that all of our [\ion{O}{iii}] lines are well resolved with the smallest velocity width in our sample being $w_{80, \rm{min}} = 600 \kmps$ prior to applying the correction. 

\begin{figure}
    \includegraphics[width=\columnwidth]{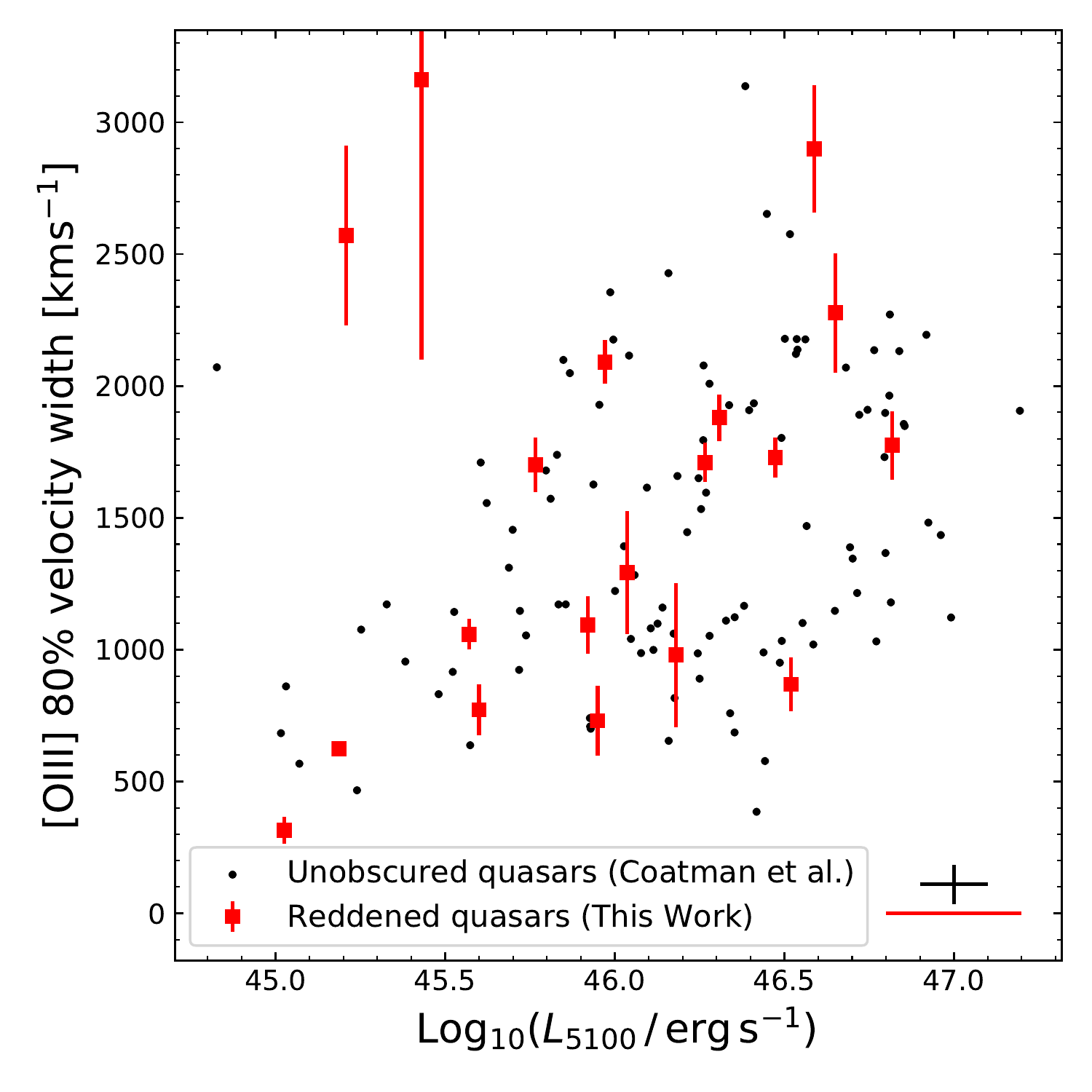}
    \caption{The [\ion{O}{iii}] 80 per cent velocity widths in our reddened sample and our comparison sample, plotted against the dust-corrected monochromatic 5100\,\AA\ luminosity. Typical uncertainties are shown in the bottom right. Using KS and AD tests, we find no significant difference between the \Oiii velocity widths in the two populations.}
    \label{fig:w80s}
\end{figure}

The [\ion{O}{iii}] velocity widths in our reddened sample and in the comparison sample of unobscured quasars are shown in Fig.~\ref{fig:w80s}. We find a weak correlation between $w_{80}$ and continuum luminosity, as is also found in unobscured quasars at the same redshifts and luminosities \citep[][]{Netzer04, Shen16, Coatman19}. There is no such correlation between $w_{80}$ and either the black hole mass, the Eddington ratio or the $E(B-V)$. Using the KS and AD tests, we find no evidence to reject the null hypothesis that the two samples are drawn from the same underlying distribution of velocity widths. As the velocity widths in both samples are significantly larger than the velocity dispersion expected from virialised gas motions, it is reasonable to assume that most, it not all, of our velocity width measurements are dominated by emission from outflowing gas. Our results therefore suggest that the ionised gas outflows in the host galaxies of our reddened quasars are, on average, no faster or slower than those in the unobscured population.

In Section \ref{sec:Oiii}, we found that around 42 per cent of the quasars in both the reddened and unobscured samples show evidence for asymmetric \Oiii emission, in that they are not adequately described by a single Gaussian line profile, and require a second Gaussian component in the fit to [\ion{O}{iii}]. For these objects, we quantify the asymmetry of the $\lambda$5008 line using $(v_{90}-v_{50})/(v_{50}-v_{10})$, and find no significant difference between the two samples in terms of the distribution of the amount of $\lambda$5008 line asymmetry.

\subsection{The equivalent width of the [\ion{O}{iii}] lines}
\label{sec:EWs}
Due to the uncertainty in the location and physical scale of the [\ion{O}{iii}]  emitting material 
and also the uncertainty in the location of the obscuring dust in heavily reddened quasars, the amount by which the [\ion{O}{iii}]  emission in our sample is attenuated is unknown. 
We can assume that the [\ion{O}{iii}] attenuation is at least zero, but no more than the attenuation of the continuum: the obscuring dust might be located entirely in front of the [\ion{O}{iii}]  emitting region, or behind the [\ion{O}{iii}]  emitting region but in front of the continuum source.  This unknown factor in the geometry of these objects leads to an uncertainty when deriving any measure of the intrinsic strength of the [\ion{O}{iii}]  emission.

With this uncertainty in mind, we derive the equivalent width (EW) of the [\ion{O}{iii}]\,$\lambda$5008 line under two different assumptions: that the emission line is reddened by the same amount as the continuum, and that the continuum is reddened but that the emission line itself is completely unobscured. The first case will produce the largest possible EW, and the second will give the smallest possible EW, as de-reddening the continuum while leaving the emission line unchanged increases the flux in the continuum and hence lowers the  EW. Any amount of `intermediate' de-reddening of the [\ion{O}{iii}] emission will produce an EW somewhere between these two extremes.

\begin{figure}
    \includegraphics[width=\columnwidth]{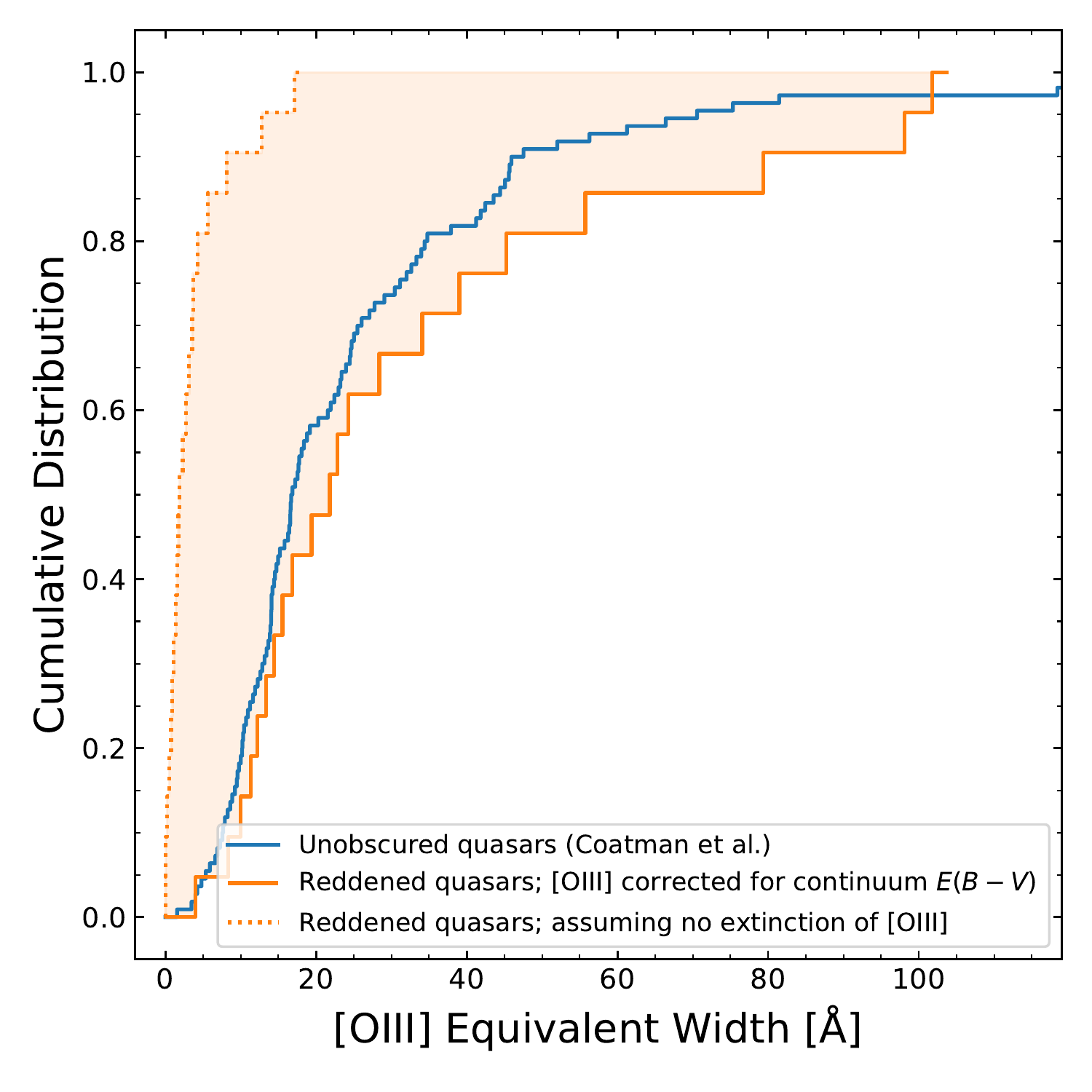}
    \caption{The [\ion{O}{iii}] EW distribution for our reddened quasars. When the continuum is corrected for extinction but the emission line is assumed to be completely unobscured (dotted orange), the \Oiii EW distribution is not consistent with that of the comparison sample (shown in blue). However, the \Oiii EWs in our reddened sample are consistent with having been drawn from the same distribution as that of the unobscured sample, if we assume that the emission line is subject to the same amount of extinction as the continuum (solid orange), suggesting that the obscuring dust is located outside the \Oiii emitting region.}
    \label{fig:EW_CDFs}
\end{figure}

In both cases, the cumulative distribution function (CDF) of the \Oiii EWs in our sample is formed and shown in Fig.~\ref{fig:EW_CDFs}. 
For the case where the [\ion{O}{iii}] emission is unobscured and so does not need to be corrected for extinction, we find that we can reject the null hypothesis with $p<0.001$, meaning that such a distribution of [\ion{O}{iii}] EWs is inconsistent with having been drawn from the same population as the unobscured quasars in \citet{Coatman19}. 

However, we are unable to reject the null hypothesis ($p>0.3$) in the case that the obscuring dust is located completely in front of the  [\ion{O}{iii}] emitting gas, meaning that our results are consistent with a scenario in which the intrinsic distribution of  [\ion{O}{iii}] EWs in our reddened sample is the same as that of the unobscured population, and where the amount of obscuration in our objects along the line of sight towards the [\ion{O}{iii}] emitting gas is is the same as the amount of obscuration along the line of sight to the continuum source.

\subsection{Degeneracy between the amount of ionised gas and the location of the reddening dust}
\label{sec:EWs2}
 When deriving any measure of the strength of the [\ion{O}{iii}] emission line, such as the equivalent width, there is a degeneracy between the intrinsic amount of line emission and the amount of obscuration along the line of sight towards the emitting material. The measured equivalent width of the emission line will always depend on both of these factors. For the heavily dust-reddened objects presented in this paper, we have no independent constraints on the location of the obscuring material - more specifically, the dust could lie inside or outside the [\ion{O}{iii}] emitting region, or could be co-located with it.
Due to this degeneracy,  we must make an assumption in order to draw any further conclusion about the relative strength of the ionised gas emission in reddened quasars compared to the unobscured population in the same redshift and luminosity range.

As the velocity widths of the  [\ion{O}{iii}]  lines, the continuum luminosities, and the  H\,$\alpha$ equivalent widths
 in our sample are consistent with having been drawn from the same distributions as those of the unobscured population, it is reasonable to assume that the amount of ionised gas in our reddened quasars is the same as in the unobscured population and so the intrinsic  [\ion{O}{iii}] emission is not significantly stronger or weaker in the reddened population.

\begin{figure}
    \includegraphics[width=\columnwidth]{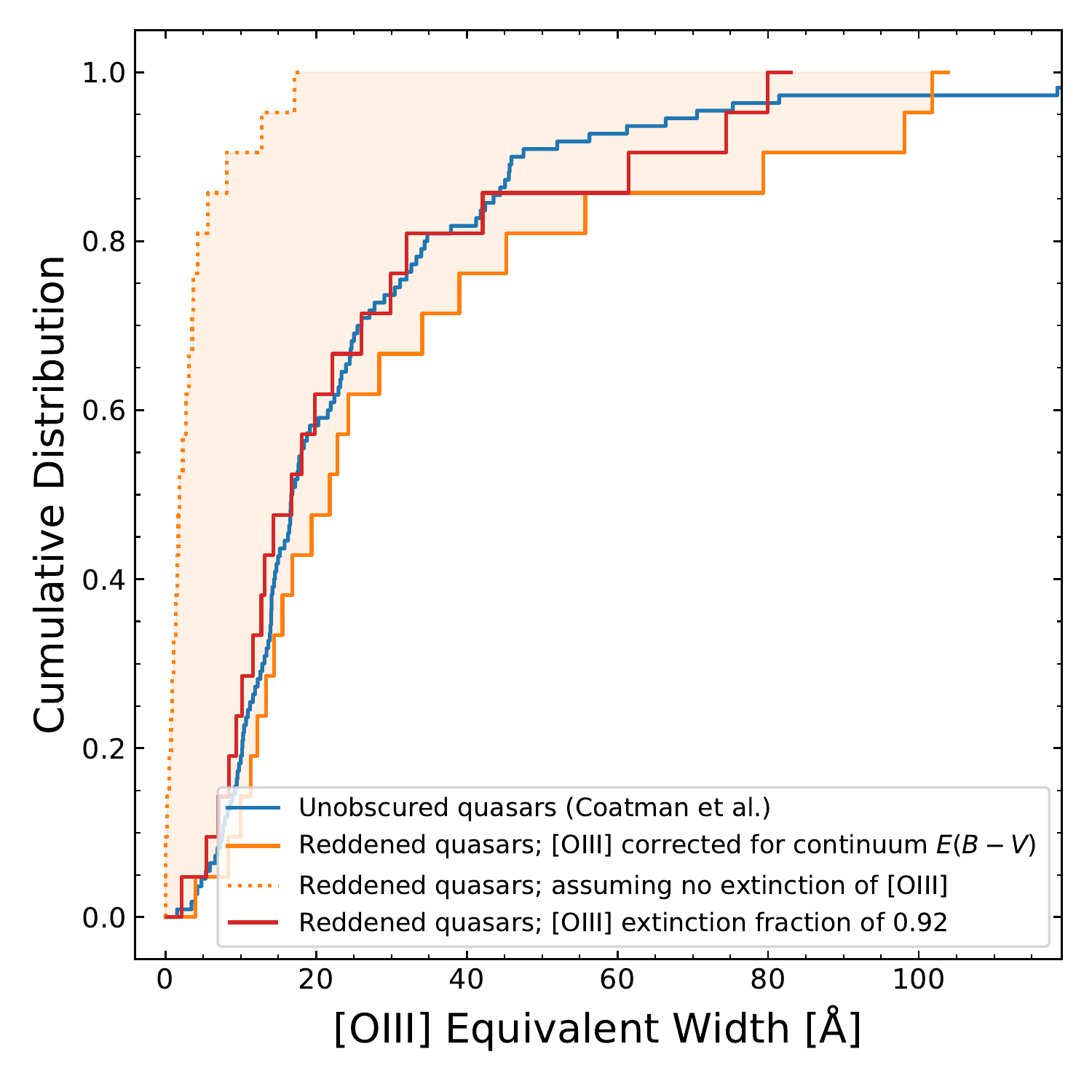}
    \caption{As Fig.~\ref{fig:EW_CDFs}, with (in red) the  CDF  when we assume that the intrinsic distribution of \Oiii EWs is the same in both the reddened and unobscured samples, and vary the amount of reddening required in the [\ion{O}{iii}] emission line to match the CDF of the unobscured sample. We find the `best fit' \Oiii extinction to be 92 per cent that of the continuum $E(B-V)$.}
    \label{fig:EW_CDF2}
\end{figure}

If we make this assumption, i.e.~that there is no difference in the distribution of the amount of ionised gas in the two populations and hence that the two populations have the same intrinsic [\ion{O}{iii}] EW distribution, then we can fit the observed CDF of our reddened sample to that of the unobscured population, as shown in Fig.~\ref{fig:EW_CDF2}.  We find that the amount of extinction required on average in the \Oiii lines of our reddened objects to best fit the EW distribution is 0.92 times the $E(B-V)$ derived from the near infra-red photometry in Section \ref{sec:HighzBHmass}:  on average, just 8 per cent of the continuum extinction arises inside the [\ion{O}{iii}] emitting region. However, it is worth reemphasising the result in Section~\ref{sec:EWs}, in that, under the assumption that the two populations have the same \Oiii~EW distribution, our reddened quasars are not inconsistent with having the obscuring material located entirely outside the [\ion{O}{iii}] emitting region.

We note that the two most luminous quasars in our reddened sample (ULAS~J1216-0313 and ULAS~J1234+0907) are the two most highly reddened objects and also have very weak \Oiii emission. Even if we correct the observed \Oiii for the largest possible extinction, the EWs in these two objects are still less than half the median EW in the rest of the reddened sample, suggesting  that the weakness of the observed \Oiii emission in these objects cannot be explained purely by obscuration, and that the \Oiii emission is intrinsically weak in these objects. 
This result is very similar to the fraction ($\sim10$ per cent) of unobscured quasars at these luminosities and redshifts which have very weak \Oiii emission \citep[i.e.~EW$<1$\AA;][]{Coatman19}, and we suggest that it is consistent with a scenario in which the most luminous quasars show weaker emission from the narrow line region \citep{Netzer04}, either by overionising or by physically removing the emitting gas.

\section{Discussion}
\label{sec:discuss}

\subsection{Ionised gas kinematics}

In Section \ref{sec:w80s}, we find no significant difference between the kinematics of the \Oiii emitting gas in the heavily reddened and unobscured quasar populations at $z>2$, as traced by either the velocity width or the asymmetry of the \Oiii line profile, although we do note that there would need to be a large dissimilarity between the two populations in order to find a statistically significant difference between our samples, due to the small sample sizes. Thus, while the kinematics of the \Oiii emission are not observed to be significantly different, we cannot rule out the possibility of smaller differences in the ionised gas emission properties of unobscured and heavily reddened quasars. We also show that the \Oiii velocity width tends to increase with increasing AGN luminosity in both populations, a trend that has also been seen in lower redshift AGN \citep{Zakamska14, Harrison16}.

\citet{Wu18} report the detection of \Oiii$\lambda\lambda$4960,5008 in W1136+4236, a \textit{WISE}-selected Hot Dust-Obscured Galaxy (HotDOG). W1136+4263 is at $z=2.41$, with a dust-corrected luminosity of $L_{5100}\approx 10^{46.5}$\ergps and an extinction of $E(B-V)=2.5$: while it is more reddened than the reddened quasars in our sample, it lies in the same luminosity and redshift range. This HotDOG has an \Oiii velocity width of $w_{80}\approx 2500\kmps$, placing it within the range spanned by our data in Fig.~\ref{fig:w80s}. It is hard to draw conclusions from this single object, which is the least obscured HotDOG studied by \citeauthor{Wu18}, but there is currently no evidence to suggest that the ionised gas kinematics in HotDOGs are any more extreme than those in our heavily reddened quasar population.
Similarly, the red quasars presented by \citet{Urrutia12} with $L_{\rm bol}\approx10^{46}$\ergps show \Oiii emission with the width of the broad component up to 1600\kmps \citep{Brusa15}, which coincides with the area of parameter space populated by our heavily reddened quasars.

Comparing to type 2 quasars, we note that the ionised gas kinematics in our reddened quasars are similar to the \Oiii velocity widths in the sample of type 2 quasars at $z<0.8$ studied by \citet{Zakamska14},
who find $w_{80}=280$-3000\kmps. These quasars have
$L_{\Oiii}= 10^{42-43.5}$\ergps. However, the broad line and continuum emitting regions in these objects are obscured by dust on nuclear scales which is unlikely to be obscuring the \Oiii emitting region, and so it is hard to directly compare the luminosities of these quasars with the objects in our sample.

Our results complement those of \citet{Harrison16}, who examined the \Oiii properties of an X-ray selected sample of optically unobscured AGN at $1.1<z<1.7$.
The AGN in their sample have  \Oiii $w_{80}$ in the range 200-1000\kmps, $L_{\Oiii} \approx 10^{42} \ergps$ and $L_X \approx 10^{44}$\ergps. They are therefore around an order of magnitude fainter than our heavily reddened quasars (Lansbury et al. in prep.) and also appear to have narrower \Oiii line profiles in comparison to our quasars. \citet{Harrison16} found no significant difference between the ionised gas kinematics in X-ray obscured and X-ray unobscured objects. Taken with our results, this supports a scenario where the most powerful ionised gas outflows in massive galaxies are driven by the luminosity of the central quasar, independently of the column density of any obscuring material along the line of sight to the observer.

By contrast, \citet{DiPompeo18} do find a difference in the \Oiii properties of red and blue colour-selected AGN, with the redder AGN having broader, more blueshifted \Oiii emission on average. However, the \citeauthor{DiPompeo18} AGN are significantly less luminous and at much lower redshifts than the quasars presented in this work. The two results are therefore not inconsistent and could indicate different redshift evolution in the outflow properties of reddened and unobscured AGN.

We note that the heavily reddened quasars in our sample do not show the same extreme \Oiii kinematics as the four rare (i.e. out of 97 ERQs; \citealt{Hamann17}) objects presented in \citet{Zakamska16}, which are in the same luminosity and redshift range as our heavily reddened quasars.
However, the \citeauthor{Zakamska16} quasars were also selected to have high equivalent width \ion{C}{iv} emission, and it is conceivable that the extreme \Oiii kinematics seen in that sample could be linked to this selection. Alternatively, \citet{Coatman19} find similarly rare (18/354) unobscured quasars with ionised gas kinematics that are almost as extreme, and so it could be that $\lesssim 5$\,per cent of the most luminous quasars at $z>2$ have ionised gas outflows of up to $5000$\kmps, irrespective of reddening, and that if we were to increase our sample of luminous heavily dust-reddened quasars by a factor of $\sim5$ we might also observe such extreme outflows.

If the objects in our reddened sample do indeed represent the `blow-out' phase for luminous quasars at redshifts $z\approx2$, then, as the signatures of fast outflows are also seen in the unobscured population at the same redshift, our results suggest that the outflowing gas persists  after the obscuring dust has been cleared from the line of sight. In particular, the asymmetries and large velocity widths seen in the \Oiii emission in our comparison sample of unobscured quasars would be due to outflows which have already cleared any obscuring material away from the line of sight to the continuum source.

\subsection{Obscuring dust}

In Sections \ref{sec:EWs} and \ref{sec:EWs2}, we found that the obscuring dust in our high redshift sample of heavily reddened quasars was most likely to be located almost entirely outside the \Oiii emitting region, suggesting that the obscuration in heavily reddened quasars arises on kiloparsec scales. Using data from ALMA, \citet{Banerji18} find large (i.e.~galaxy-scale) dust emitting regions in at least some heavily reddened quasars, consistent with this result. Deeper observations providing spatially resolved information about either the location of the dust or the location of the ionised gas are necessary to place firmer constraints on the physical scales and covering factor of the dust relative to the ionised gas in our sample.

We note that our results could be interpreted with a model whereby our reddened quasars are in fact unobscured quasars which are viewed along a line of sight that is skimming the plane of the host galaxy. However, we disfavour such a  model as it would (a) require a significant disc component to the host galaxy, which would not be consistent with a merger-driven AGN triggering event, and (b) require the \Oiii emitting gas to be located very close in to the centre of the galaxy, in contrast  with local AGN (see Section \ref{sec:intro}).

Our results are consistent with a model of galaxy formation and evolution where outflows persist long after the obscuring dust has cleared from the line of sight. There are also  examples of unobscured, UV-luminous quasars at high redshifts with significant amounts of cool dust inferred from their FIR and sub-millimeter continuum emission \citep[][]{Harris16, Pitchford16}. The general consensus is that this cool dust is being heated by star formation in the quasar host galaxy rather than the quasar itself (although see \citealt{Symeonidis17} for an alternative explanation). Thus, star formation in luminous quasar hosts can also persist after the dust has been cleared from the line of sight. Using a population of SDSS quasars, \citet{Maddox17} find that strong FIR emission (and therefore presumably star formation) is associated with strong nuclear outflows as traced by large \ion{C}{iv} blueshifts. Taken together with the results of \citet{Coatman19}, where a correlation between the blueshift of \Oiii and the blueshift of the \ion{C}{iv} emission line is seen, this implies that FIR-luminous quasars also exhibit stronger galaxy-scale outflows compared to their FIR-faint counterparts. The FIR bright quasars studied by \citet{Maddox17} are not reddened, but outflows and starburst activity are nevertheless being seen in this population. There is likely to be a considerable overlap between star formation, outflows, and the presence and clearing of dust in massive (active) galaxies, and obtaining observations at multiple wavelengths of the same objects in order to quantify  these processes is the only way to move to a complete understanding of the interplay between SMBH accretion, outflows, and star formation, and their impact on galaxies at $z>2$.

\section{Summary}

From 26 candidates, selected by virtue of their red colours in the near infra-red, we have spectroscopically confirmed 25 new heavily dust-reddened broad line quasars at $0.7<z<2.6$, which extend the bright and faint ends of the luminosity function presented in \citet{Banerji15}. 

By combining with the objects presented in \citet{Banerji12}, we construct a sample of 22 heavily reddened quasars with  extinctions $0.5<E(B-V)<2.0$,  luminosities $45<{\rm Log}(L_{5100} /\erg\s^{-1})<48$, redshifts $2.0<z<2.6$,
and spectra covering the H\,$\alpha$, H\,$\beta$, and [\ion{O}{iii}] emission lines. Using the \Oiii line, we derive parameters describing the ionised gas kinematics and compare to a large sample of 111 unobscured quasars in the same redshift and luminosity range. This is the first time that the ionised gas kinematics in the reddened and unobscured quasar populations have been systematically compared at redshifts $z>2$.  Our main results are:

\smallskip
\noindent
(i)\, There is no significant difference between the two populations in the velocity widths or asymmetries of their \Oiii lines, suggesting that ionised gas outflows are moving, on average, no faster or slower in the reddened population than in the unobscured population.

\medskip
\noindent
(ii)\, As our samples are drawn from the same luminosity and redshift range, this suggests that the outflow driving mechanisms are not significantly different in the reddened and unobscured populations.

\medskip
\noindent
(iii)\, The \Oiii equivalent width distributions in the two populations are consistent with a scenario in which the reddened quasars have the same strength of ionised gas emission as the unobscured quasars, and the \Oiii emission in the reddened objects is subject to the same amount of extinction as the continuum. In other words, the obscuring dust is most probably located outside the \Oiii emitting region on kpc-scale distances from the central SMBH.

\medskip

We suggest that our results are best explained by a model in which, following a starburst episode,  quasar-driven ionised gas outflows persist for some time after any obscuring dust has been cleared from the line of sight. Such an episode of star formation would most likely be triggered by a major galaxy merger, which would also provide fuel for black hole accretion and hence the trigger for quasar activity.

\begin{table*}
 \centering
\caption{\small Derived properties from our SINFONI observations of the 12 new quasars at $z<2$.  Black hole mass estimates have uncertainties of $\sim0.4$ dex.}
\begin{tabular}{l |c  c  c c}
\hline
Name & Redshift & $(H-K)_{\rm{AB, SINFONI}}$  & $E(B-V)$ & $\textrm{Log}_{10}(\textrm{M}_{BH}/\Msun)$ \\
\hline
ULAS J1415+0836 & 1.120 & 1.73  & 3.0 & 9.03 \\
VIK J2214-3100 & 1.069 & 1.48  & 2.5 & 8.80 \\
VIK J2230-2956 & 1.319 & 1.04  & 2.0 & 8.63 \\
VIK J2238-2836 & 1.231 & 0.87  & 1.4 & 8.99 \\
VIK J2241-3006 & 0.720 & 1.20  & 2.0 & 7.56 \\
VIK J2245-3516 & 1.335 & 1.66  & 3.0 & 9.59 \\
VIK J2251-3433 & 1.693 & 1.05  & 1.9 & 9.23 \\
VIK J2258-3219 & 0.879 & 0.84  & 1.3 & 7.52 \\
VIK J2306-3050 & 1.060 & 1.07  & 1.8 & 8.34 \\
ULAS J2312+0454 & 0.700 & 1.36 & 2.4 & 8.24 \\
VIK J2313-2904 & 1.851 & 1.10  & 1.4 & 9.58 \\
VIK J2357-3024 & 1.129 & 1.36  & 2.2 & 8.66 \\
\hline
\end{tabular}
\label{tab:lowz}
\end{table*}

\begin{table*}
 \centering
\caption{\small Derived properties for the 22  quasars at $z>2$ for which we have coverage of H\,$\alpha$, H\,$\beta$, and [\ion{O}{iii}]. Redshifts are taken from the broad component of the fit to the H\,$\alpha$ line. 
The $E(B-V)$ is estimated from the near infra-red photometry, and has an uncertainty of $\sim 0.15$ magnitudes.
Rest frame 5100\,\AA\ luminosities have been corrected for dust extinction using the $E(B-V)$, giving rise to an uncertainty of $\sim 0.2$ dex. Black hole masses are derived using the prescription from \citet{vp06}, and have an uncertainty of $\sim 0.4$ dex.
Eddington ratios are estimated assuming $L = 8 \times L_{5100}$, and have an associated uncertainty of a factor of 4.
H\,$\alpha$ equivalent widths are given in the rest frame.
[\ion{O}{iii}]\,$\lambda$5008 line luminosities  are given in the observed frame with no correction for extinction. 80 per cent velocity widths have been corrected for instrumental broadening of $\sigma=200\kmps$. 
\newline
*Indicates that object is consistent with having no [\ion{O}{iii}] emission, and an upper bound on the line luminosity has been estimated from a template fit.}
\begin{tabular}{l | c  c c c c c c c c}
\hline
Name & Redshift  & $E(B-V)$ &  $\textrm{Log}_{10}(\textrm{M}_{BH})$ & $\textrm{Log}_{10}(L_{5100})$ & $L/L_{\rm Edd}$ & EW(H\,$\alpha$)  & $\textrm{Log}_{10}(L_{[\ion{O}{iii}]})$& [\ion{O}{iii}]~$w_{80}$  \\
 &  &  [magnitudes] & [\Msun] & [\ergps] & & [\AA] & [\ergps] & [\kmps]  \\
\hline
ULAS J0016-0038 & 2.176 & 0.55 & 9.20 & 45.95 & 0.36 & 883 & 43.22$\pm$0.07 & 730$\pm$131 \\
ULAS J0041-0021 & 2.510 & 0.81 & 9.52 & 46.52 & 0.64 & 281 & 43.18$\pm$0.06 & 868$\pm$101 \\
ULAS J0141+0101 & 2.562 & 0.58 & 9.43 & 45.97 & 0.22 & 490 & 43.79$\pm$0.05 & 2091$\pm$83 \\
ULAS J0144-0014 & 2.504 & 0.68 & 9.58 & 46.40 & 0.42 & 286 & - & -\\
ULAS J0144+0036 & 2.281 & 0.74 & 9.65 & 46.31 & 0.29 & 377 & 43.31$\pm$0.06 & 1880$\pm$89 \\
ULAS J0221-0019 & 2.248 & 0.58 & 8.86 & 45.92 & 0.73 & 656 & 43.24$\pm$0.06 & 1093$\pm$108 \\
VHS J1117-1528 & 2.428 & 1.25 & 9.52 & 46.65 & 0.87 & 342 & 43.16$\pm$0.06 & 2277$\pm$227 \\
VHS J1122-1919 & 2.464 & 1.09 & 9.24 & 46.82 & 2.39 & 513 & 43.81$\pm$0.05 & 1775$\pm$130 \\
ULAS J1216-0313 & 2.574 & 1.62 & 10.07 & 47.51 & 1.73 & 229 & $<42.60$* & -\\
ULAS J1234+0907 & 2.502 & 1.94 & 10.48 & 47.64 & 0.91 & 432 & $<42.20$* & -\\
VHS J1301-1624 & 2.138 & 1.24 & 9.40 & 46.59 & 0.99 & 364 & 43.41$\pm$0.06 & 2898$\pm$241 \\
ULAS J2200+0056 & 2.541 & 0.55 & 9.10 & 46.47 & 1.49 & 461 & 44.15$\pm$0.05 & 1729$\pm$76 \\
VIK J2205-3132 & 2.307 & 0.58 & 8.79 & 45.60 & 0.41 & 485 & 42.71$\pm$0.07 & 772$\pm$96 \\
ULAS J2224-0015 & 2.219 & 0.61 & 8.98 & 46.04 & 0.73 & 624 & 43.20$\pm$0.07 & 1292$\pm$233 \\
VIK J2228-3205 & 2.364 & 0.57 & 8.48 & 45.19 & 0.32 & 560 & 43.24$\pm$0.05 & 624$\pm$28 \\
VIK J2232-2844 & 2.292 & 0.65 & 9.02 & 45.43 & 0.16 & 762 & 42.47$\pm$0.09 & 3161$\pm$1062 \\
VIK J2243-3504 & 2.085 & 1.05 & 8.99 & 46.27 & 1.21 & 445 & 43.68$\pm$0.05 &  1710$\pm$75 \\
VIK J2256-3114 & 2.329 & 0.69 & 9.16 & 45.77 & 0.26 & 394 & 43.18$\pm$0.05 & 1702$\pm$103 \\
VIK J2309-3433 & 2.159 & 0.52 & 9.39 & 45.03 & 0.03 & 647 & 42.54$\pm$0.06 & 315$\pm$51 \\
VIK J2314-3459 & 2.325 & 0.53 & 8.41 & 45.21 & 0.40 & 414 & 43.32$\pm$0.05 & 2572$\pm$341 \\
VIK J2323-3222 & 2.191 & 0.99 & 9.09 & 46.18 & 0.79 & 475 & 42.98$\pm$0.06 & 979$\pm$273 \\
VIK J2350-3019 & 2.324 & 0.58 & 8.58 & 45.57 & 0.62 & 337 & 43.03$\pm$0.05 & 1059$\pm$59 \\
\hline
\end{tabular}
\label{tab:Oiii_objects}
\end{table*}

\section*{Acknowledgements}


We thank George Lansbury and Richard McMahon for useful discussions.
An anonymous referee provided a comprehensive review of the paper for which we are grateful.
MJT thanks the Science and Technology Facilities Council (STFC) for the award of a studentship. MB  acknowledges  funding  from  the Royal Society via a University Research Fellowship. PCH acknowledges  funding from  STFC  via  the  Institute  of  Astronomy, Cambridge, Consolidated Grant. CP thanks the Alexander von Humboldt Foundation for the granting of a Bessel Research Award held at MPA. CP is also grateful to the ESO and the DFG cluster of excellence `Origin and Structure of the Universe' for support.

This work is based on observations collected at the European Southern Observatory under ESO programme 095.A-0094(A), and made use of Astropy \citep{astropy:2013, astropy:2018} and Matplotlib \citep{Hunter:2007}. 




\bibliographystyle{mnras}
\bibliography{Paper_refs}

\begin{thebibliography}{}
\makeatletter
\relax
\def\mn@urlcharsother{\let\do\@makeother \do\$\do\&\do\#\do\^\do\_\do\%\do\~}
\def\mn@doi{\begingroup\mn@urlcharsother \@ifnextchar [ {\mn@doi@}
  {\mn@doi@[]}}
\def\mn@doi@[#1]#2{\def\@tempa{#1}\ifx\@tempa\@empty \href
  {http://dx.doi.org/#2} {doi:#2}\else \href {http://dx.doi.org/#2} {#1}\fi
  \endgroup}
\def\mn@eprint#1#2{\mn@eprint@#1:#2::\@nil}
\def\mn@eprint@arXiv#1{\href {http://arxiv.org/abs/#1} {{\tt arXiv:#1}}}
\def\mn@eprint@dblp#1{\href {http://dblp.uni-trier.de/rec/bibtex/#1.xml}
  {dblp:#1}}
\def\mn@eprint@#1:#2:#3:#4\@nil{\def\@tempa {#1}\def\@tempb {#2}\def\@tempc
  {#3}\ifx \@tempc \@empty \let \@tempc \@tempb \let \@tempb \@tempa \fi \ifx
  \@tempb \@empty \def\@tempb {arXiv}\fi \@ifundefined
  {mn@eprint@\@tempb}{\@tempb:\@tempc}{\expandafter \expandafter \csname
  mn@eprint@\@tempb\endcsname \expandafter{\@tempc}}}

\bibitem[\protect\citeauthoryear{Anderson \& Darling}{Anderson \&
  Darling}{1952}]{Anderson52}
Anderson T.~W.,  Darling D.~A.,  1952, \mn@doi [Ann. Math. Statist.]
  {10.1214/aoms/1177729437}, 23, 193

\bibitem[\protect\citeauthoryear{{Astropy Collaboration} et~al.,}{{Astropy
  Collaboration} et~al.}{2013}]{astropy:2013}
{Astropy Collaboration} et~al., 2013, \mn@doi [\aap]
  {10.1051/0004-6361/201322068}, \href
  {http://adsabs.harvard.edu/abs/2013A%26A...558A..33A} {558, A33}

\bibitem[\protect\citeauthoryear{{Banerji}, {McMahon}, {Hewett},
  {Alaghband-Zadeh}, {Gonzalez-Solares}, {Venemans}  \& {Hawthorn}}{{Banerji}
  et~al.}{2012}]{Banerji12}
{Banerji} M.,  {McMahon} R.~G.,  {Hewett} P.~C.,  {Alaghband-Zadeh} S.,
  {Gonzalez-Solares} E.,  {Venemans} B.~P.,   {Hawthorn} M.~J.,  2012, \mn@doi
  [\mnras] {10.1111/j.1365-2966.2012.22099.x}, \href
  {http://adsabs.harvard.edu/abs/2012MNRAS.427.2275B} {427, 2275}

\bibitem[\protect\citeauthoryear{{Banerji}, {McMahon}, {Hewett},
  {Gonzalez-Solares}  \& {Koposov}}{{Banerji} et~al.}{2013}]{Banerji13}
{Banerji} M.,  {McMahon} R.~G.,  {Hewett} P.~C.,  {Gonzalez-Solares} E.,
  {Koposov} S.~E.,  2013, \mn@doi [\mnras] {10.1093/mnrasl/sls023}, \href
  {http://adsabs.harvard.edu/abs/2013MNRAS.429L..55B} {429, L55}

\bibitem[\protect\citeauthoryear{{Banerji}, {Alaghband-Zadeh}, {Hewett}  \&
  {McMahon}}{{Banerji} et~al.}{2015}]{Banerji15}
{Banerji} M.,  {Alaghband-Zadeh} S.,  {Hewett} P.~C.,   {McMahon} R.~G.,  2015,
  \mn@doi [\mnras] {10.1093/mnras/stu2649}, \href
  {http://adsabs.harvard.edu/abs/2015MNRAS.447.3368B} {447, 3368}

\bibitem[\protect\citeauthoryear{{Banerji}, {Carilli}, {Jones}, {Wagg},
  {McMahon}, {Hewett}, {Alaghband-Zadeh}  \& {Feruglio}}{{Banerji}
  et~al.}{2017}]{Banerji17}
{Banerji} M.,  {Carilli} C.~L.,  {Jones} G.,  {Wagg} J.,  {McMahon} R.~G.,
  {Hewett} P.~C.,  {Alaghband-Zadeh} S.,   {Feruglio} C.,  2017, \mn@doi
  [\mnras] {10.1093/mnras/stw3019}, \href
  {http://adsabs.harvard.edu/abs/2017MNRAS.465.4390B} {465, 4390}

\bibitem[\protect\citeauthoryear{{Banerji}, {Jones}, {Wagg}, {Carilli},
  {Bisbas}  \& {Hewett}}{{Banerji} et~al.}{2018}]{Banerji18}
{Banerji} M.,  {Jones} G.~C.,  {Wagg} J.,  {Carilli} C.~L.,  {Bisbas} T.~G.,
  {Hewett} P.~C.,  2018, \mn@doi [\mnras] {10.1093/mnras/sty1443}, \href
  {http://adsabs.harvard.edu/abs/2018MNRAS.479.1154B} {479, 1154}

\bibitem[\protect\citeauthoryear{{Baron} \& {Netzer}}{{Baron} \&
  {Netzer}}{2019}]{Baron19}
{Baron} D.,  {Netzer} H.,  2019, \mn@doi [\mnras] {10.1093/mnras/sty2935},
  \href {http://adsabs.harvard.edu/abs/2019MNRAS.482.3915B} {482, 3915}

\bibitem[\protect\citeauthoryear{{Baskin} \& {Laor}}{{Baskin} \&
  {Laor}}{2005}]{Baskin05}
{Baskin} A.,  {Laor} A.,  2005, \mn@doi [\mnras]
  {10.1111/j.1365-2966.2005.08841.x}, \href
  {http://adsabs.harvard.edu/abs/2005MNRAS.358.1043B} {358, 1043}

\bibitem[\protect\citeauthoryear{{Bischetti} et~al.,}{{Bischetti}
  et~al.}{2017}]{Bischetti17}
{Bischetti} M.,  et~al., 2017, \mn@doi [\aap] {10.1051/0004-6361/201629301},
  \href {http://adsabs.harvard.edu/abs/2017A%26A...598A.122B} {598, A122}

\bibitem[\protect\citeauthoryear{{Boroson} \& {Green}}{{Boroson} \&
  {Green}}{1992}]{BG92}
{Boroson} T.~A.,  {Green} R.~F.,  1992, \mn@doi [\apjs] {10.1086/191661}, \href
  {http://adsabs.harvard.edu/abs/1992ApJS...80..109B} {80, 109}

\bibitem[\protect\citeauthoryear{{Brusa} et~al.,}{{Brusa}
  et~al.}{2015}]{Brusa15}
{Brusa} M.,  et~al., 2015, \mn@doi [\mnras] {10.1093/mnras/stu2117}, \href
  {http://adsabs.harvard.edu/abs/2015MNRAS.446.2394B} {446, 2394}

\bibitem[\protect\citeauthoryear{{Coatman}, {Hewett}, {Banerji}, {Richards},
  {Hennawi}  \& {Prochaska}}{{Coatman} et~al.}{2017}]{Coatman17}
{Coatman} L.,  {Hewett} P.~C.,  {Banerji} M.,  {Richards} G.~T.,  {Hennawi}
  J.~F.,   {Prochaska} J.~X.,  2017, \mn@doi [\mnras] {10.1093/mnras/stw2797},
  \href {http://adsabs.harvard.edu/abs/2017MNRAS.465.2120C} {465, 2120}

\bibitem[\protect\citeauthoryear{Coatman, Hewett, Banerji, Richards, Hennawi
  \& Xavier~Prochaska}{Coatman et~al.}{2019}]{Coatman19}
Coatman L.,  Hewett P.~C.,  Banerji M.,  Richards G.~T.,  Hennawi J.~F.,
  Xavier~Prochaska J.,  2019, \mn@doi [\mnras] {10.1093/mnras/stz1167}

\bibitem[\protect\citeauthoryear{Darling}{Darling}{1957}]{Darling57}
Darling D.~A.,  1957, \mn@doi [Ann. Math. Statist.] {10.1214/aoms/1177706788},
  28, 823

\bibitem[\protect\citeauthoryear{{Di Matteo}, {Springel}  \& {Hernquist}}{{Di
  Matteo} et~al.}{2005}]{DiMatteo05}
{Di Matteo} T.,  {Springel} V.,   {Hernquist} L.,  2005, \mn@doi [\nat]
  {10.1038/nature03335}, \href
  {http://adsabs.harvard.edu/abs/2005Natur.433..604D} {433, 604}

\bibitem[\protect\citeauthoryear{{DiPompeo}, {Hickox}, {Carroll}, {Runnoe},
  {Mullaney}  \& {Fischer}}{{DiPompeo} et~al.}{2018}]{DiPompeo18}
{DiPompeo} M.~A.,  {Hickox} R.~C.,  {Carroll} C.~M.,  {Runnoe} J.~C.,
  {Mullaney} J.~R.,   {Fischer} T.~C.,  2018, \mn@doi [\apj]
  {10.3847/1538-4357/aab365}, \href
  {http://adsabs.harvard.edu/abs/2018ApJ...856...76D} {856, 76}

\bibitem[\protect\citeauthoryear{{Dopita} \& {Sutherland}}{{Dopita} \&
  {Sutherland}}{2003}]{BalmerDecValue}
{Dopita} M.~A.,  {Sutherland} R.~S.,  2003, {Astrophysics of the diffuse
  universe}.
Springer, Astronomy and astrophysics library, ISBN 3540433627

\bibitem[\protect\citeauthoryear{{Edge}, {Sutherland}, {Kuijken}, {Driver},
  {McMahon}, {Eales}  \& {Emerson}}{{Edge} et~al.}{2013}]{Edge13}
{Edge} A.,  {Sutherland} W.,  {Kuijken} K.,  {Driver} S.,  {McMahon} R.,
  {Eales} S.,   {Emerson} J.~P.,  2013, The Messenger, \href
  {http://adsabs.harvard.edu/abs/2013Msngr.154...32E} {154, 32}

\bibitem[\protect\citeauthoryear{{Fabian}}{{Fabian}}{2012}]{Fabian12}
{Fabian} A.~C.,  2012, \mn@doi [\araa] {10.1146/annurev-astro-081811-125521},
  \href {http://adsabs.harvard.edu/abs/2012ARA%26A..50..455F} {50, 455}

\bibitem[\protect\citeauthoryear{{Greene} \& {Ho}}{{Greene} \&
  {Ho}}{2005}]{GreeneHo}
{Greene} J.~E.,  {Ho} L.~C.,  2005, \mn@doi [\apj] {10.1086/431897}, \href
  {http://adsabs.harvard.edu/abs/2005ApJ...630..122G} {630, 122}

\bibitem[\protect\citeauthoryear{{Greene}, {Zakamska}, {Ho}  \&
  {Barth}}{{Greene} et~al.}{2011}]{Greene11}
{Greene} J.~E.,  {Zakamska} N.~L.,  {Ho} L.~C.,   {Barth} A.~J.,  2011, \mn@doi
  [\apj] {10.1088/0004-637X/732/1/9}, \href
  {http://adsabs.harvard.edu/abs/2011ApJ...732....9G} {732, 9}

\bibitem[\protect\citeauthoryear{{Greene}, {Zakamska}  \& {Smith}}{{Greene}
  et~al.}{2012}]{Greene12}
{Greene} J.~E.,  {Zakamska} N.~L.,   {Smith} P.~S.,  2012, \mn@doi [\apj]
  {10.1088/0004-637X/746/1/86}, \href
  {http://adsabs.harvard.edu/abs/2012ApJ...746...86G} {746, 86}

\bibitem[\protect\citeauthoryear{{Hamann} et~al.,}{{Hamann}
  et~al.}{2017}]{Hamann17}
{Hamann} F.,  et~al., 2017, \mn@doi [\mnras] {10.1093/mnras/stw2387}, \href
  {http://adsabs.harvard.edu/abs/2017MNRAS.464.3431H} {464, 3431}

\bibitem[\protect\citeauthoryear{{Harris} et~al.,}{{Harris}
  et~al.}{2016}]{Harris16}
{Harris} K.,  et~al., 2016, \mn@doi [\mnras] {10.1093/mnras/stw286}, \href
  {http://adsabs.harvard.edu/abs/2016MNRAS.457.4179H} {457, 4179}

\bibitem[\protect\citeauthoryear{{Harrison}}{{Harrison}}{2017}]{Harrison17}
{Harrison} C.~M.,  2017, \mn@doi [Nature Astronomy] {10.1038/s41550-017-0165},
  \href {http://adsabs.harvard.edu/abs/2017NatAs...1E.165H} {1, 0165}

\bibitem[\protect\citeauthoryear{{Harrison}, {Alexander}, {Mullaney}  \&
  {Swinbank}}{{Harrison} et~al.}{2014}]{Harrison14}
{Harrison} C.~M.,  {Alexander} D.~M.,  {Mullaney} J.~R.,   {Swinbank} A.~M.,
  2014, \mn@doi [\mnras] {10.1093/mnras/stu515}, \href
  {http://adsabs.harvard.edu/abs/2014MNRAS.441.3306H} {441, 3306}

\bibitem[\protect\citeauthoryear{{Harrison} et~al.,}{{Harrison}
  et~al.}{2016}]{Harrison16}
{Harrison} C.~M.,  et~al., 2016, \mn@doi [\mnras] {10.1093/mnras/stv2727},
  \href {http://adsabs.harvard.edu/abs/2016MNRAS.456.1195H} {456, 1195}

\bibitem[\protect\citeauthoryear{{Harrison}, {Costa}, {Tadhunter},
  {Fl{\"u}tsch}, {Kakkad}, {Perna}  \& {Vietri}}{{Harrison}
  et~al.}{2018}]{Harrison18}
{Harrison} C.~M.,  {Costa} T.,  {Tadhunter} C.~N.,  {Fl{\"u}tsch} A.,  {Kakkad}
  D.,  {Perna} M.,   {Vietri} G.,  2018, \mn@doi [Nature Astronomy]
  {10.1038/s41550-018-0403-6}, \href
  {http://adsabs.harvard.edu/abs/2018NatAs...2..198H} {2, 198}

\bibitem[\protect\citeauthoryear{{Hopkins}, {Hernquist}, {Cox}  \& {Kere{\v
  s}}}{{Hopkins} et~al.}{2008}]{Hopkins08}
{Hopkins} P.~F.,  {Hernquist} L.,  {Cox} T.~J.,   {Kere{\v s}} D.,  2008,
  \mn@doi [\apjs] {10.1086/524362}, \href
  {http://adsabs.harvard.edu/abs/2008ApJS..175..356H} {175, 356}

\bibitem[\protect\citeauthoryear{Hunter}{Hunter}{2007}]{Hunter:2007}
Hunter J.~D.,  2007, \mn@doi [Computing In Science \& Engineering]
  {10.1109/MCSE.2007.55}, 9, 90

\bibitem[\protect\citeauthoryear{{Ishibashi} \& {Fabian}}{{Ishibashi} \&
  {Fabian}}{2015}]{Ishibashi15}
{Ishibashi} W.,  {Fabian} A.~C.,  2015, \mn@doi [\mnras]
  {10.1093/mnras/stv944}, \href
  {http://adsabs.harvard.edu/abs/2015MNRAS.451...93I} {451, 93}

\bibitem[\protect\citeauthoryear{{Ishibashi}, {Banerji}  \&
  {Fabian}}{{Ishibashi} et~al.}{2017}]{Ishibashi17}
{Ishibashi} W.,  {Banerji} M.,   {Fabian} A.~C.,  2017, \mn@doi [\mnras]
  {10.1093/mnras/stx921}, \href
  {http://adsabs.harvard.edu/abs/2017MNRAS.469.1496I} {469, 1496}

\bibitem[\protect\citeauthoryear{{Karouzos}, {Woo}  \& {Bae}}{{Karouzos}
  et~al.}{2016}]{Karouzos16}
{Karouzos} M.,  {Woo} J.-H.,   {Bae} H.-J.,  2016, \mn@doi [\apj]
  {10.3847/0004-637X/819/2/148}, \href
  {http://adsabs.harvard.edu/abs/2016ApJ...819..148K} {819, 148}

\bibitem[\protect\citeauthoryear{{Kim}, {Im}  \& {Kim}}{{Kim}
  et~al.}{2010}]{Kim10}
{Kim} D.,  {Im} M.,   {Kim} M.,  2010, \mn@doi [\apj]
  {10.1088/0004-637X/724/1/386}, \href
  {http://adsabs.harvard.edu/abs/2010ApJ...724..386K} {724, 386}

\bibitem[\protect\citeauthoryear{{Konigl} \& {Kartje}}{{Konigl} \&
  {Kartje}}{1994}]{Konigl94}
{Konigl} A.,  {Kartje} J.~F.,  1994, \mn@doi [\apj] {10.1086/174746}, \href
  {http://adsabs.harvard.edu/abs/1994ApJ...434..446K} {434, 446}

\bibitem[\protect\citeauthoryear{{Korista} \& {Goad}}{{Korista} \&
  {Goad}}{2004}]{Korista04}
{Korista} K.~T.,  {Goad} M.~R.,  2004, \mn@doi [\apj] {10.1086/383193}, \href
  {http://adsabs.harvard.edu/abs/2004ApJ...606..749K} {606, 749}

\bibitem[\protect\citeauthoryear{{Kova{\v c}evi{\'c}}, {Popovi{\'c}}  \&
  {Dimitrijevi{\'c}}}{{Kova{\v c}evi{\'c}} et~al.}{2010}]{Kovacevic10}
{Kova{\v c}evi{\'c}} J.,  {Popovi{\'c}} L.~{\v C}.,   {Dimitrijevi{\'c}} M.~S.,
   2010, \mn@doi [\apjs] {10.1088/0067-0049/189/1/15}, \href
  {http://adsabs.harvard.edu/abs/2010ApJS..189...15K} {189, 15}

\bibitem[\protect\citeauthoryear{{Kulkarni}, {Worseck}  \&
  {Hennawi}}{{Kulkarni} et~al.}{2018}]{Kulkarni18}
{Kulkarni} G.,  {Worseck} G.,   {Hennawi} J.~F.,  2018, preprint, \href
  {http://adsabs.harvard.edu/abs/2018arXiv180709774K} {} (\mn@eprint {arXiv}
  {1807.09774})

\bibitem[\protect\citeauthoryear{{Liu}, {Zakamska}, {Greene}, {Nesvadba}  \&
  {Liu}}{{Liu} et~al.}{2013}]{Liu13}
{Liu} G.,  {Zakamska} N.~L.,  {Greene} J.~E.,  {Nesvadba} N.~P.~H.,   {Liu} X.,
   2013, \mn@doi [\mnras] {10.1093/mnras/stt1755}, \href
  {http://adsabs.harvard.edu/abs/2013MNRAS.436.2576L} {436, 2576}

\bibitem[\protect\citeauthoryear{{Madau} \& {Dickinson}}{{Madau} \&
  {Dickinson}}{2014}]{MD14}
{Madau} P.,  {Dickinson} M.,  2014, \mn@doi [\araa]
  {10.1146/annurev-astro-081811-125615}, \href
  {http://adsabs.harvard.edu/abs/2014ARA%26A..52..415M} {52, 415}

\bibitem[\protect\citeauthoryear{{Maddox}, {Hewett}, {Warren}  \&
  {Croom}}{{Maddox} et~al.}{2008}]{Maddox08}
{Maddox} N.,  {Hewett} P.~C.,  {Warren} S.~J.,   {Croom} S.~M.,  2008, \mn@doi
  [\mnras] {10.1111/j.1365-2966.2008.13138.x}, \href
  {http://adsabs.harvard.edu/abs/2008MNRAS.386.1605M} {386, 1605}

\bibitem[\protect\citeauthoryear{{Maddox} et~al.,}{{Maddox}
  et~al.}{2017}]{Maddox17}
{Maddox} N.,  et~al., 2017, \mn@doi [\mnras] {10.1093/mnras/stx1416}, \href
  {http://adsabs.harvard.edu/abs/2017MNRAS.470.2314M} {470, 2314}

\bibitem[\protect\citeauthoryear{{Magorrian} et~al.,}{{Magorrian}
  et~al.}{1998}]{Magorrian98}
{Magorrian} J.,  et~al., 1998, \mn@doi [\aj] {10.1086/300353}, \href
  {http://adsabs.harvard.edu/abs/1998AJ....115.2285M} {115, 2285}

\bibitem[\protect\citeauthoryear{{McConnell} \& {Ma}}{{McConnell} \&
  {Ma}}{2013}]{McConnell13}
{McConnell} N.~J.,  {Ma} C.-P.,  2013, \mn@doi [\apj]
  {10.1088/0004-637X/764/2/184}, \href
  {http://adsabs.harvard.edu/abs/2013ApJ...764..184M} {764, 184}

\bibitem[\protect\citeauthoryear{{Murray}, {Chiang}, {Grossman}  \&
  {Voit}}{{Murray} et~al.}{1995}]{Murray95}
{Murray} N.,  {Chiang} J.,  {Grossman} S.~A.,   {Voit} G.~M.,  1995, \mn@doi
  [\apj] {10.1086/176238}, \href
  {http://adsabs.harvard.edu/abs/1995ApJ...451..498M} {451, 498}

\bibitem[\protect\citeauthoryear{{Narayanan} et~al.,}{{Narayanan}
  et~al.}{2010}]{Narayanan10}
{Narayanan} D.,  et~al., 2010, \mn@doi [\mnras]
  {10.1111/j.1365-2966.2010.16997.x}, \href
  {http://adsabs.harvard.edu/abs/2010MNRAS.407.1701N} {407, 1701}

\bibitem[\protect\citeauthoryear{{Nemmen} \& {Brotherton}}{{Nemmen} \&
  {Brotherton}}{2010}]{Nemmen10}
{Nemmen} R.~S.,  {Brotherton} M.~S.,  2010, \mn@doi [\mnras]
  {10.1111/j.1365-2966.2010.17224.x}, \href
  {https://ui.adsabs.harvard.edu/#abs/2010MNRAS.408.1598N} {408, 1598}

\bibitem[\protect\citeauthoryear{{Netzer}, {Shemmer}, {Maiolino}, {Oliva},
  {Croom}, {Corbett}  \& {di Fabrizio}}{{Netzer} et~al.}{2004}]{Netzer04}
{Netzer} H.,  {Shemmer} O.,  {Maiolino} R.,  {Oliva} E.,  {Croom} S.,
  {Corbett} E.,   {di Fabrizio} L.,  2004, \mn@doi [\apj] {10.1086/423608},
  \href {http://adsabs.harvard.edu/abs/2004ApJ...614..558N} {614, 558}

\bibitem[\protect\citeauthoryear{{Peacock}}{{Peacock}}{1983}]{Peacock83}
{Peacock} J.~A.,  1983, \mn@doi [\mnras] {10.1093/mnras/202.3.615}, \href
  {http://adsabs.harvard.edu/abs/1983MNRAS.202..615P} {202, 615}

\bibitem[\protect\citeauthoryear{{Pitchford} et~al.,}{{Pitchford}
  et~al.}{2016}]{Pitchford16}
{Pitchford} L.~K.,  et~al., 2016, \mn@doi [\mnras] {10.1093/mnras/stw1840},
  \href {http://adsabs.harvard.edu/abs/2016MNRAS.462.4067P} {462, 4067}

\bibitem[\protect\citeauthoryear{{Price-Whelan} et~al.,}{{Price-Whelan}
  et~al.}{2018}]{astropy:2018}
{Price-Whelan} A.~M.,  et~al., 2018, \mn@doi [\aj] {10.3847/1538-3881/aabc4f},
  \href {https://ui.adsabs.harvard.edu/#abs/2018AJ....156..123T} {156, 123}

\bibitem[\protect\citeauthoryear{{Runnoe}, {Brotherton}  \& {Shang}}{{Runnoe}
  et~al.}{2012}]{Runnoe12}
{Runnoe} J.~C.,  {Brotherton} M.~S.,   {Shang} Z.,  2012, \mn@doi [\mnras]
  {10.1111/j.1365-2966.2012.20620.x}, \href
  {https://ui.adsabs.harvard.edu/#abs/2012MNRAS.422..478R} {422, 478}

\bibitem[\protect\citeauthoryear{{Sanders}, {Soifer}, {Elias}, {Madore},
  {Matthews}, {Neugebauer}  \& {Scoville}}{{Sanders} et~al.}{1988}]{Sanders88}
{Sanders} D.~B.,  {Soifer} B.~T.,  {Elias} J.~H.,  {Madore} B.~F.,  {Matthews}
  K.,  {Neugebauer} G.,   {Scoville} N.~Z.,  1988, \mn@doi [\apj]
  {10.1086/165983}, \href {http://adsabs.harvard.edu/abs/1988ApJ...325...74S}
  {325, 74}

\bibitem[\protect\citeauthoryear{{Shen}}{{Shen}}{2016}]{Shen16}
{Shen} Y.,  2016, \mn@doi [\apj] {10.3847/0004-637X/817/1/55}, \href
  {http://adsabs.harvard.edu/abs/2016ApJ...817...55S} {817, 55}

\bibitem[\protect\citeauthoryear{{Shen} et~al.,}{{Shen} et~al.}{2011}]{Shen11}
{Shen} Y.,  et~al., 2011, \mn@doi [\apjs] {10.1088/0067-0049/194/2/45}, \href
  {http://adsabs.harvard.edu/abs/2011ApJS..194...45S} {194, 45}

\bibitem[\protect\citeauthoryear{{Sijacki}, {Vogelsberger}, {Genel},
  {Springel}, {Torrey}, {Snyder}, {Nelson}  \& {Hernquist}}{{Sijacki}
  et~al.}{2015}]{Sijacki15}
{Sijacki} D.,  {Vogelsberger} M.,  {Genel} S.,  {Springel} V.,  {Torrey} P.,
  {Snyder} G.~F.,  {Nelson} D.,   {Hernquist} L.,  2015, \mn@doi [\mnras]
  {10.1093/mnras/stv1340}, \href
  {http://adsabs.harvard.edu/abs/2015MNRAS.452..575S} {452, 575}

\bibitem[\protect\citeauthoryear{{Storey} \& {Hummer}}{{Storey} \&
  {Hummer}}{1995}]{Storey95}
{Storey} P.~J.,  {Hummer} D.~G.,  1995, \mn@doi [\mnras]
  {10.1093/mnras/272.1.41}, \href
  {http://adsabs.harvard.edu/abs/1995MNRAS.272...41S} {272, 41}

\bibitem[\protect\citeauthoryear{{Storey} \& {Zeippen}}{{Storey} \&
  {Zeippen}}{2000}]{Oiii_ratio}
{Storey} P.~J.,  {Zeippen} C.~J.,  2000, \mn@doi [\mnras]
  {10.1046/j.1365-8711.2000.03184.x}, \href
  {http://adsabs.harvard.edu/abs/2000MNRAS.312..813S} {312, 813}

\bibitem[\protect\citeauthoryear{{Symeonidis}}{{Symeonidis}}{2017}]{Symeonidis17}
{Symeonidis} M.,  2017, \mn@doi [\mnras] {10.1093/mnras/stw2784}, \href
  {http://adsabs.harvard.edu/abs/2017MNRAS.465.1401S} {465, 1401}

\bibitem[\protect\citeauthoryear{{Tadhunter} et~al.,}{{Tadhunter}
  et~al.}{2018}]{Tadhunter18}
{Tadhunter} C.,  et~al., 2018, \mn@doi [\mnras] {10.1093/mnras/sty1064}, \href
  {http://adsabs.harvard.edu/abs/2018MNRAS.tmp.1116T} {478, 1558}

\bibitem[\protect\citeauthoryear{{Urrutia}, {Lacy}, {Spoon}, {Glikman},
  {Petric}  \& {Schulz}}{{Urrutia} et~al.}{2012}]{Urrutia12}
{Urrutia} T.,  {Lacy} M.,  {Spoon} H.,  {Glikman} E.,  {Petric} A.,   {Schulz}
  B.,  2012, \mn@doi [\apj] {10.1088/0004-637X/757/2/125}, \href
  {http://adsabs.harvard.edu/abs/2012ApJ...757..125U} {757, 125}

\bibitem[\protect\citeauthoryear{{Vestergaard} \& {Peterson}}{{Vestergaard} \&
  {Peterson}}{2006}]{vp06}
{Vestergaard} M.,  {Peterson} B.~M.,  2006, \mn@doi [\apj] {10.1086/500572},
  \href {http://adsabs.harvard.edu/abs/2006ApJ...641..689V} {641, 689}

\bibitem[\protect\citeauthoryear{{Vogelsberger}, {Genel}, {Sijacki}, {Torrey},
  {Springel}  \& {Hernquist}}{{Vogelsberger} et~al.}{2013}]{Vogelsberger13}
{Vogelsberger} M.,  {Genel} S.,  {Sijacki} D.,  {Torrey} P.,  {Springel} V.,
  {Hernquist} L.,  2013, \mn@doi [\mnras] {10.1093/mnras/stt1789}, \href
  {http://adsabs.harvard.edu/abs/2013MNRAS.436.3031V} {436, 3031}

\bibitem[\protect\citeauthoryear{{Wethers} et~al.,}{{Wethers}
  et~al.}{2018}]{Wethers18}
{Wethers} C.~F.,  et~al., 2018, \mn@doi [\mnras] {10.1093/mnras/stx3332}, \href
  {http://adsabs.harvard.edu/abs/2018MNRAS.475.3682W} {475, 3682}

\bibitem[\protect\citeauthoryear{{Williams}, {Maiolino}, {Krongold},
  {Carniani}, {Cresci}, {Mannucci}  \& {Marconi}}{{Williams}
  et~al.}{2017}]{easysinf}
{Williams} R.~J.,  {Maiolino} R.,  {Krongold} Y.,  {Carniani} S.,  {Cresci} G.,
   {Mannucci} F.,   {Marconi} A.,  2017, \mn@doi [\mnras]
  {10.1093/mnras/stx311}, \href
  {http://adsabs.harvard.edu/abs/2017MNRAS.467.3399W} {467, 3399}

\bibitem[\protect\citeauthoryear{{Wu} et~al.,}{{Wu} et~al.}{2018}]{Wu18}
{Wu} J.,  et~al., 2018, \mn@doi [\apj] {10.3847/1538-4357/aa9ff3}, \href
  {http://adsabs.harvard.edu/abs/2018ApJ...852...96W} {852, 96}

\bibitem[\protect\citeauthoryear{{Zakamska} \& {Greene}}{{Zakamska} \&
  {Greene}}{2014}]{Zakamska14}
{Zakamska} N.~L.,  {Greene} J.~E.,  2014, \mn@doi [\mnras]
  {10.1093/mnras/stu842}, \href
  {http://adsabs.harvard.edu/abs/2014MNRAS.442..784Z} {442, 784}

\bibitem[\protect\citeauthoryear{{Zakamska} et~al.,}{{Zakamska}
  et~al.}{2016}]{Zakamska16}
{Zakamska} N.~L.,  et~al., 2016, \mn@doi [\mnras] {10.1093/mnras/stw718}, \href
  {http://adsabs.harvard.edu/abs/2016MNRAS.459.3144Z} {459, 3144}

\makeatother
\end{thebibliography}




\appendix

\section{Spectra and Emission Line Fits}
\label{sec:fits}

\begin{figure*}
\begin{center}
\includegraphics[width=0.666\columnwidth]{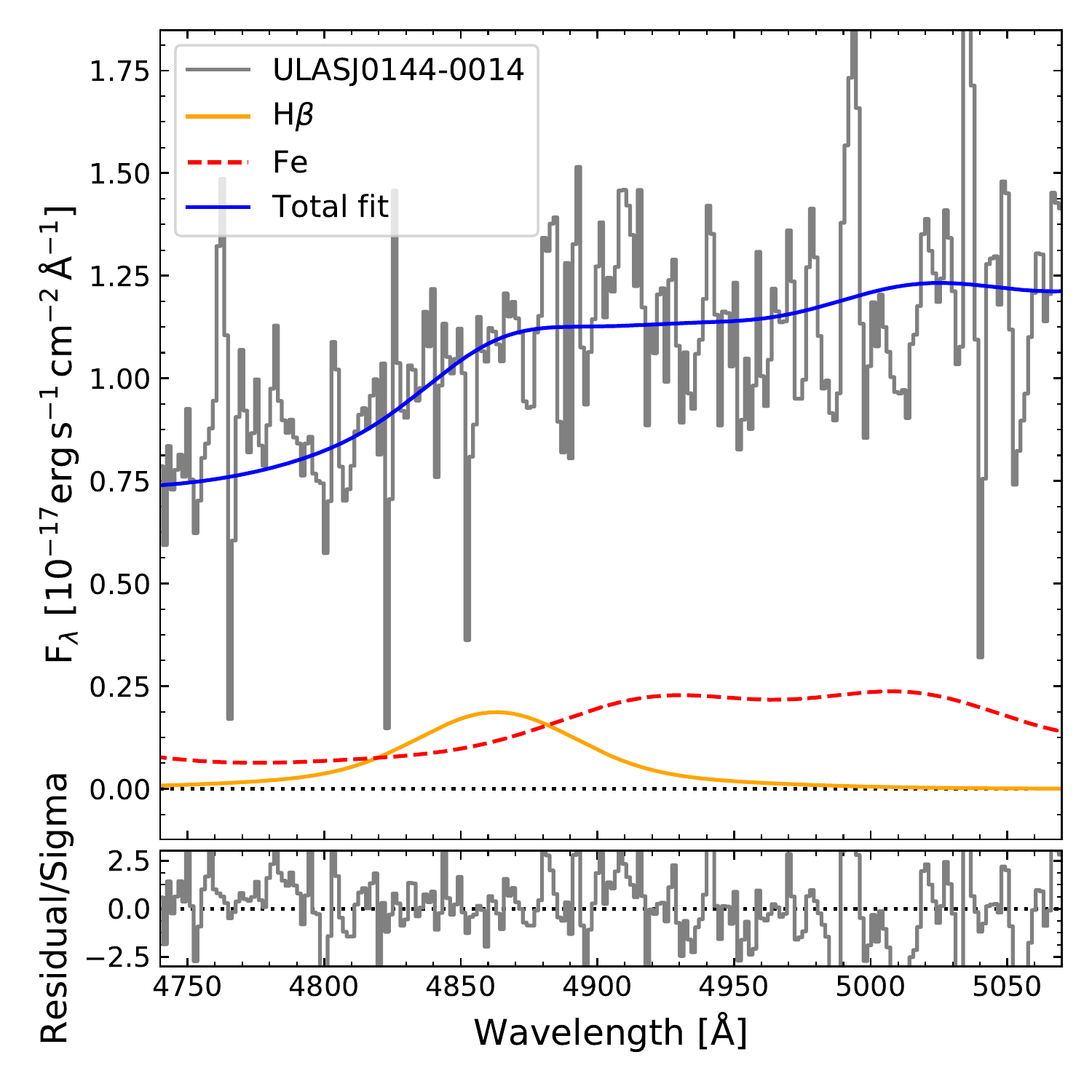}
\includegraphics[width=0.666\columnwidth]{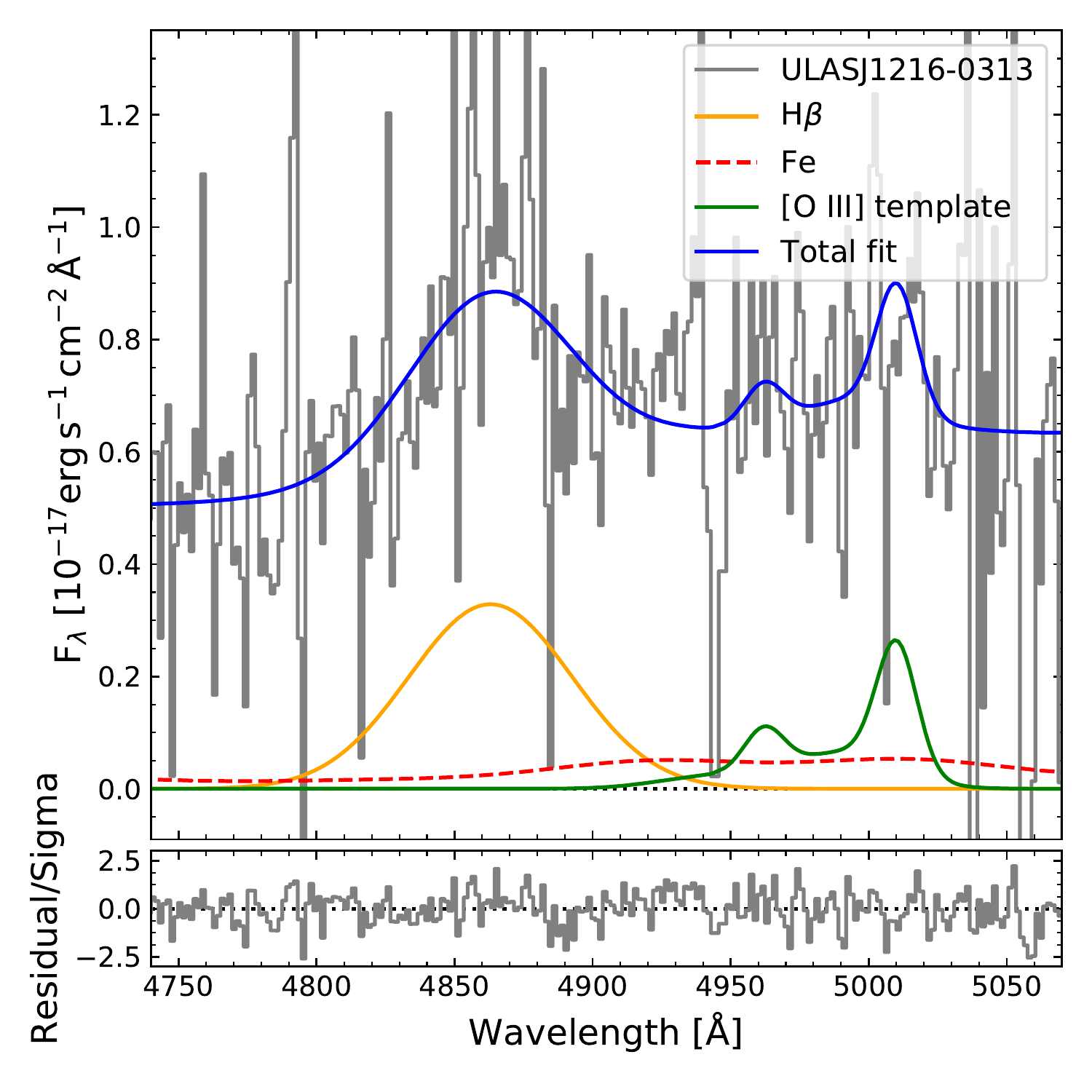}
\includegraphics[width=0.666\columnwidth]{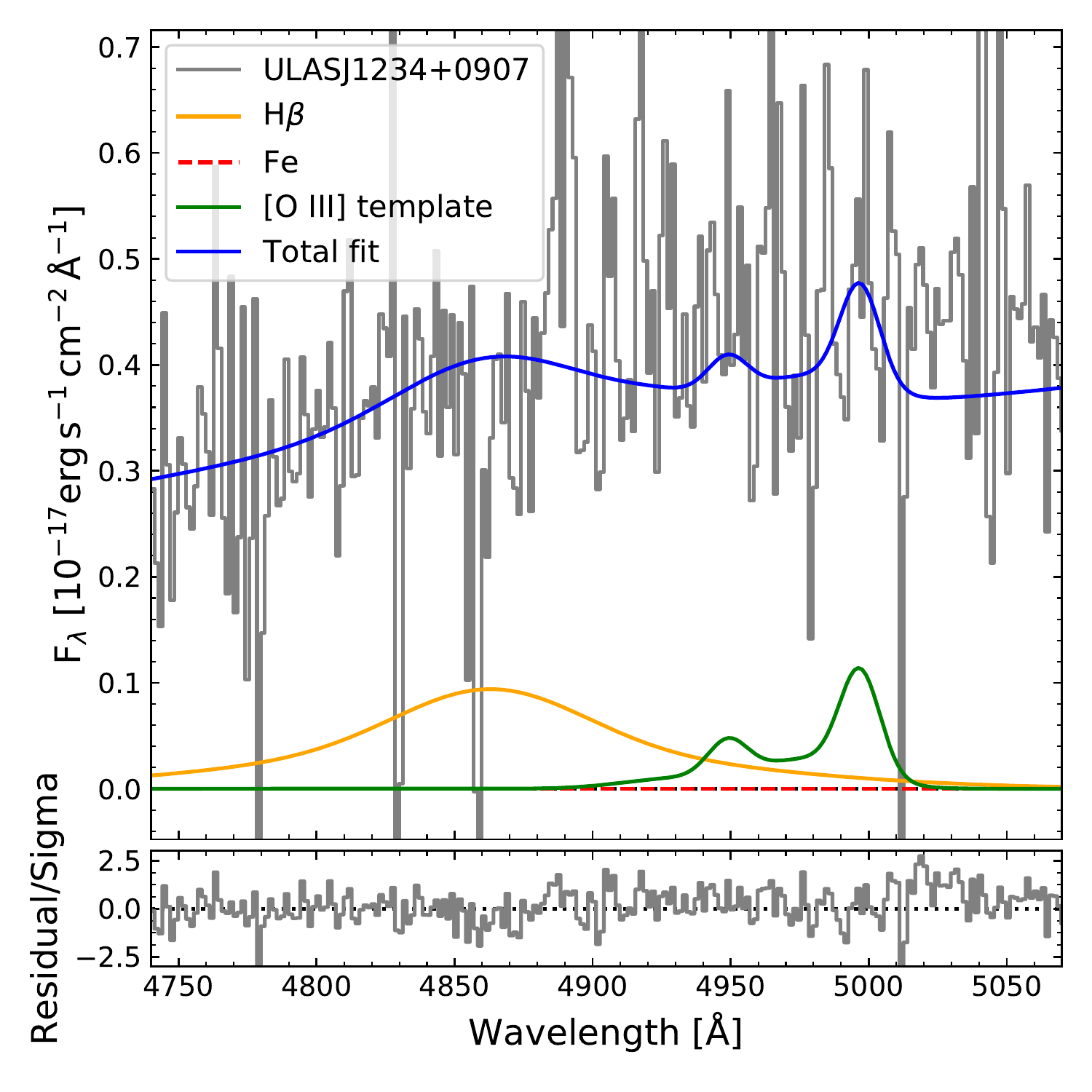}
\end{center}
    \caption{Fits to the region around H\,$\beta$ and \Oiii for the three objects with very weak \Oiii emission. An upper limit on the equivalent width of the line can be estimated using a template in ULAS~J1216-0313 and ULAS~J1234+0907. Residuals are shown in the panels below, scaled by the noise.}
    \label{fig:template}
\end{figure*}

We show the new SINFONI spectra for our 26 targets in Figure~\ref{fig:spectra}, and our fits to H\,$\alpha$, H\,$\beta$ and \Oiii in our `high redshift sample' in Figures~\ref{fig:fits},~\ref{fig:Ha}, \ref{fig:Oiii}~and~\ref{fig:template}.

\begin{figure*}
\begin{center}
\includegraphics[width=\columnwidth]{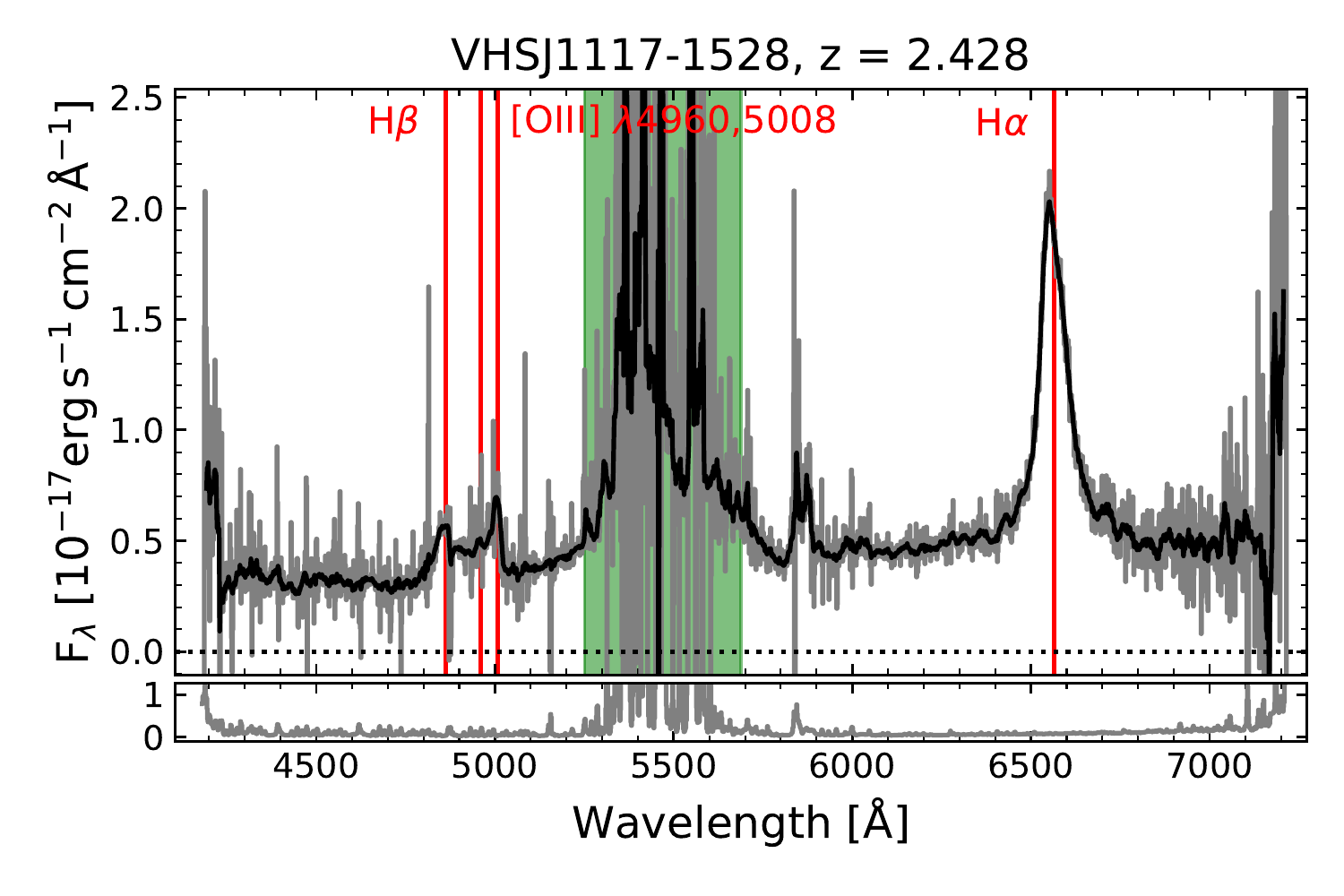}
\includegraphics[width=\columnwidth]{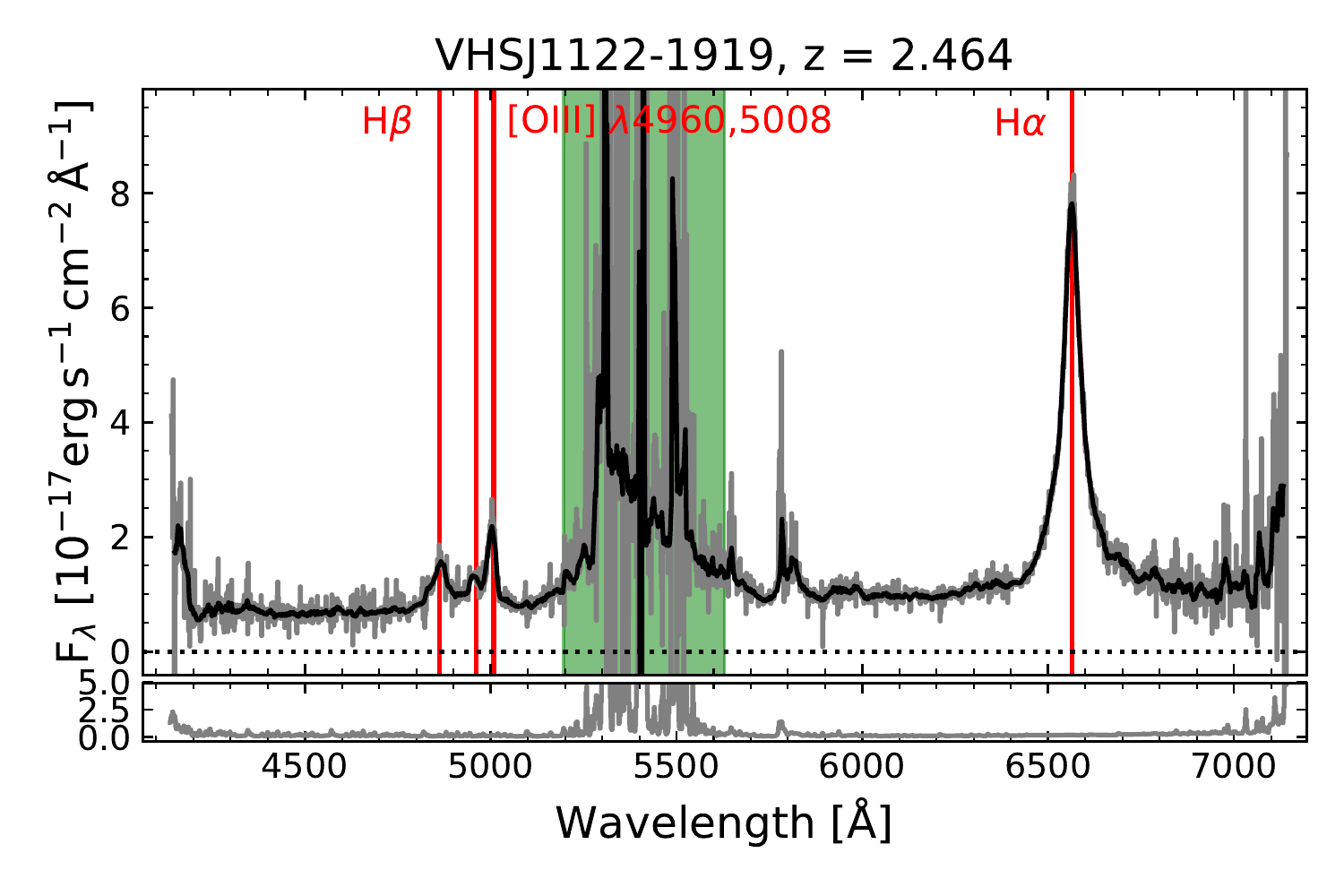}
\end{center}
    \caption{$H+K$ spectra of our 26 targets. The data are in gray, and smoothed with a 15 pixel inverse-variance weighted filter in black. The observed frame wavelength region 1.80-1.95\,$\mu\textrm{m}$, where atmospheric transparency is poor, is shaded in green. Below are plotted the noise arrays, with the same units as the flux.}
    \label{fig:spectra}
\end{figure*}
\begin{figure*}
\begin{center}
\includegraphics[width=\columnwidth]{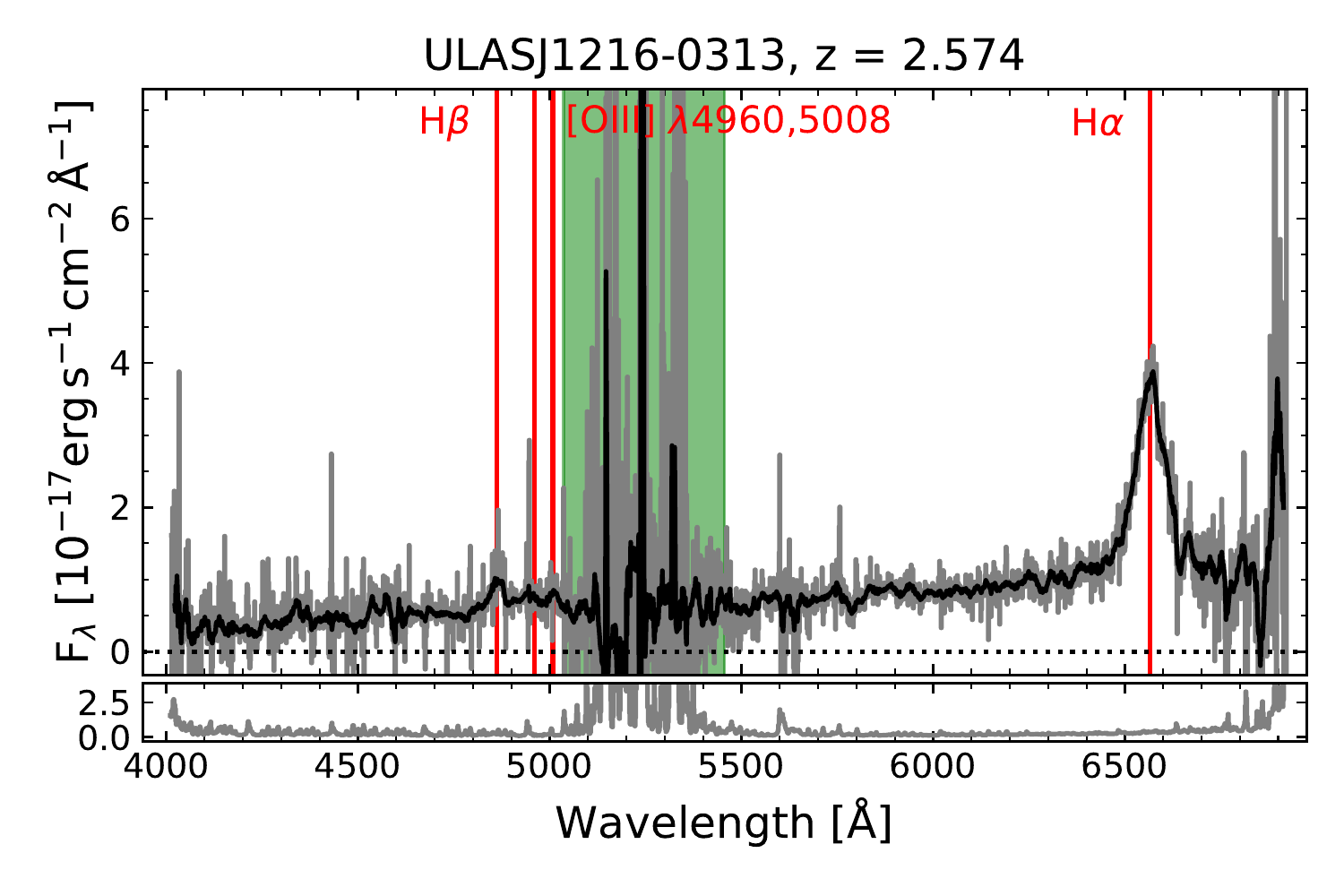}
\includegraphics[width=\columnwidth]{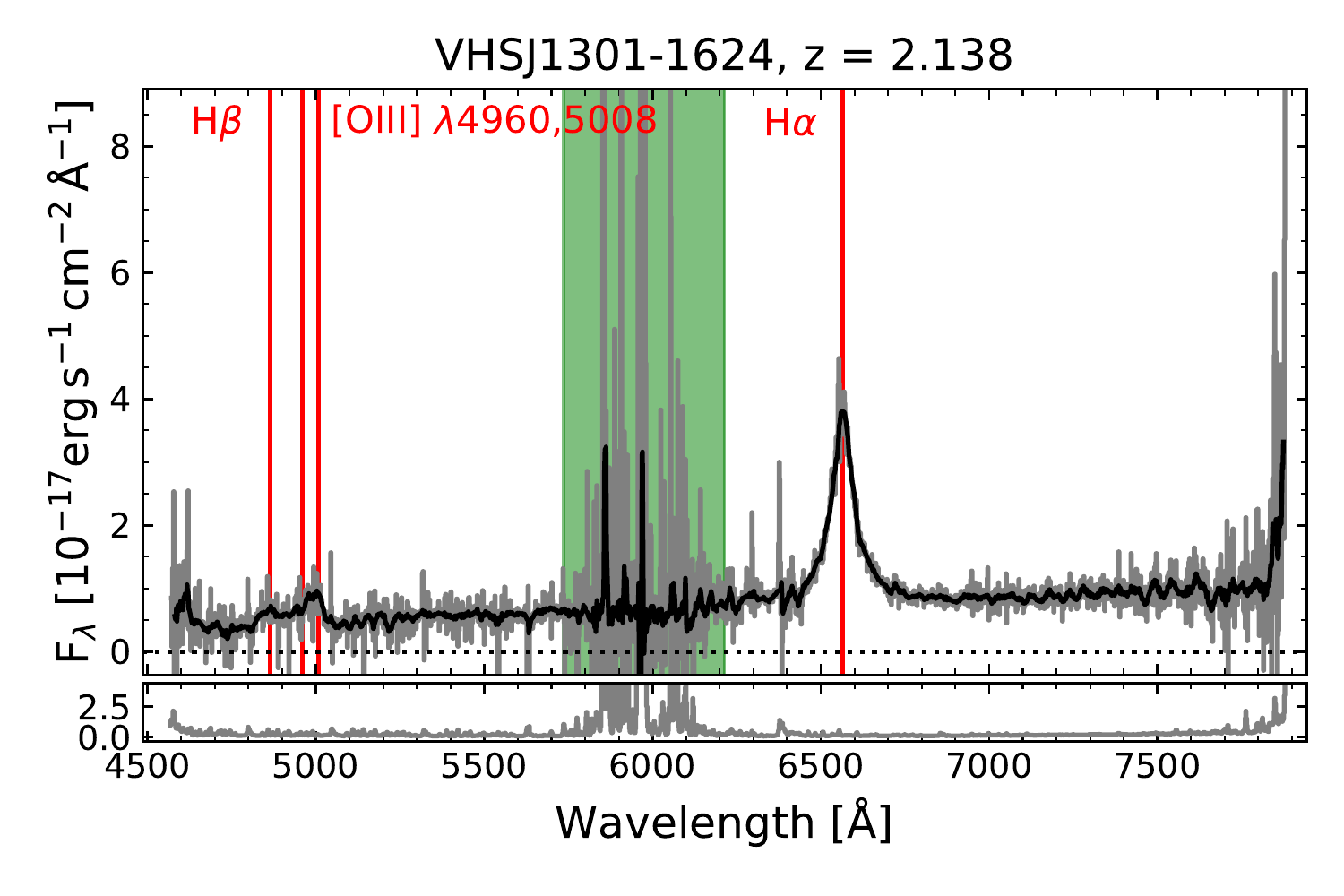}
\includegraphics[width=\columnwidth]{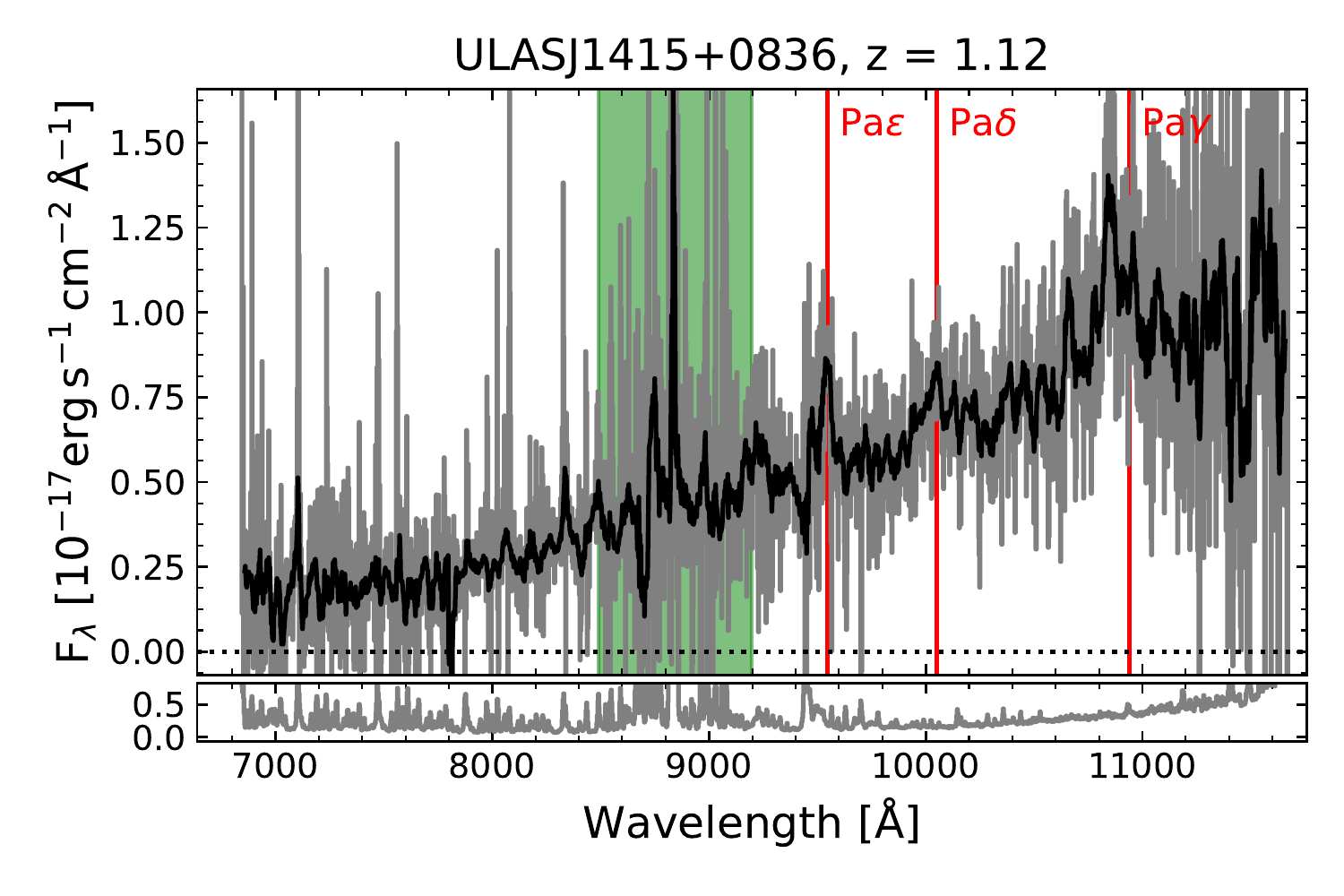}
\includegraphics[width=\columnwidth]{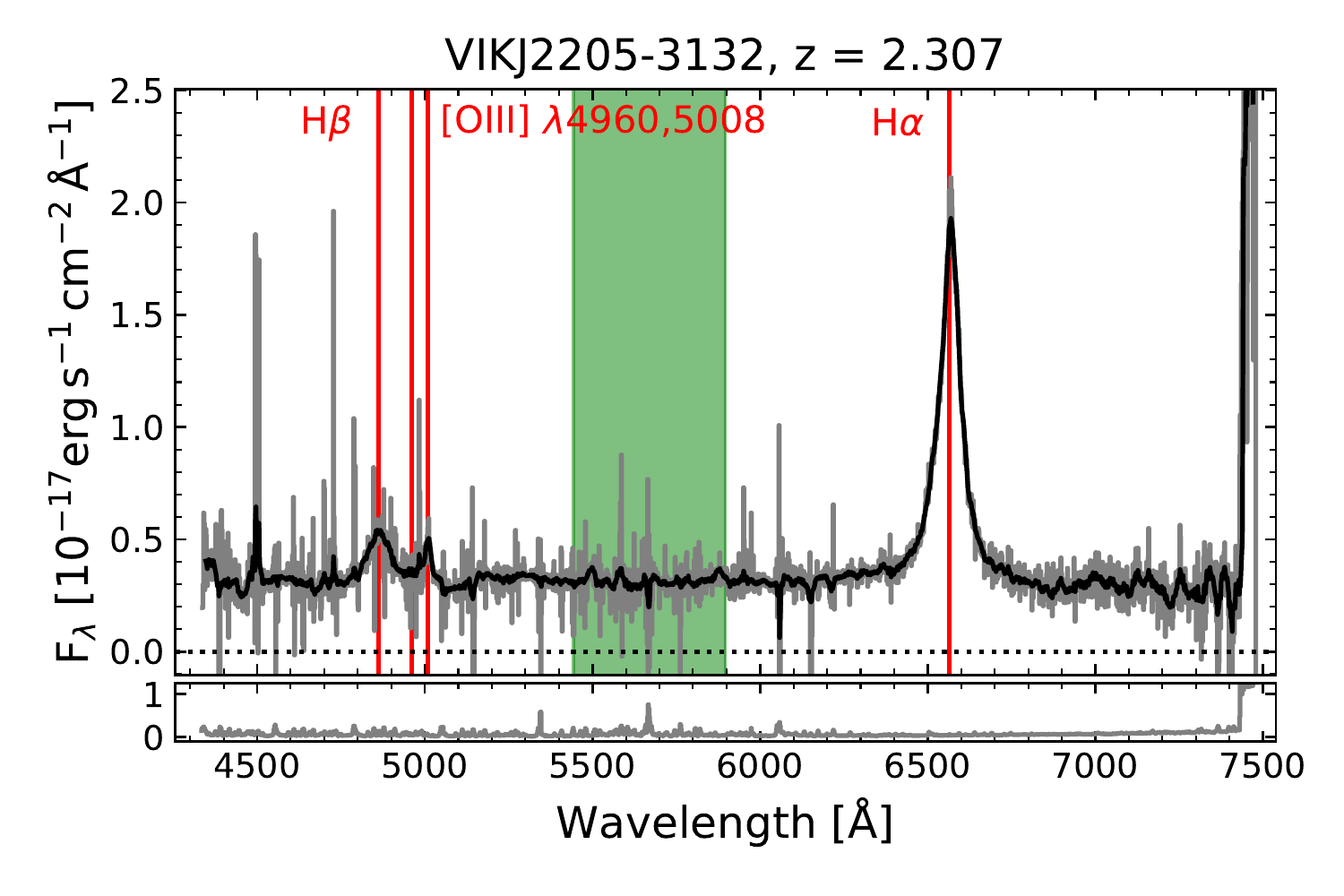}
\includegraphics[width=\columnwidth]{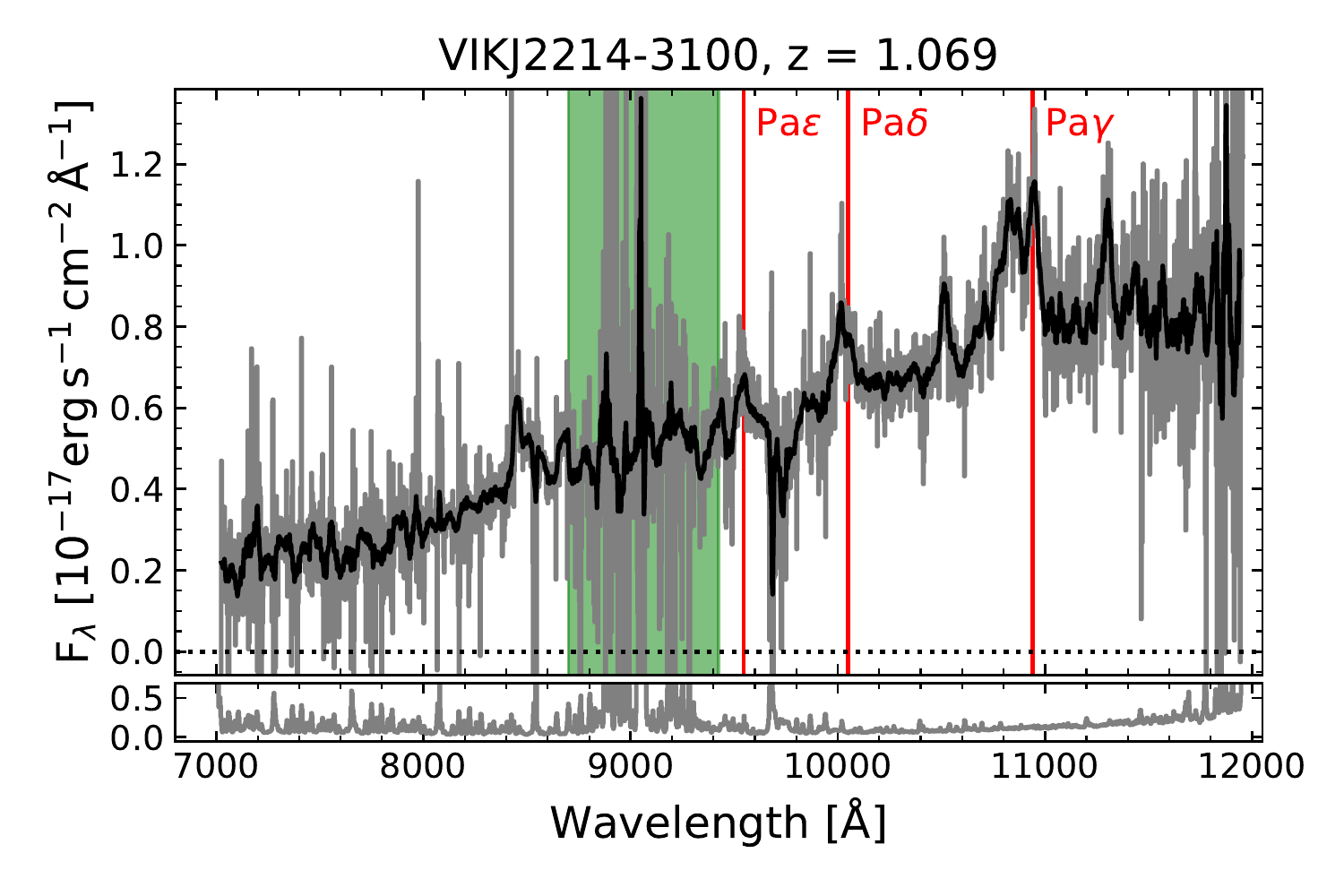}
\includegraphics[width=\columnwidth]{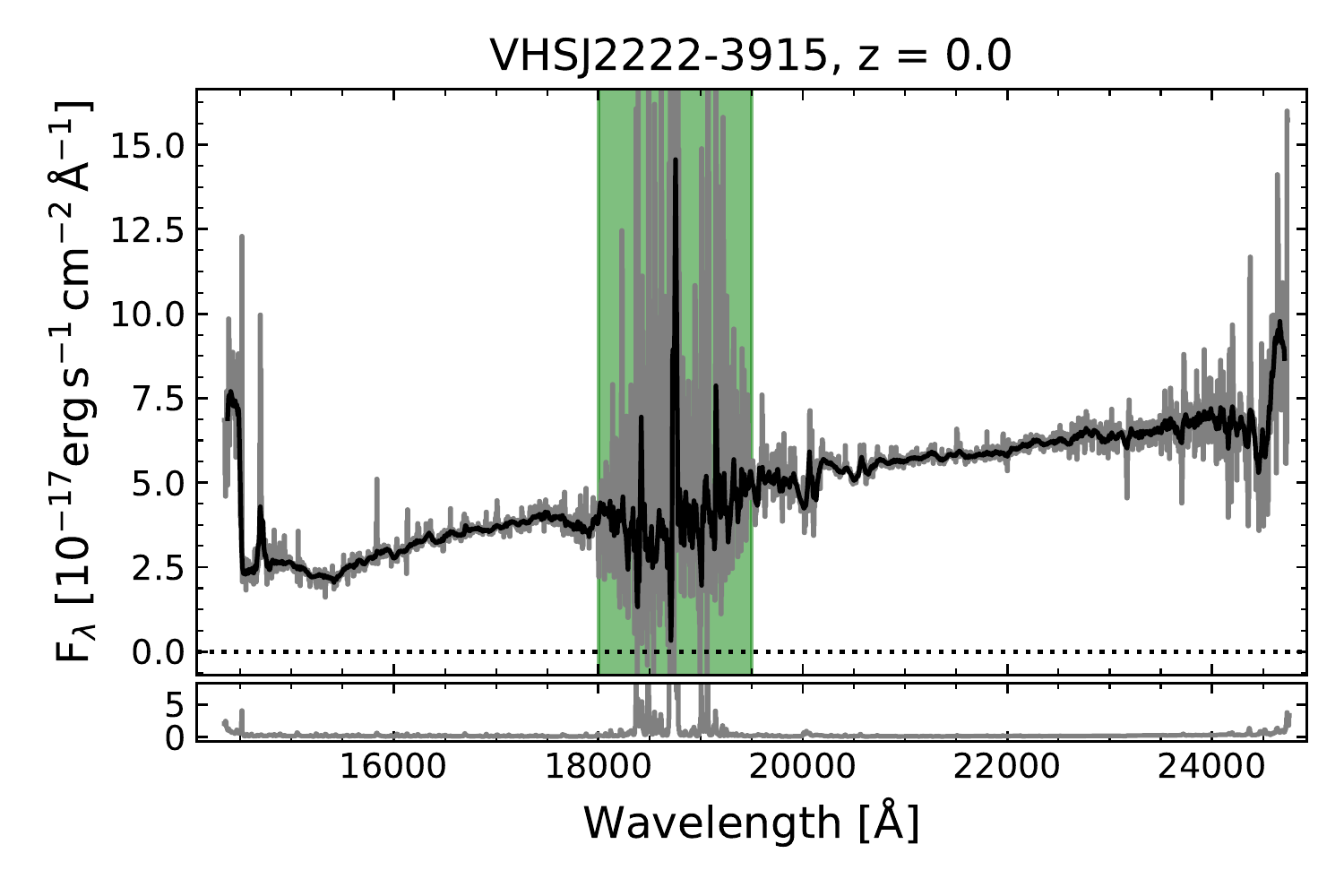}
\includegraphics[width=\columnwidth]{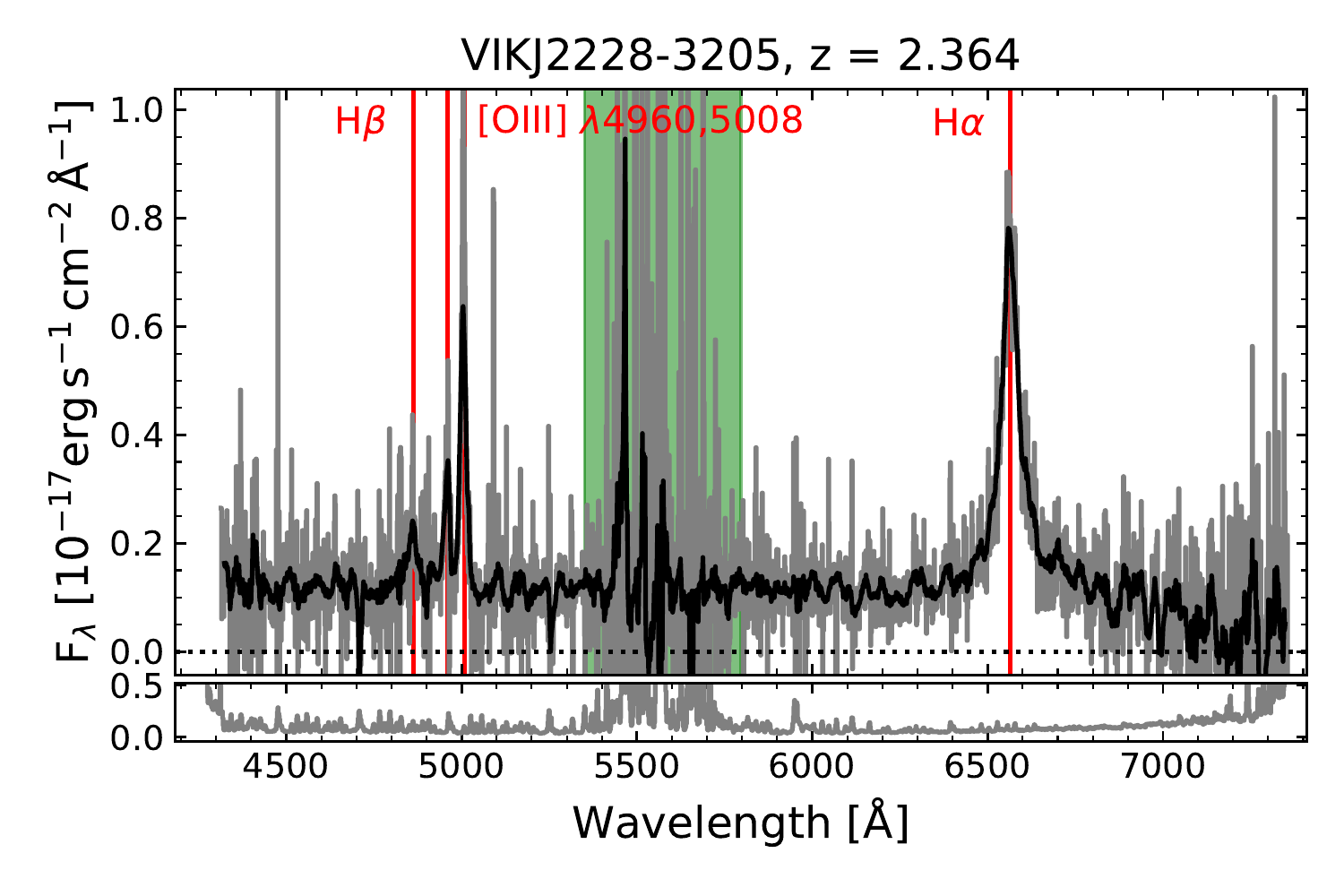}
\includegraphics[width=\columnwidth]{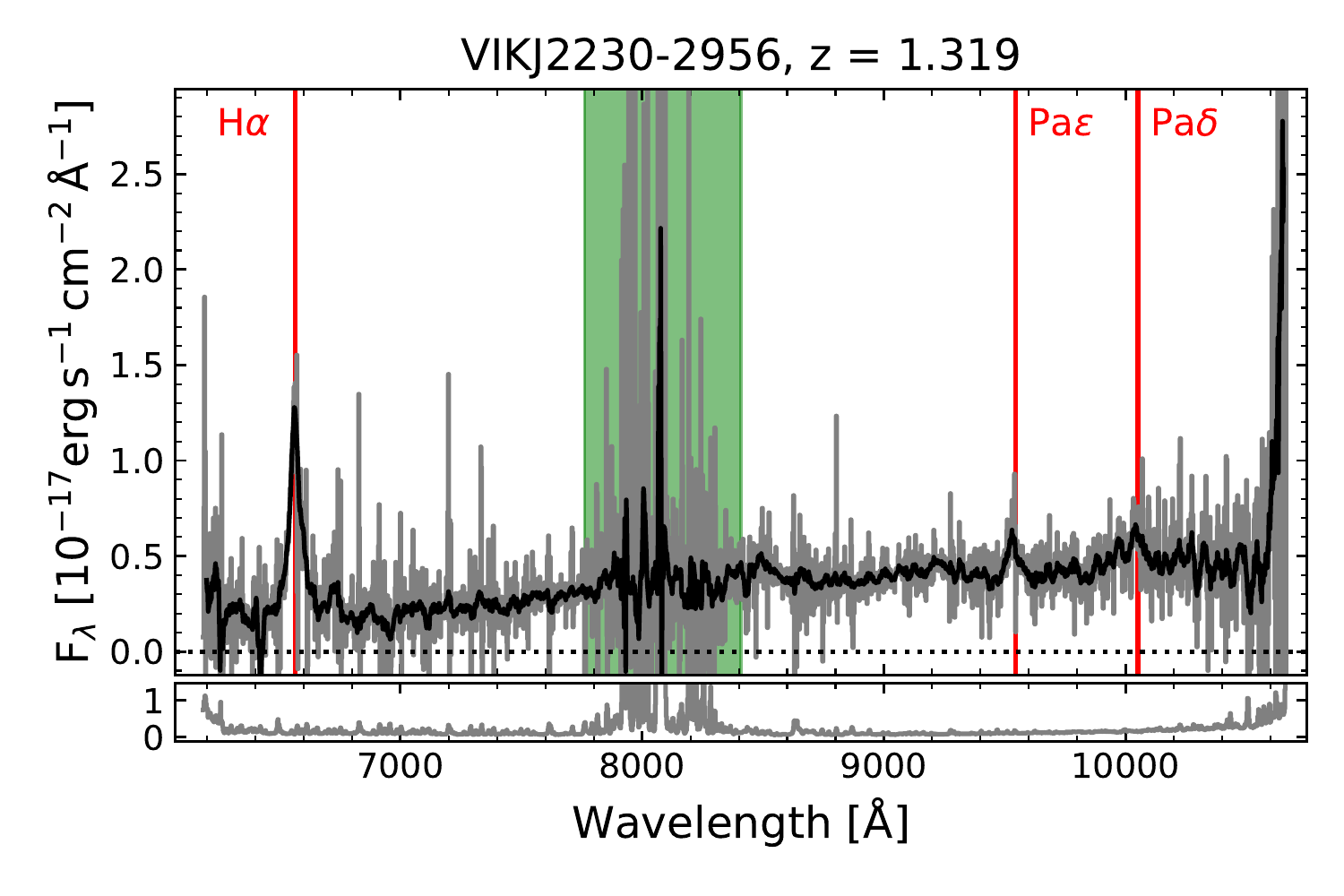}
\end{center}
\contcaption{}
\end{figure*}
\begin{figure*}
\begin{center}
\includegraphics[width=\columnwidth]{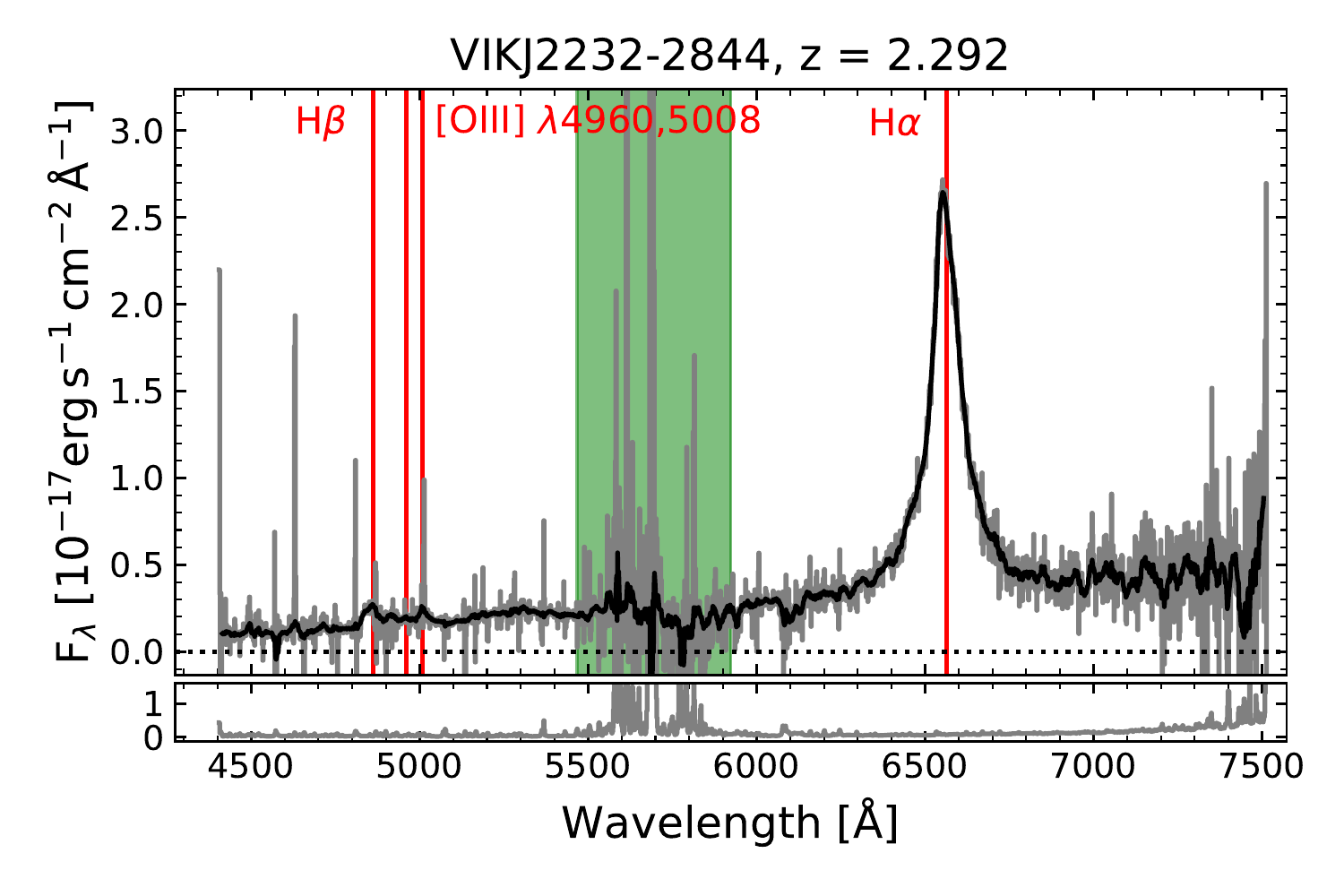}
\includegraphics[width=\columnwidth]{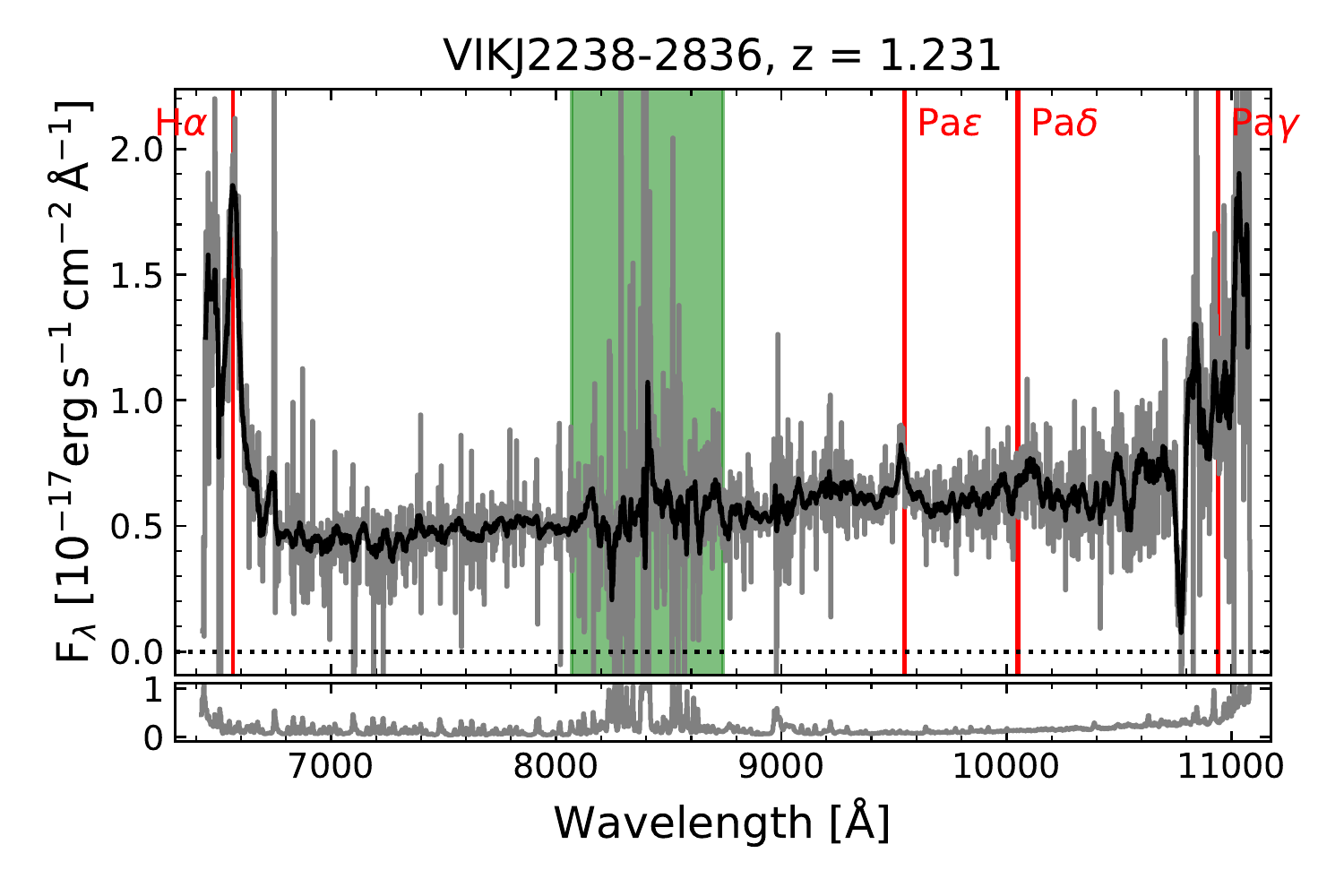}
\includegraphics[width=\columnwidth]{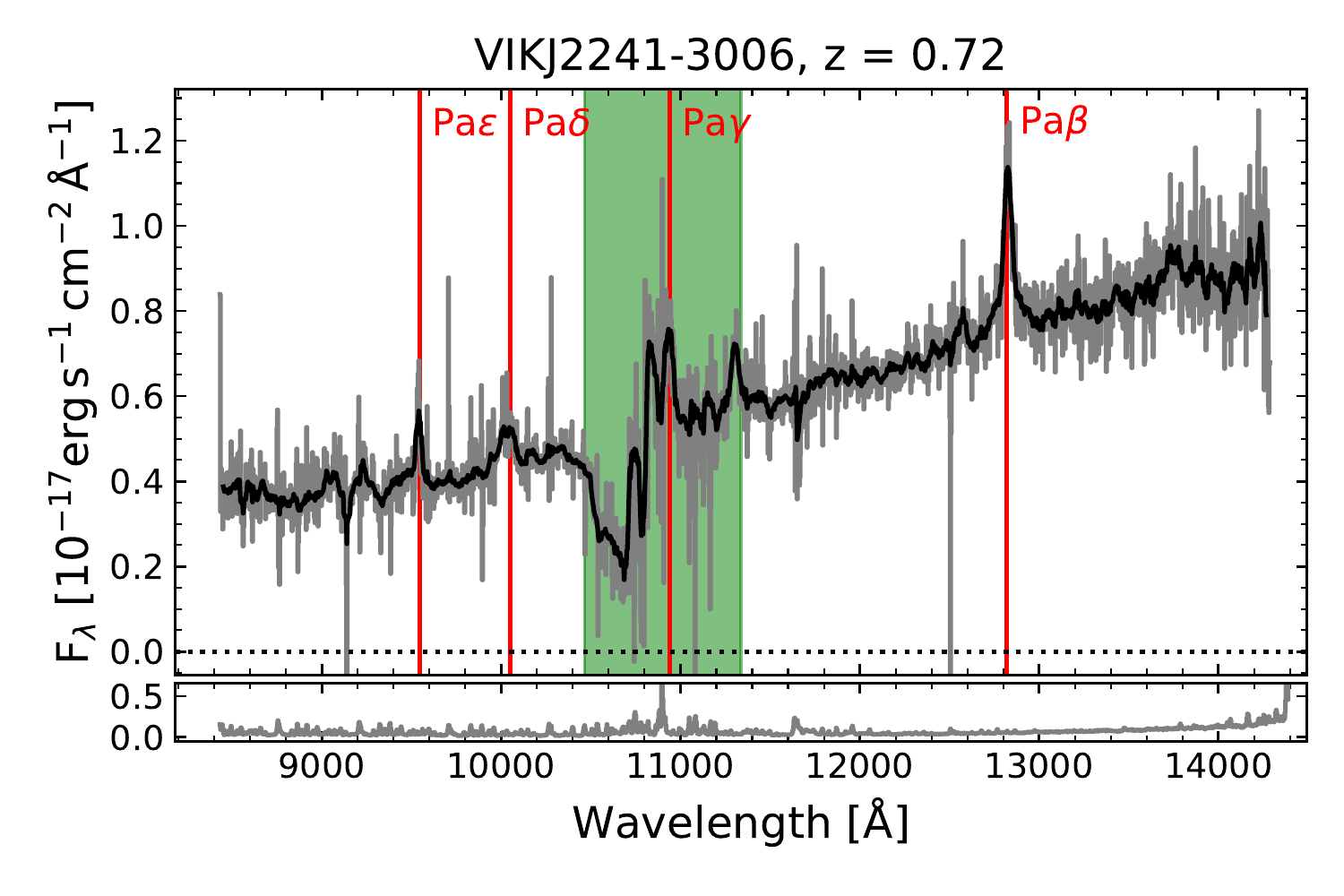}
\includegraphics[width=\columnwidth]{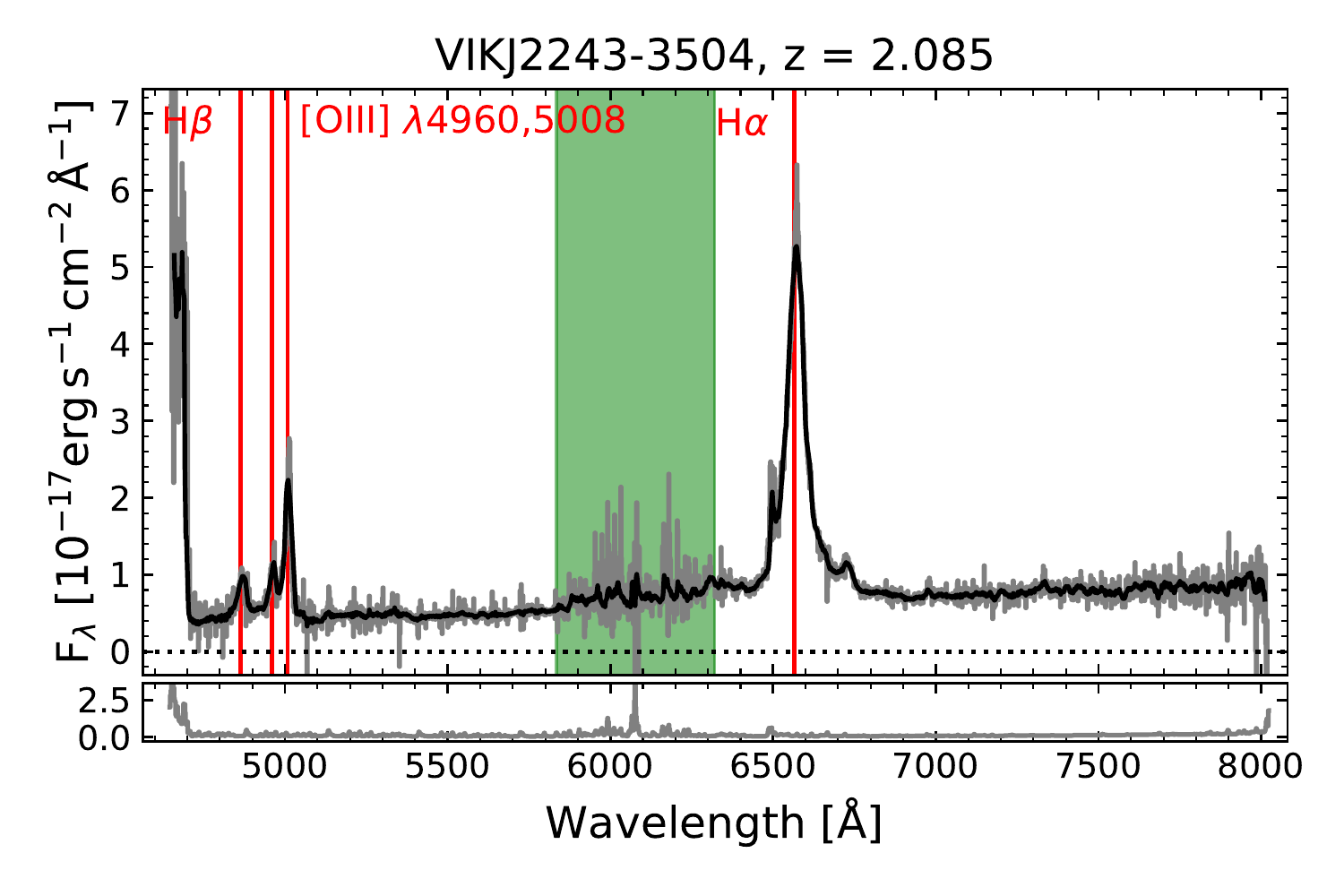}
\includegraphics[width=\columnwidth]{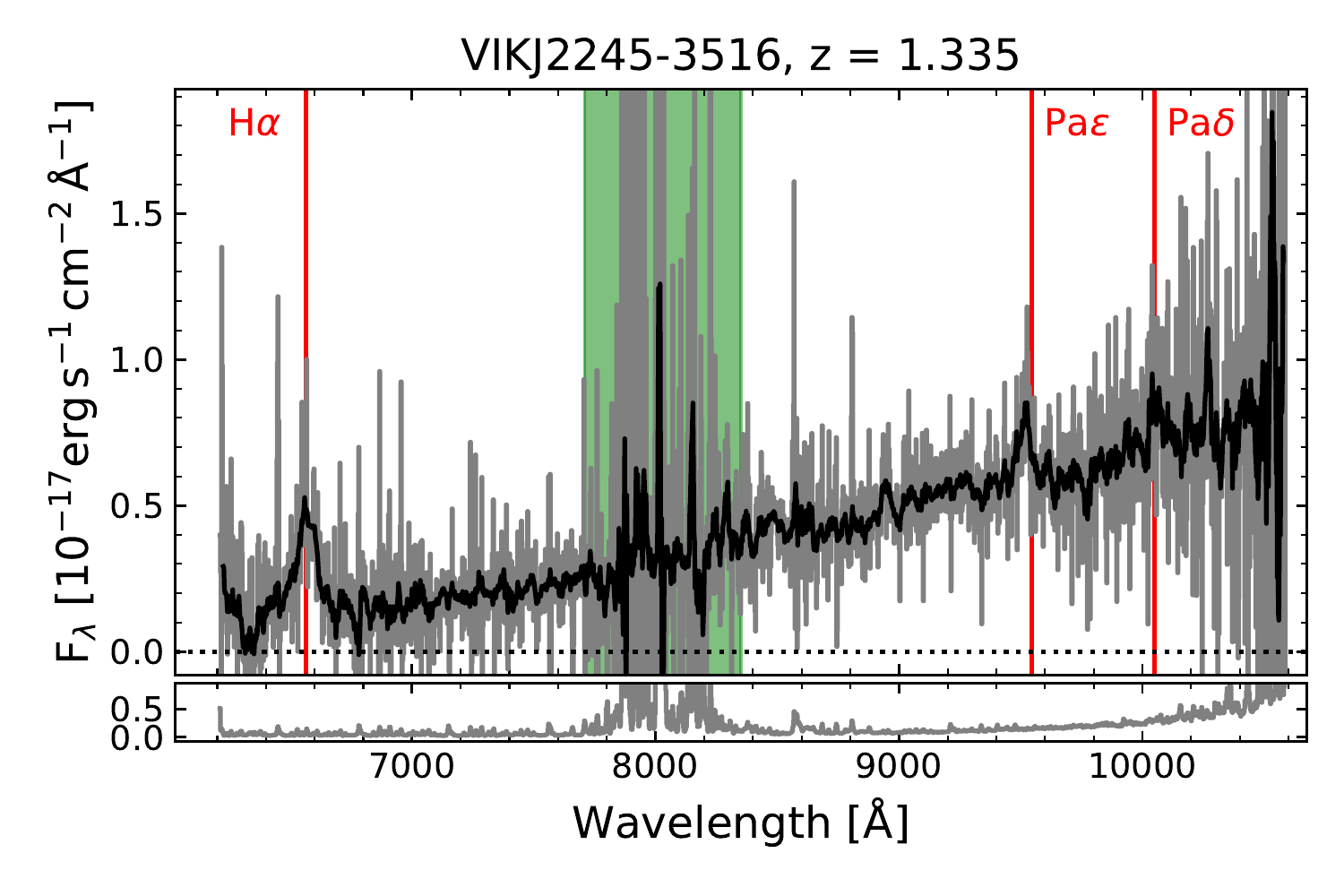}
\includegraphics[width=\columnwidth]{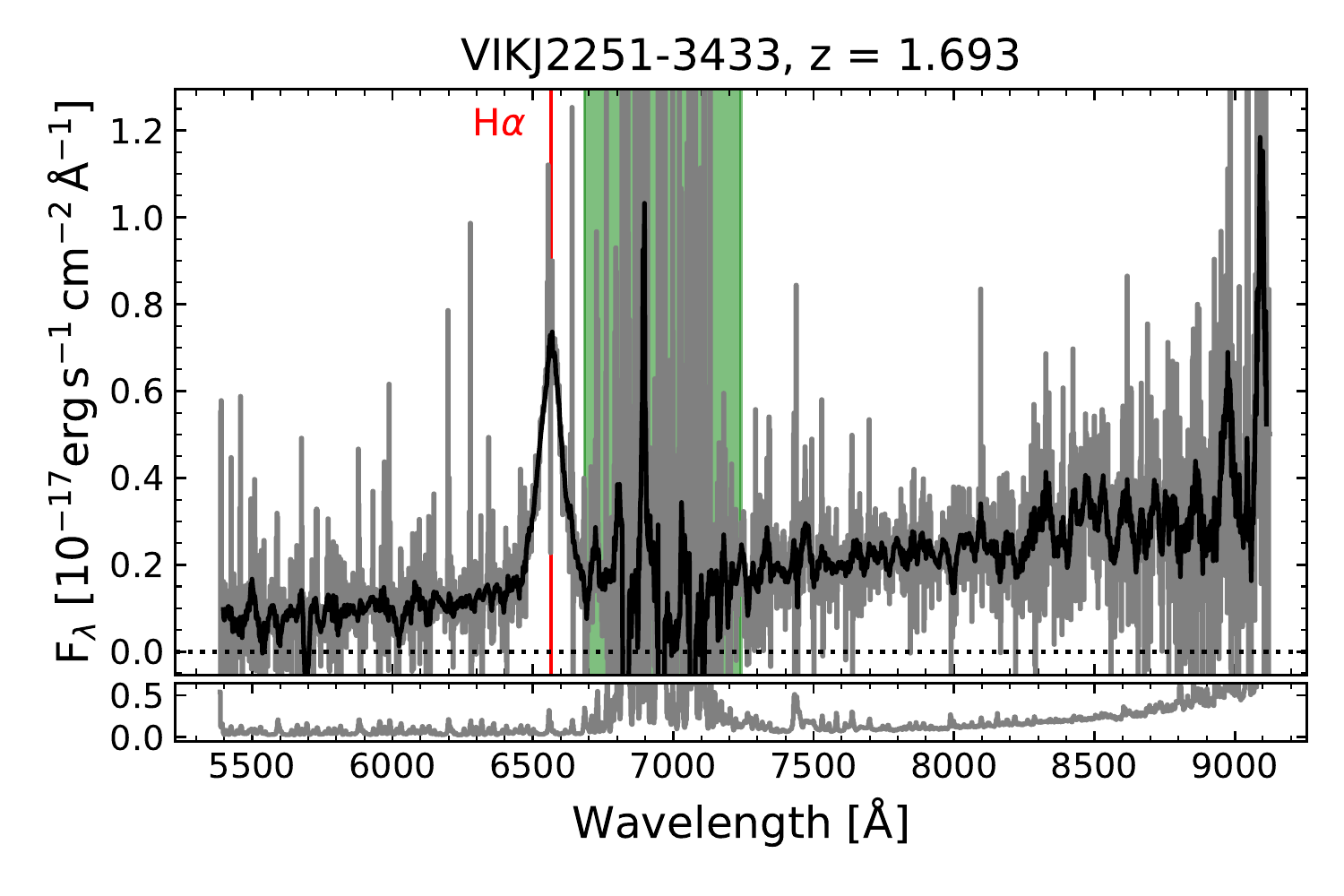}
\includegraphics[width=\columnwidth]{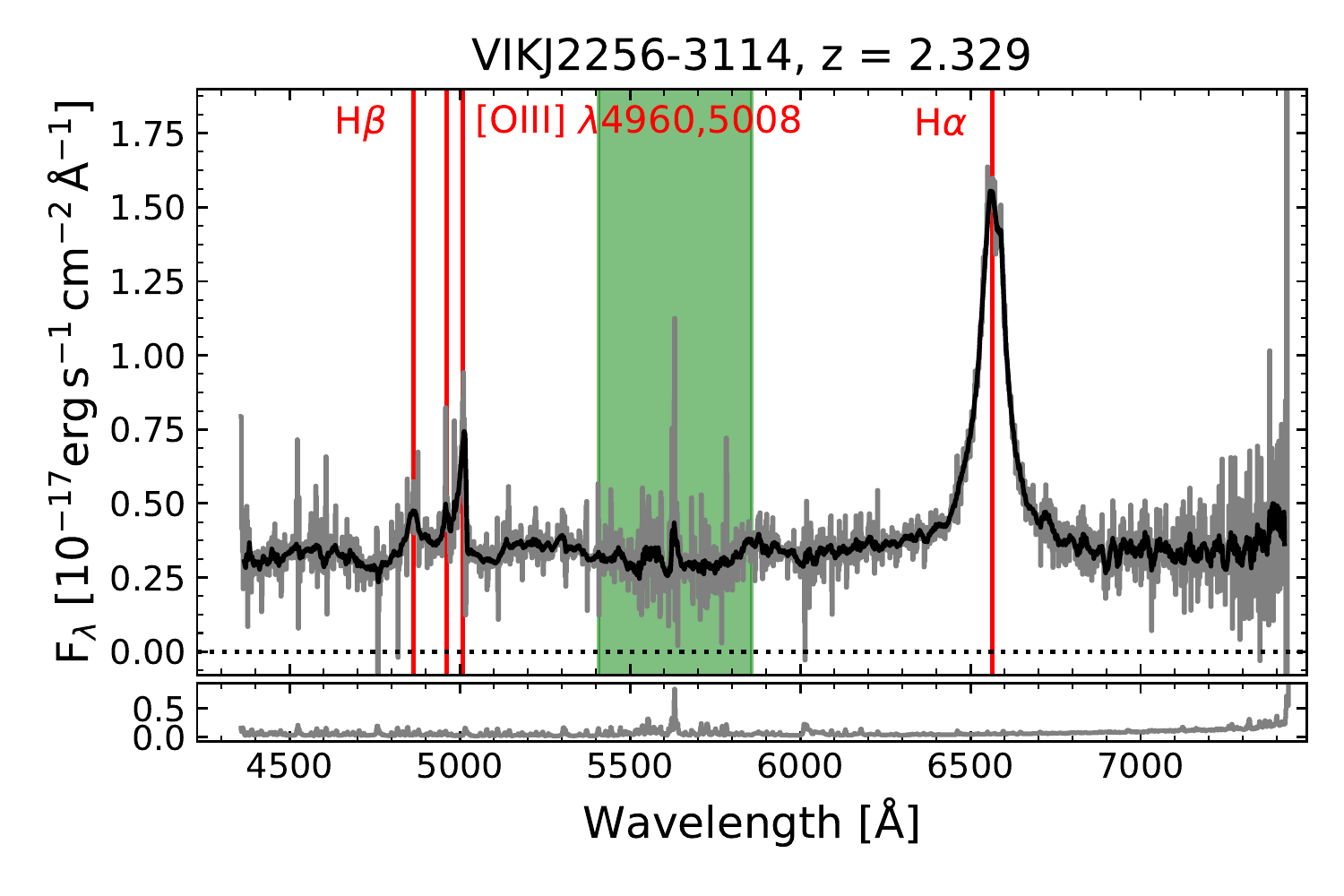}
\includegraphics[width=\columnwidth]{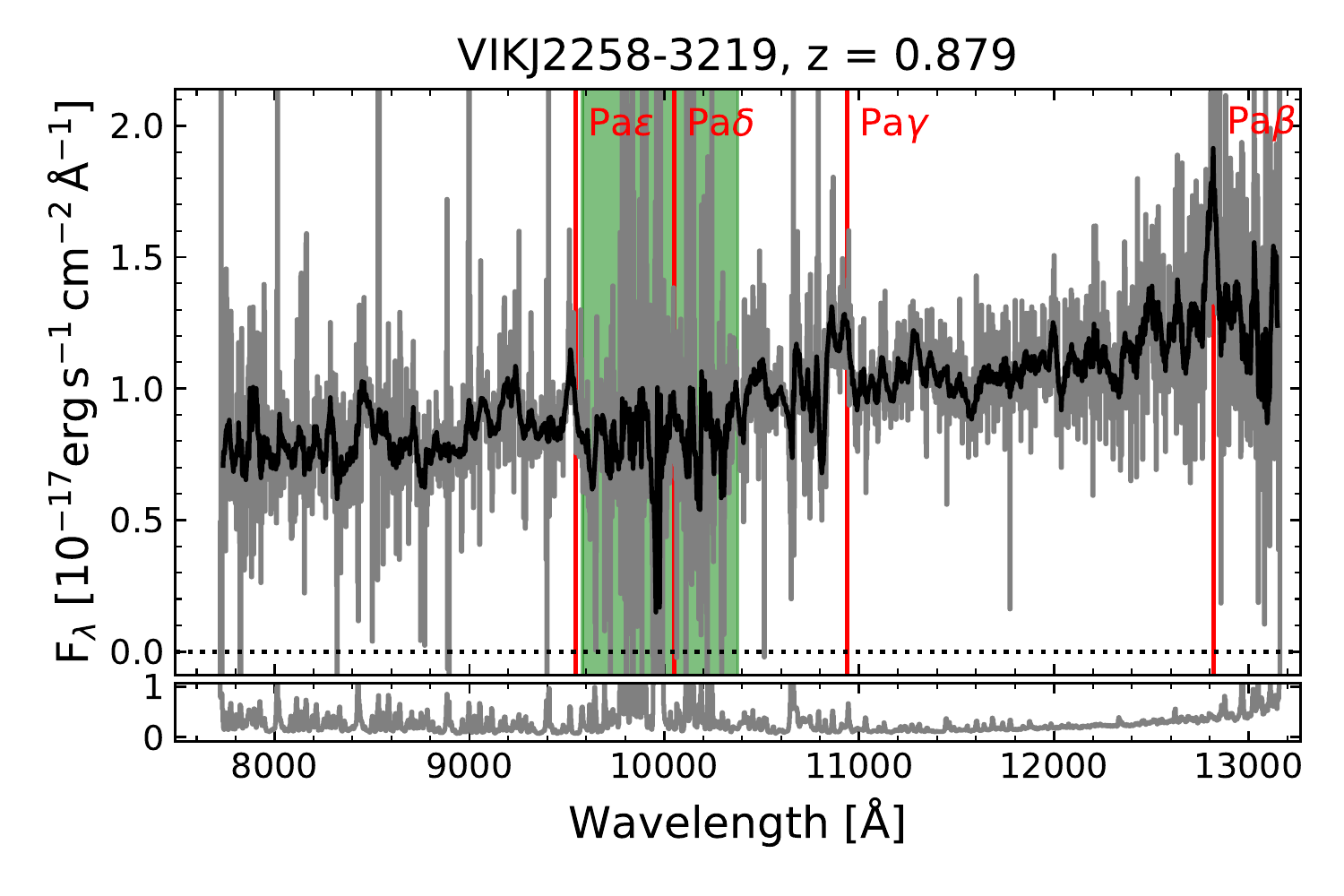}
\end{center}
\contcaption{}
\end{figure*}
\begin{figure*}
\begin{center}
\includegraphics[width=\columnwidth]{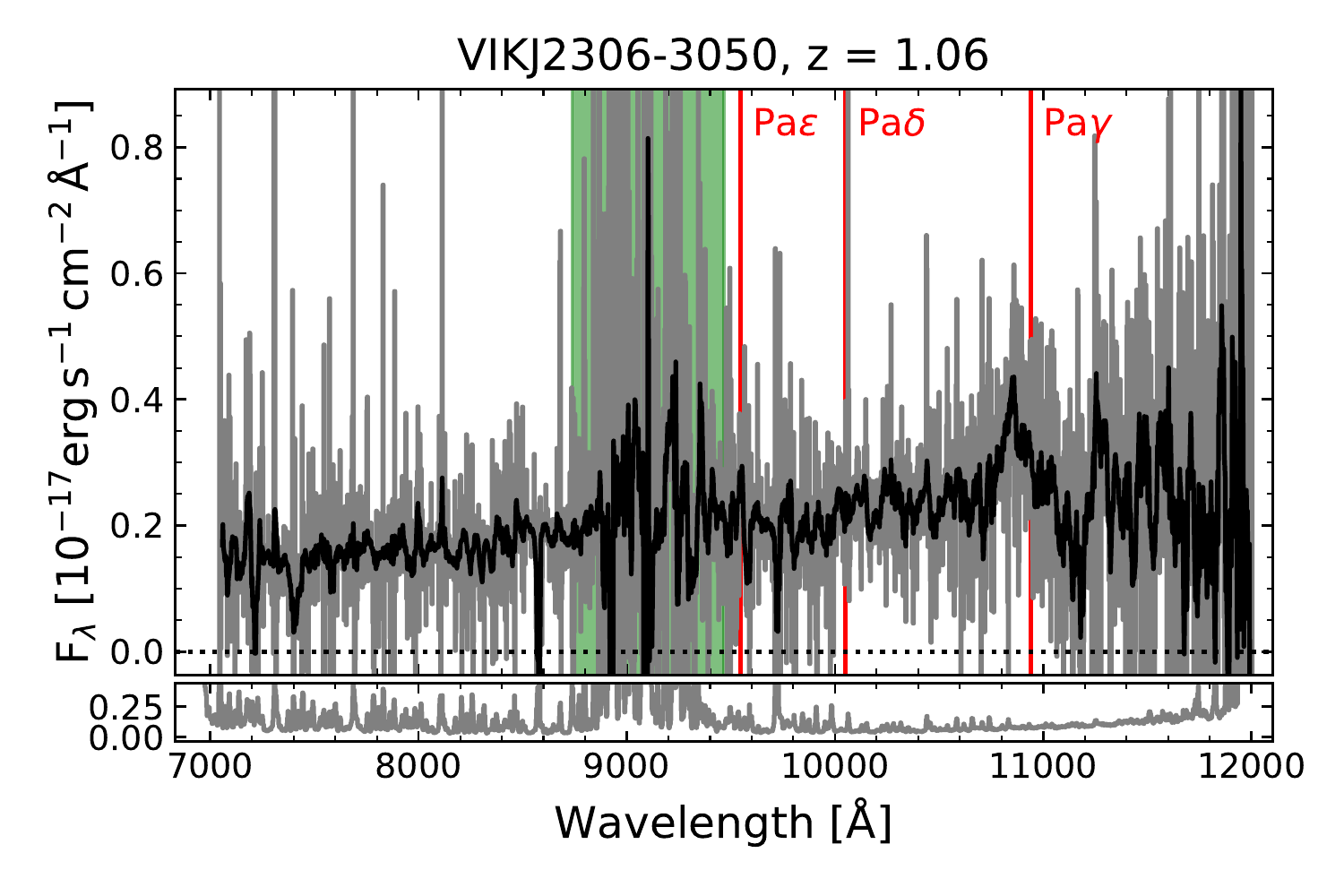}
\includegraphics[width=\columnwidth]{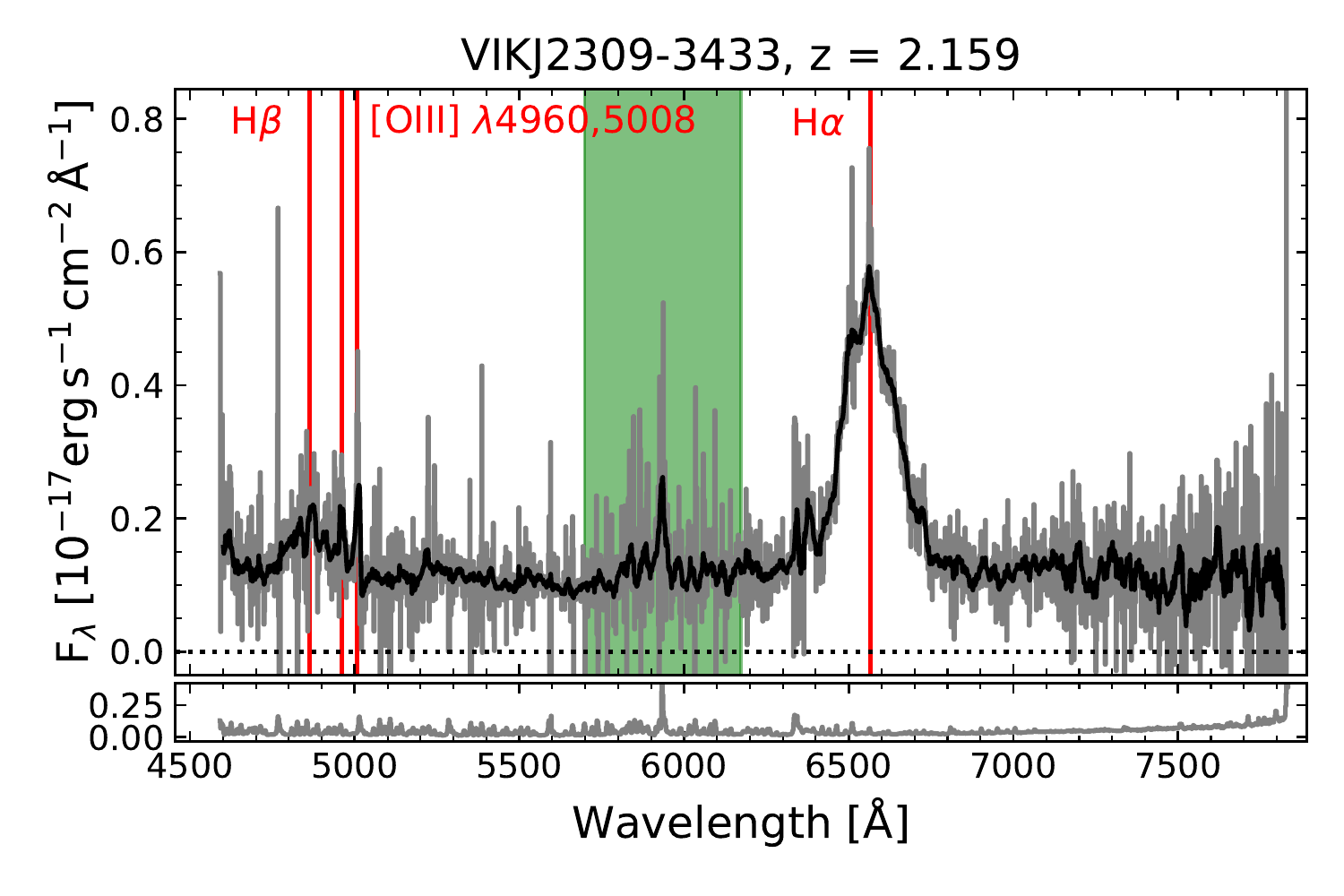}
\includegraphics[width=\columnwidth]{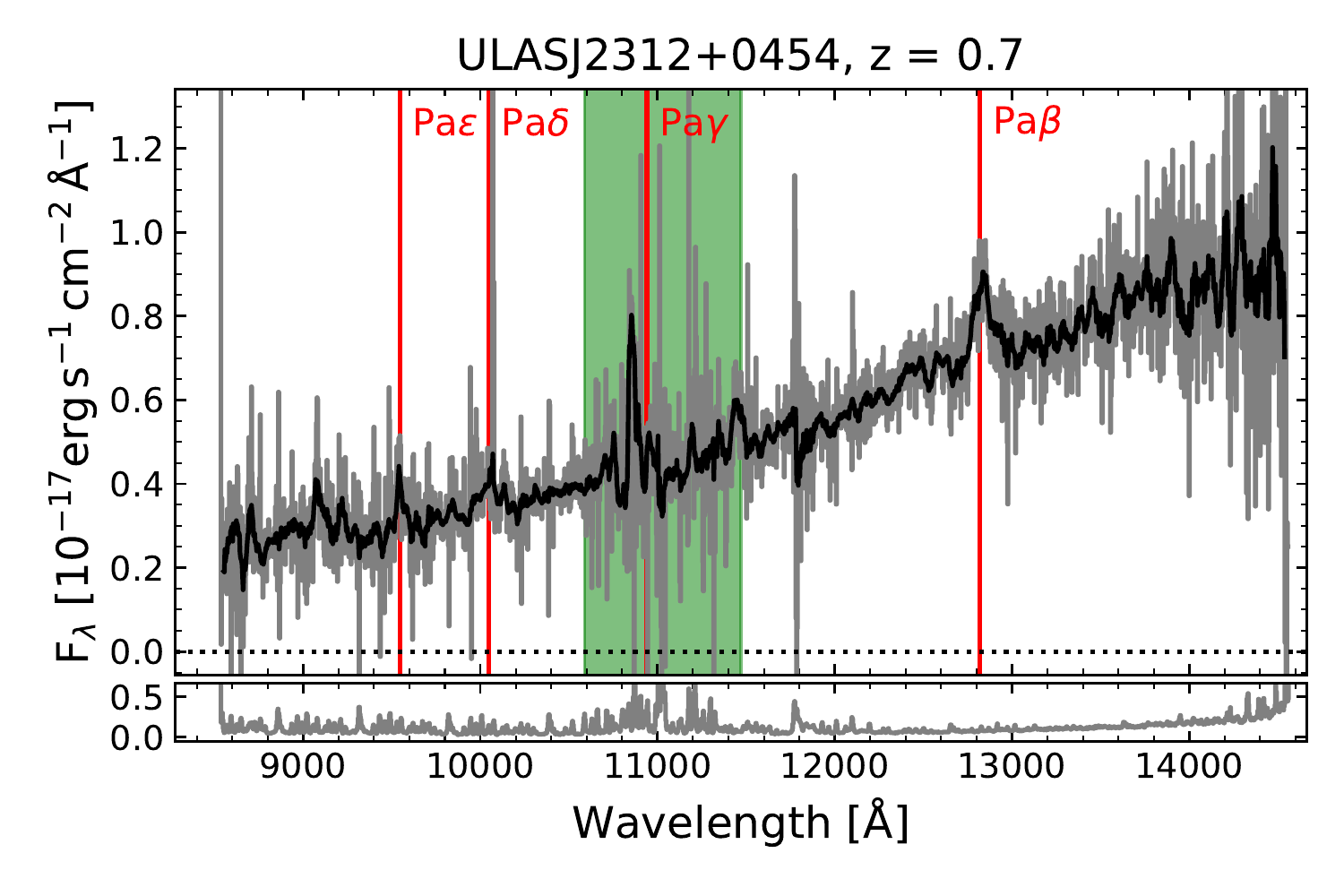}
\includegraphics[width=\columnwidth]{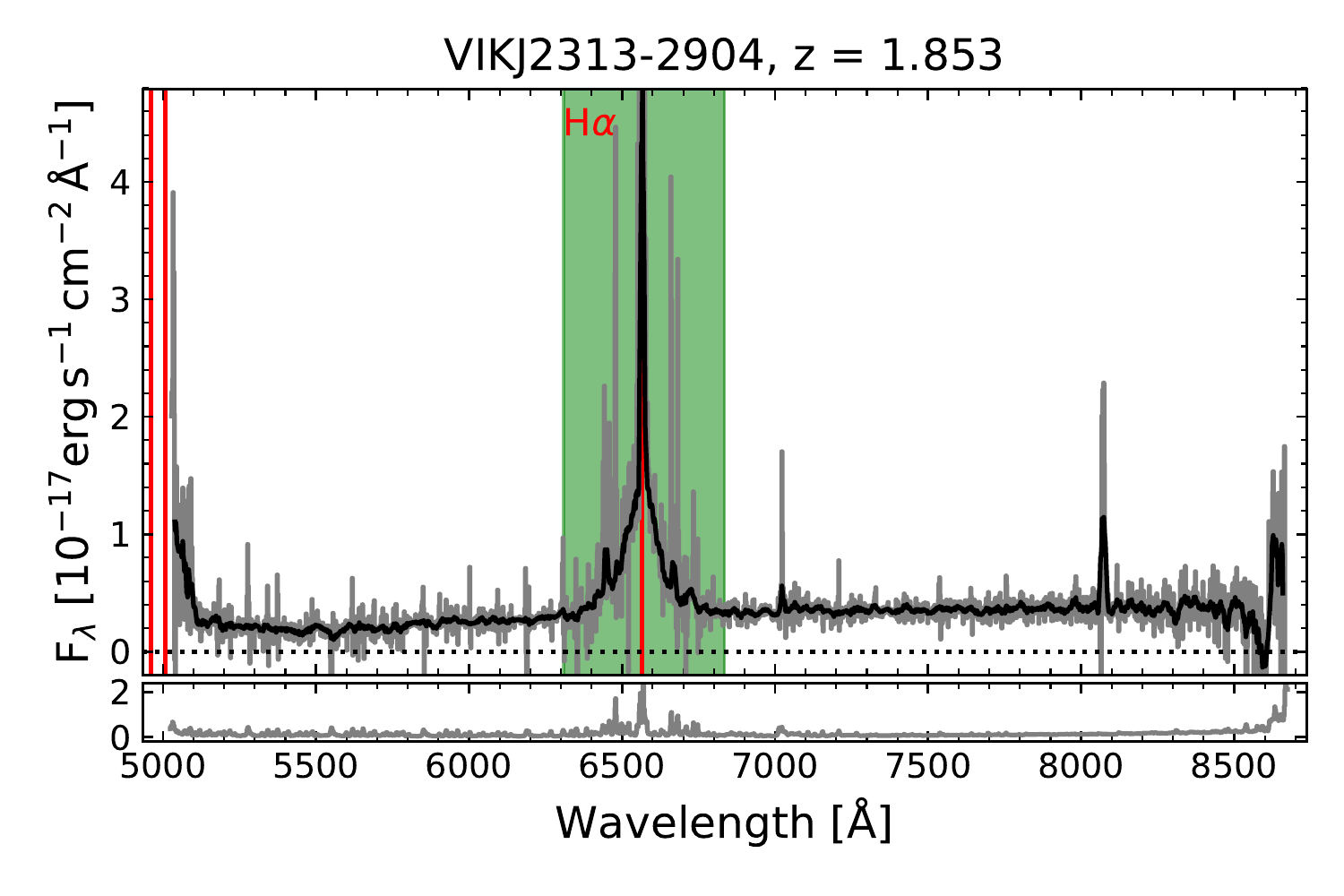}
\includegraphics[width=\columnwidth]{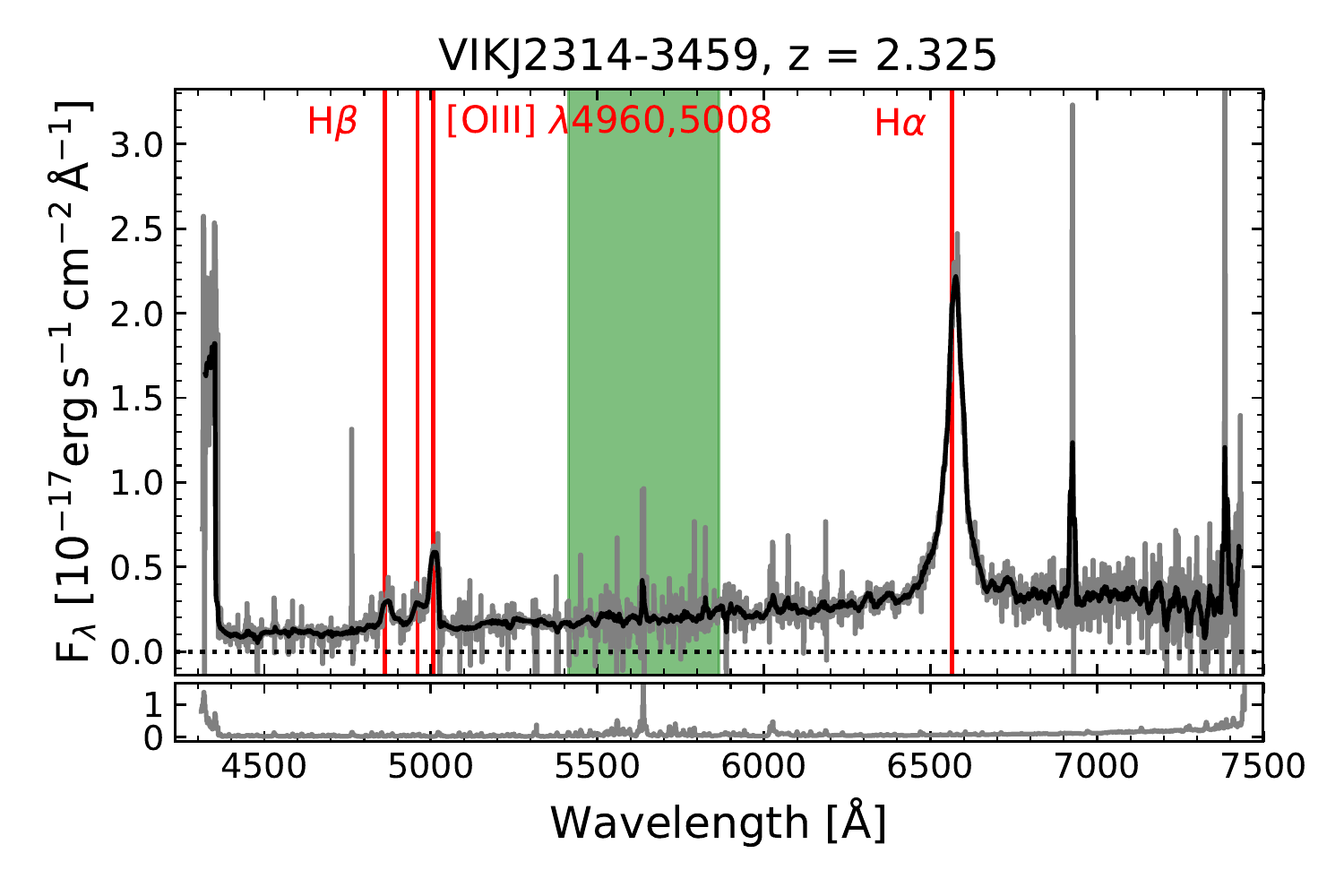}
\includegraphics[width=\columnwidth]{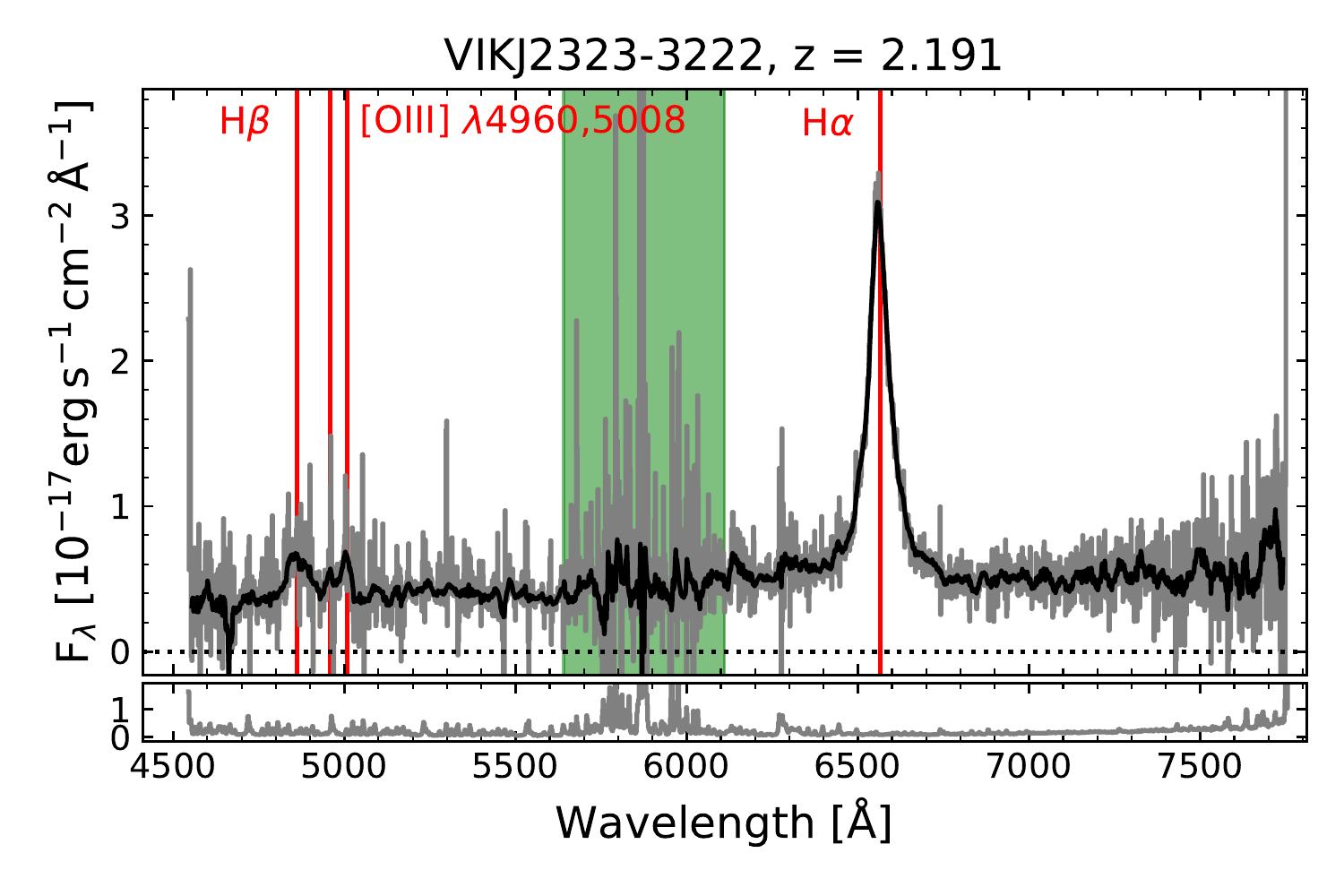}
\includegraphics[width=\columnwidth]{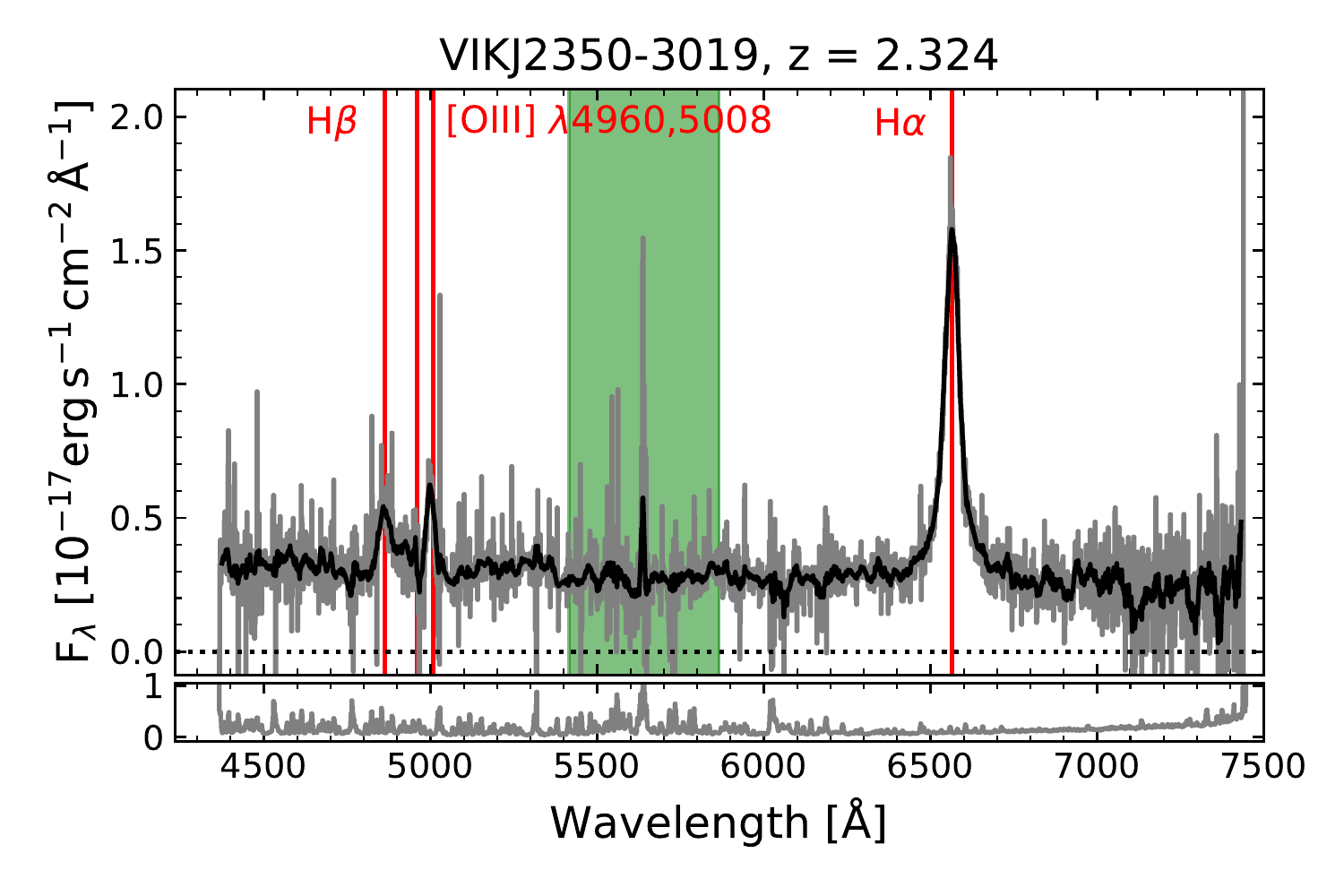}
\includegraphics[width=\columnwidth]{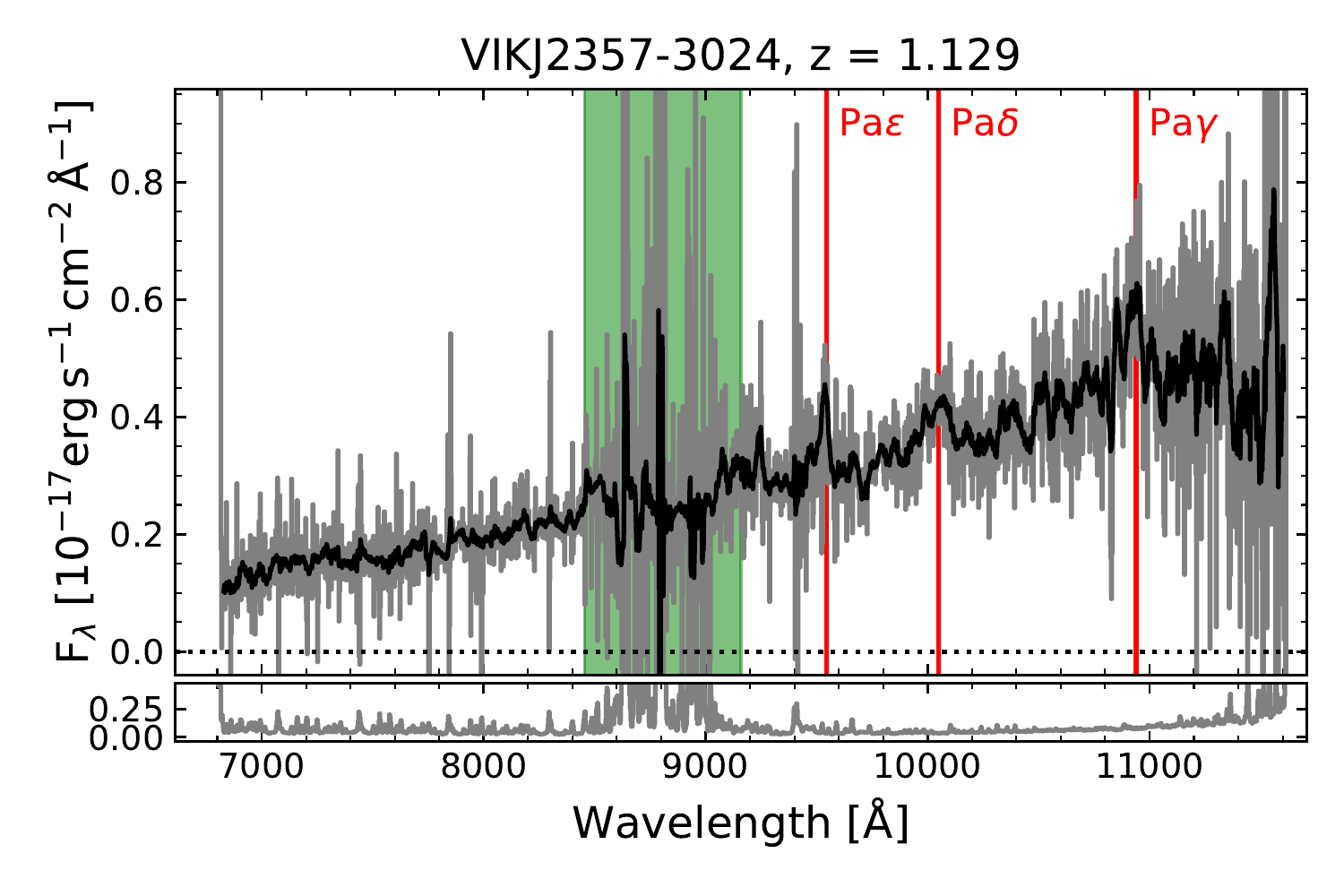}
\end{center}
\contcaption{}
\end{figure*}

\begin{figure*}
\includegraphics[width=0.666\columnwidth]{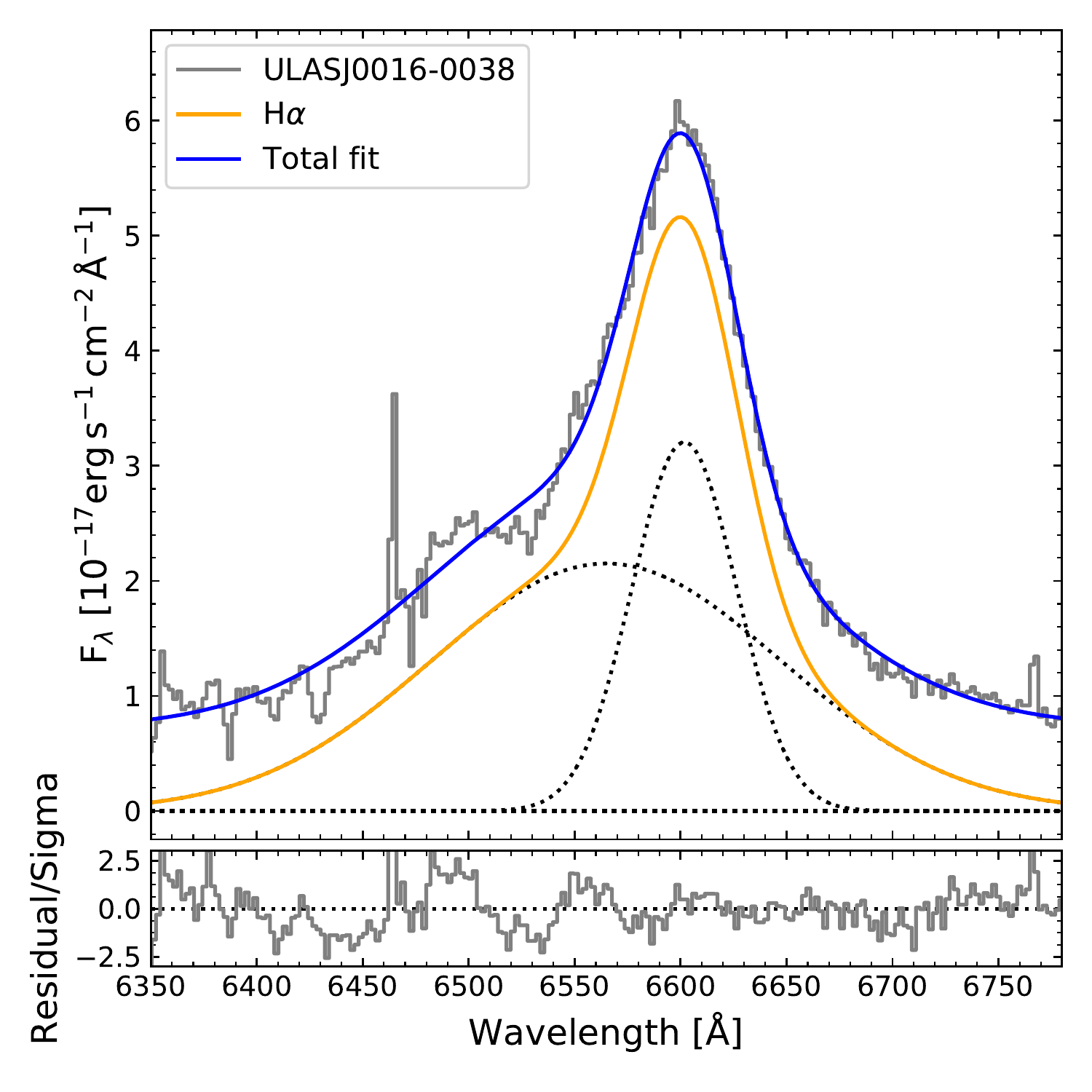}
\includegraphics[width=0.666\columnwidth]{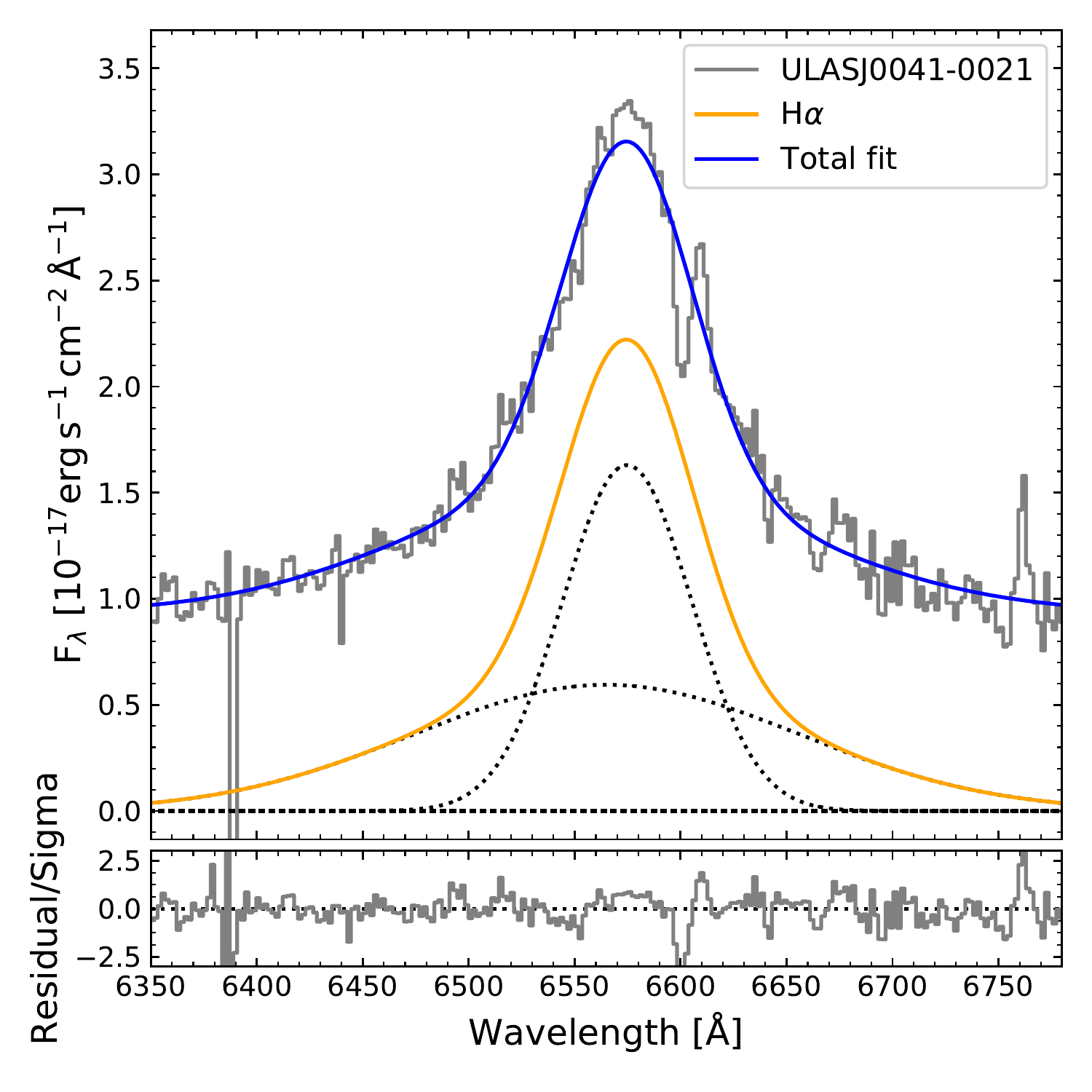}
\includegraphics[width=0.666\columnwidth]{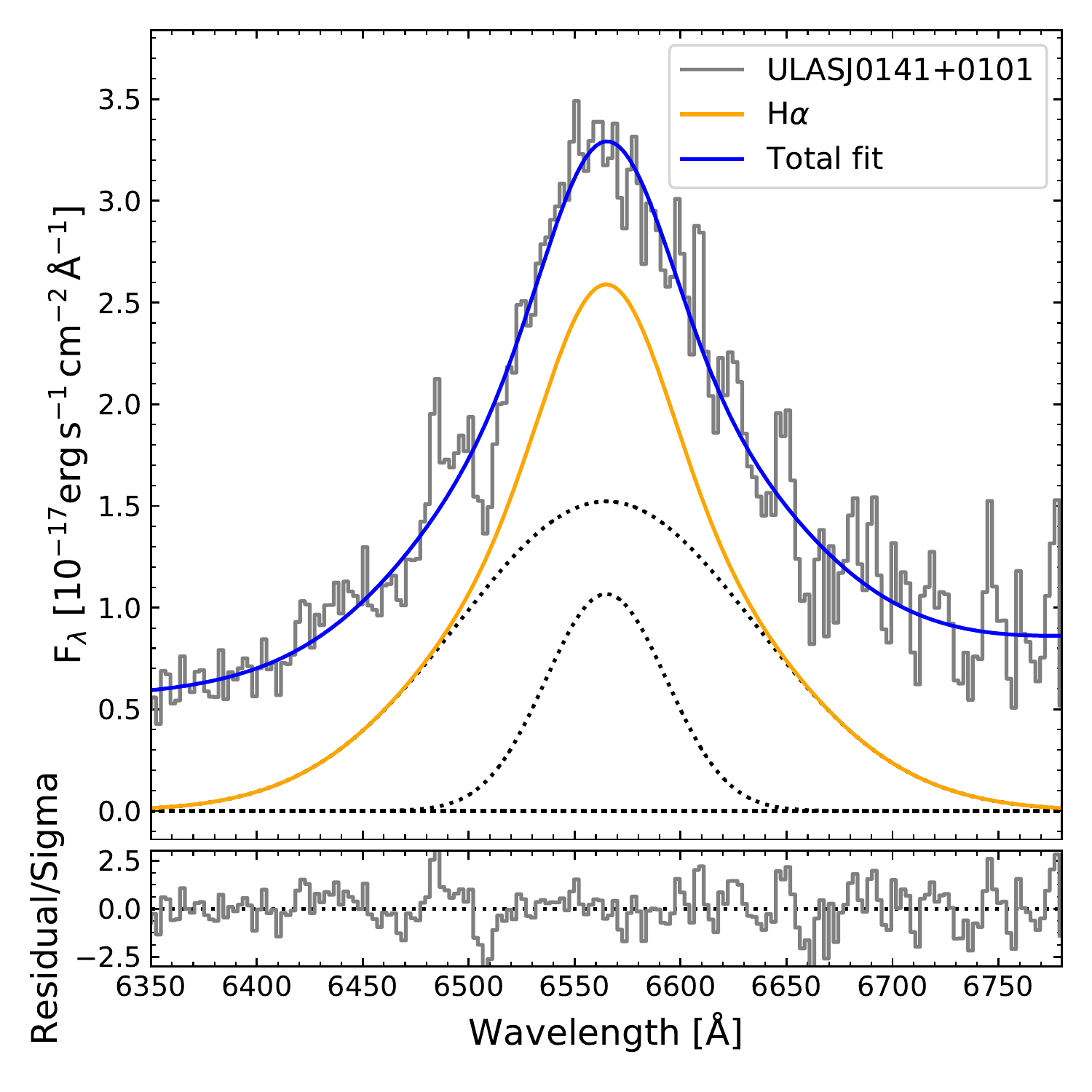}
\includegraphics[width=0.666\columnwidth]{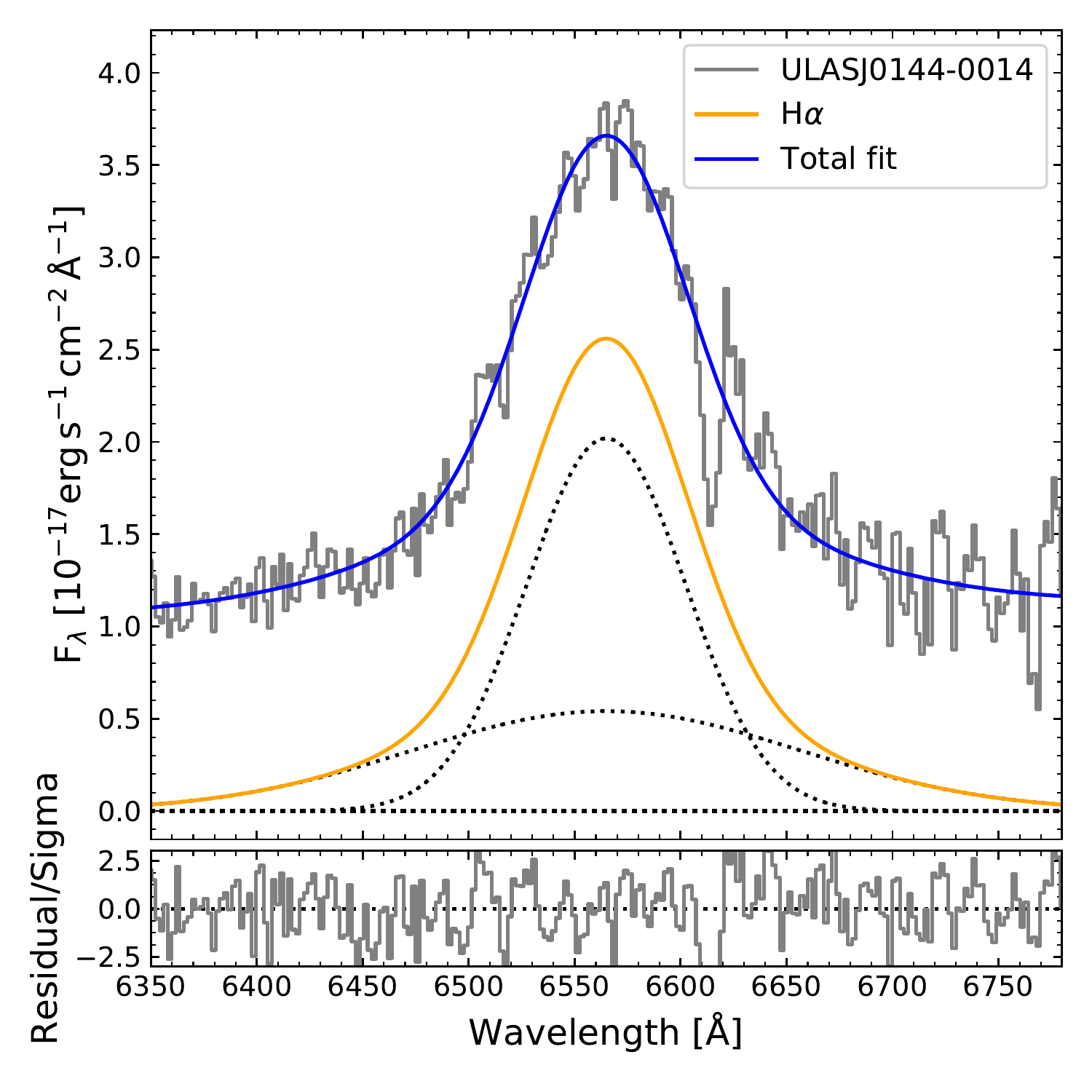}
\includegraphics[width=0.666\columnwidth]{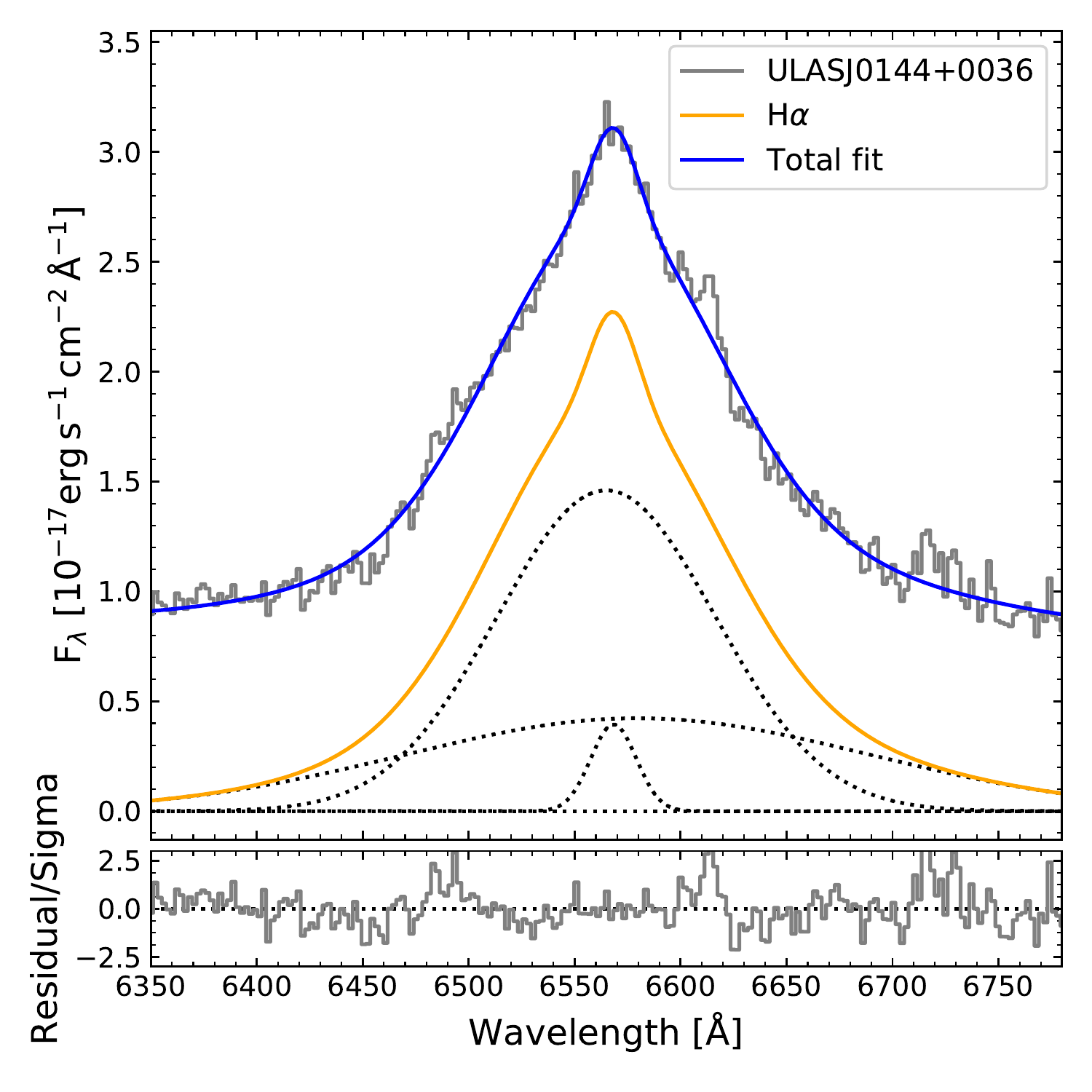}
\includegraphics[width=0.666\columnwidth]{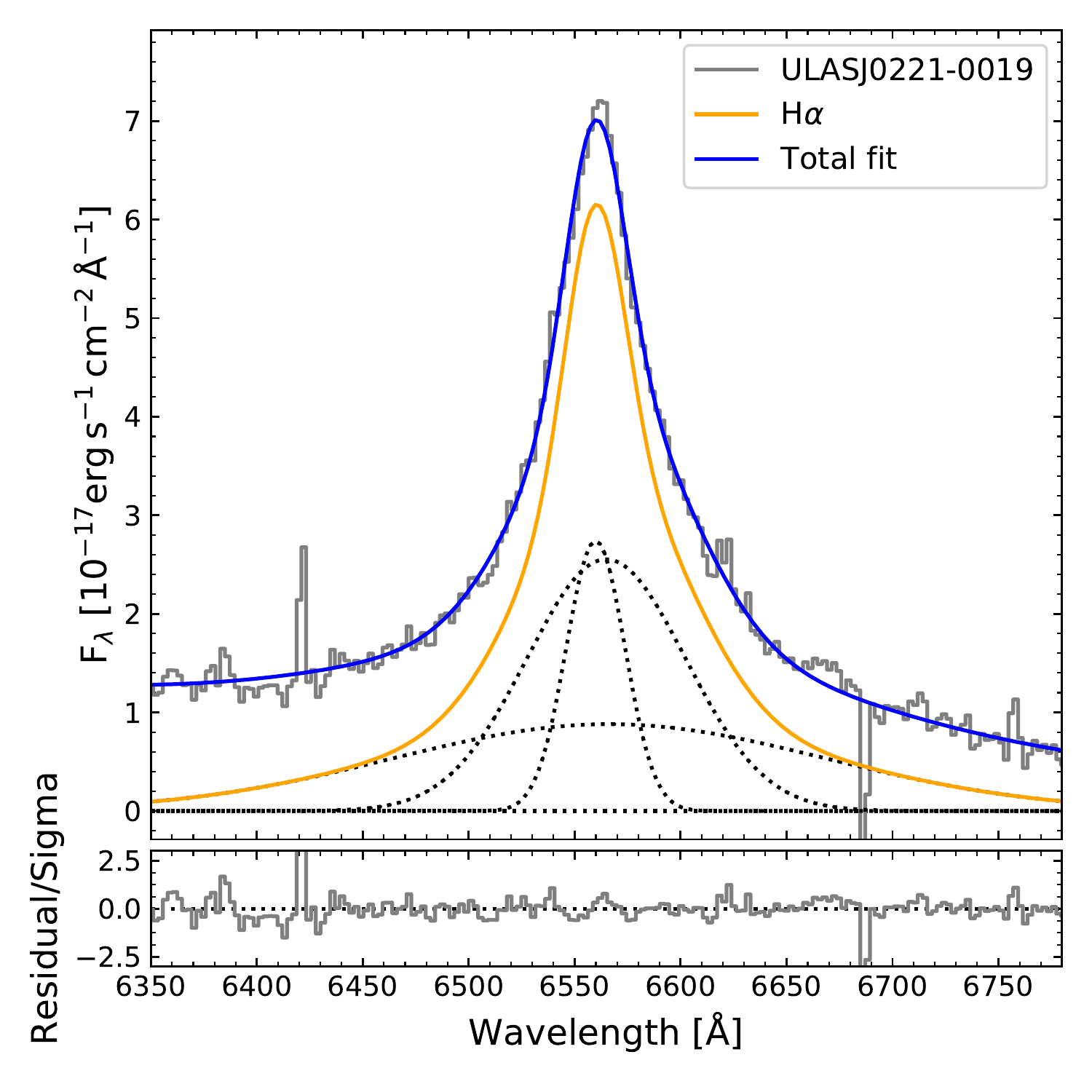}
\includegraphics[width=0.666\columnwidth]{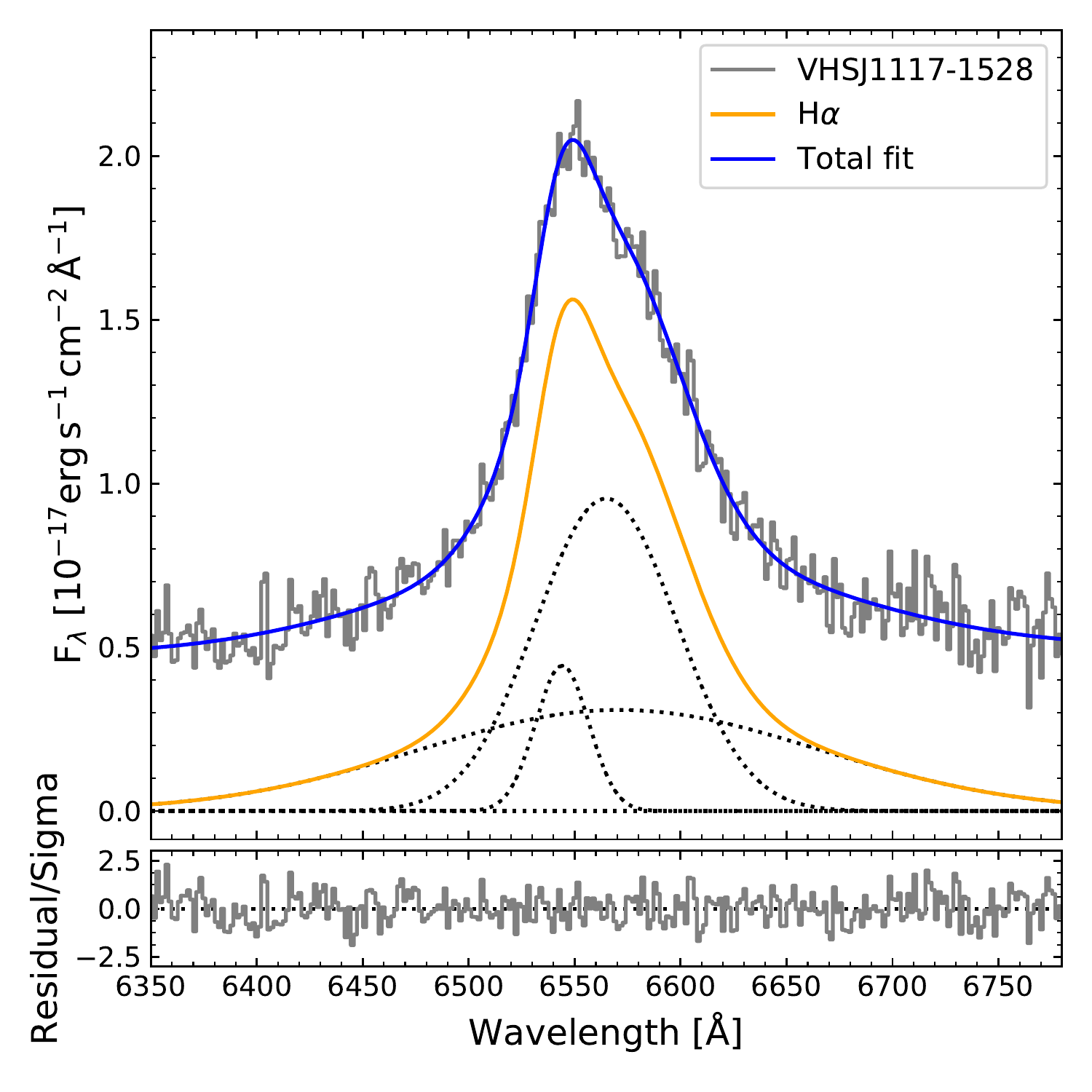}
\includegraphics[width=0.666\columnwidth]{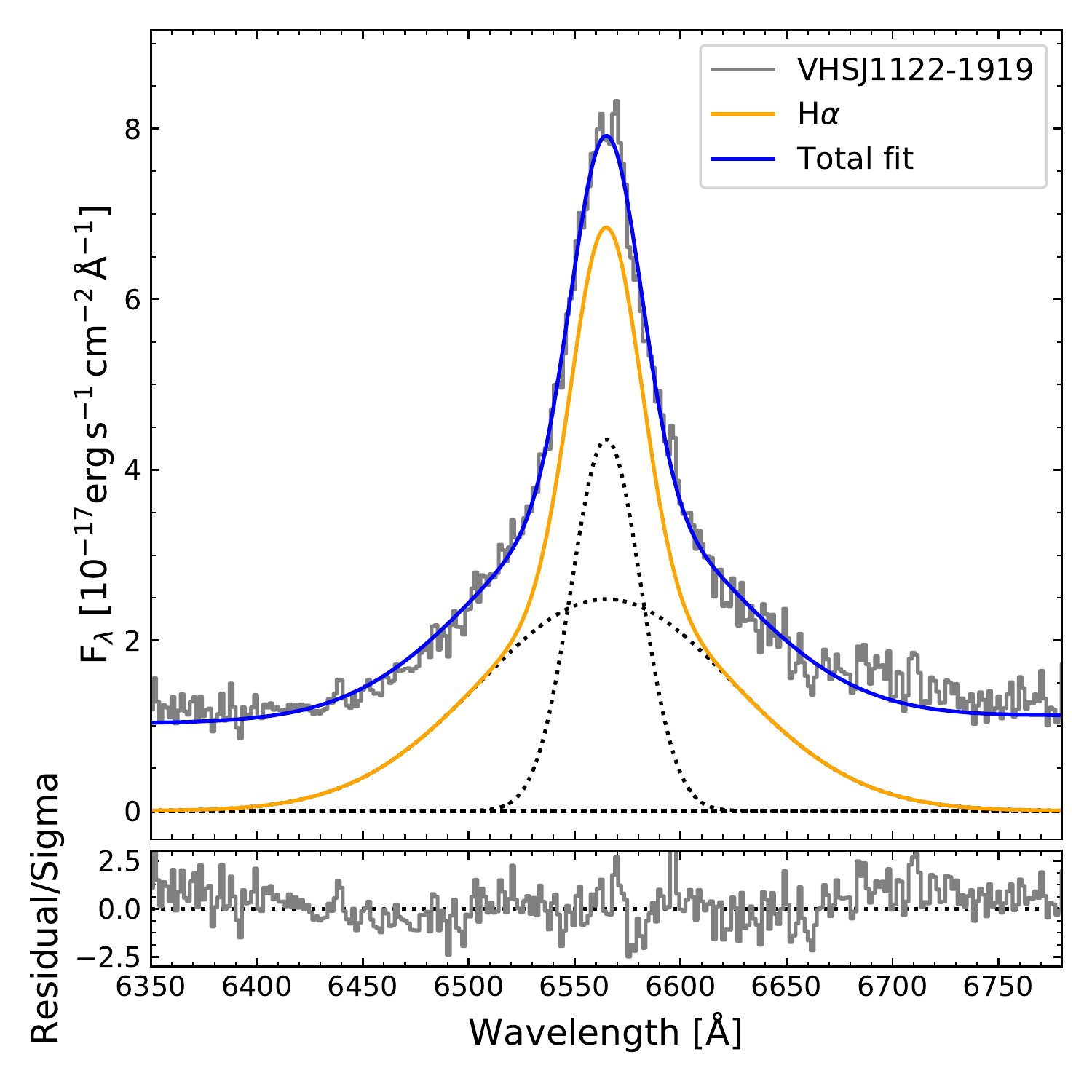}
\includegraphics[width=0.666\columnwidth]{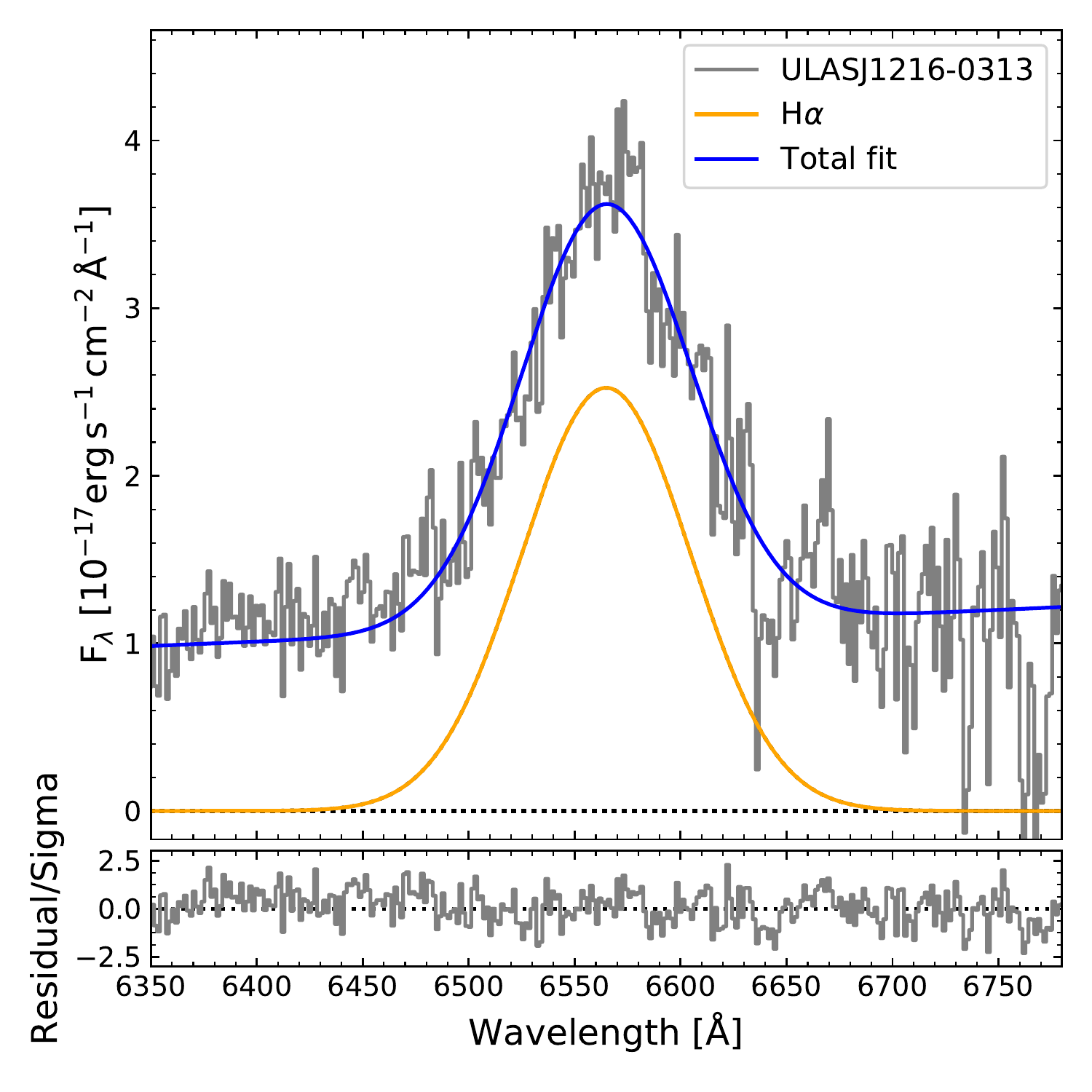}
\includegraphics[width=0.666\columnwidth]{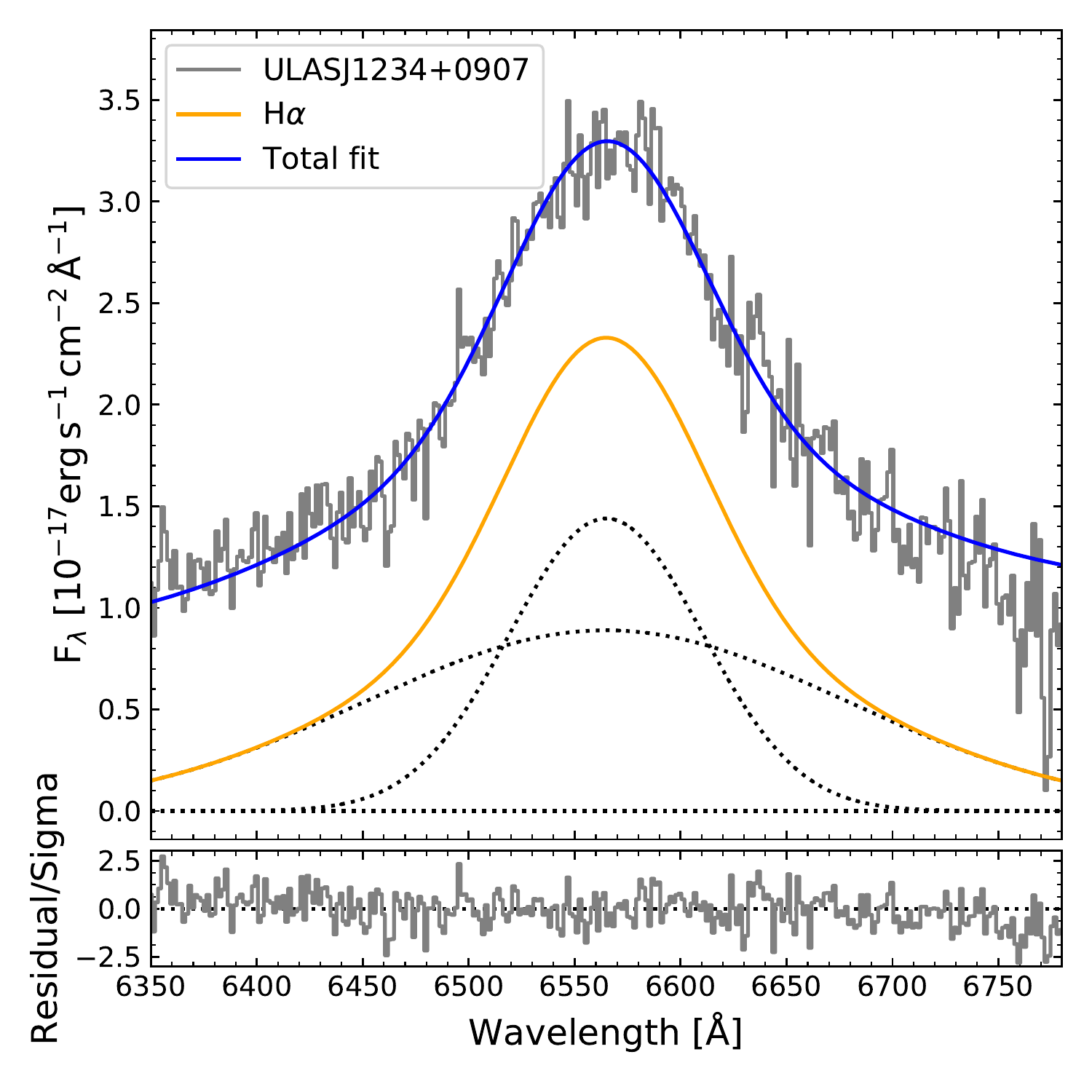}
\includegraphics[width=0.666\columnwidth]{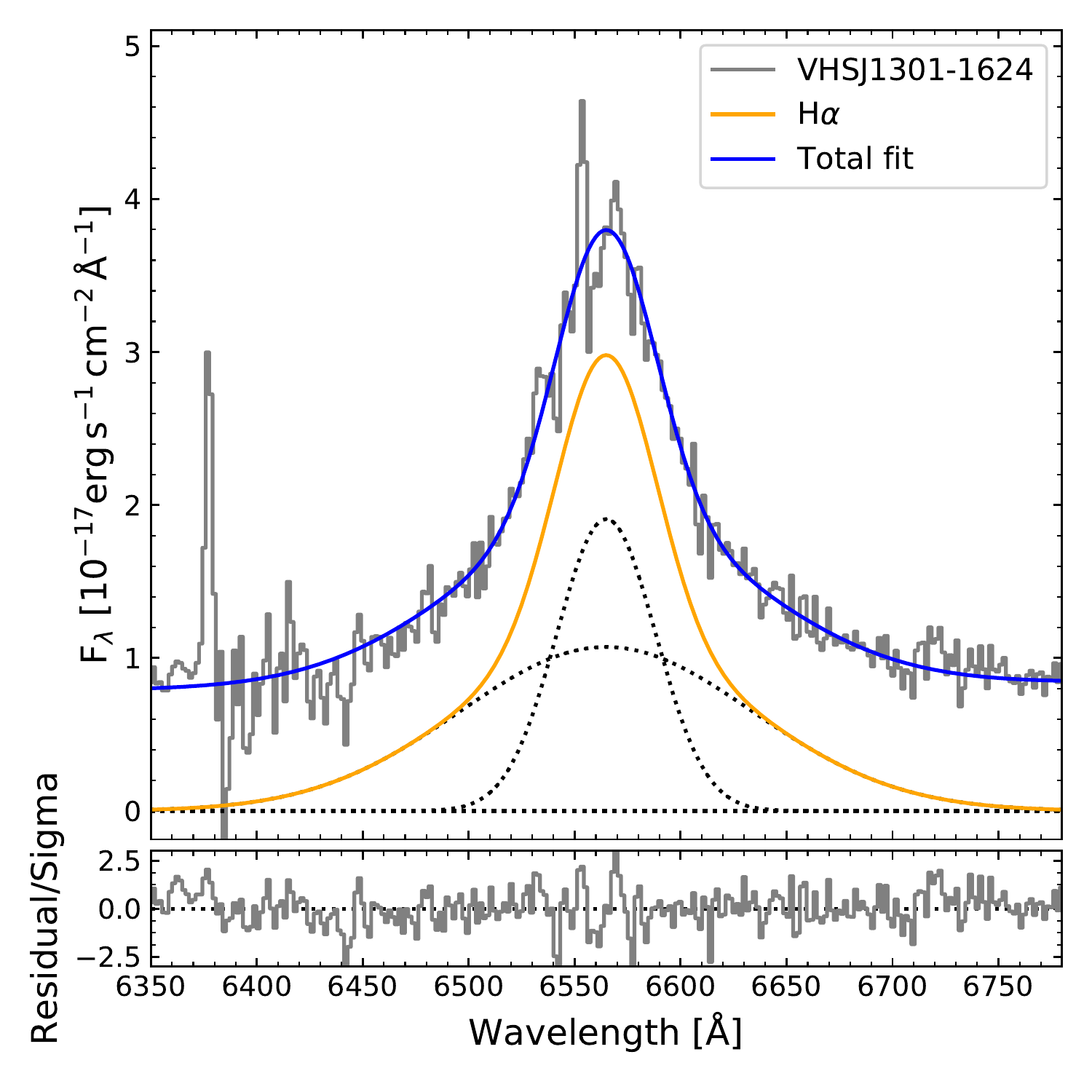}
\includegraphics[width=0.666\columnwidth]{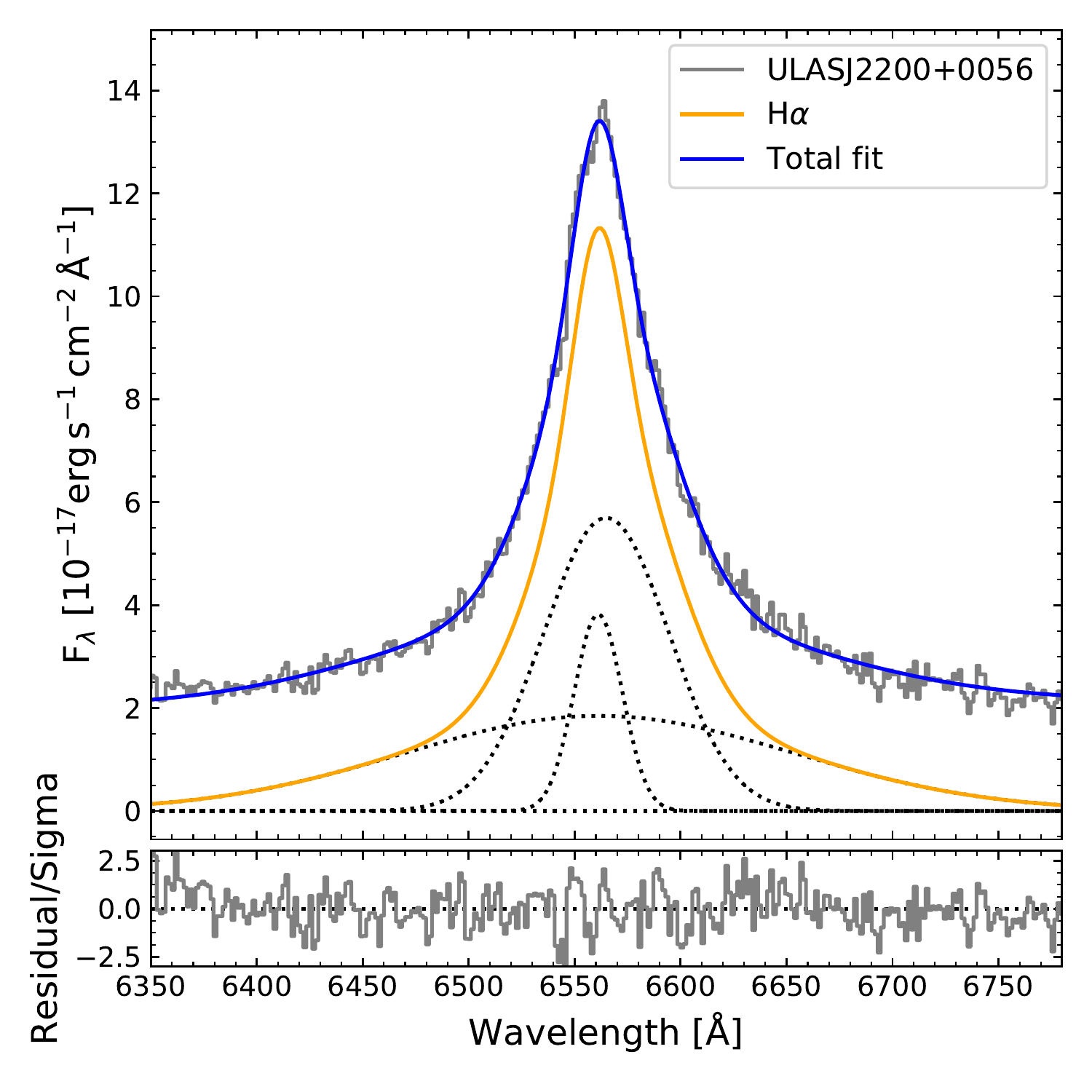}
    \caption{Fits to the H\,$\alpha$ emission line. Each Gaussian component is shown with dotted lines. Residuals are shown in the panels below, scaled by the noise.}
    \label{fig:Ha}
\end{figure*}
\begin{figure*}
\begin{center}
\includegraphics[width=0.666\columnwidth]{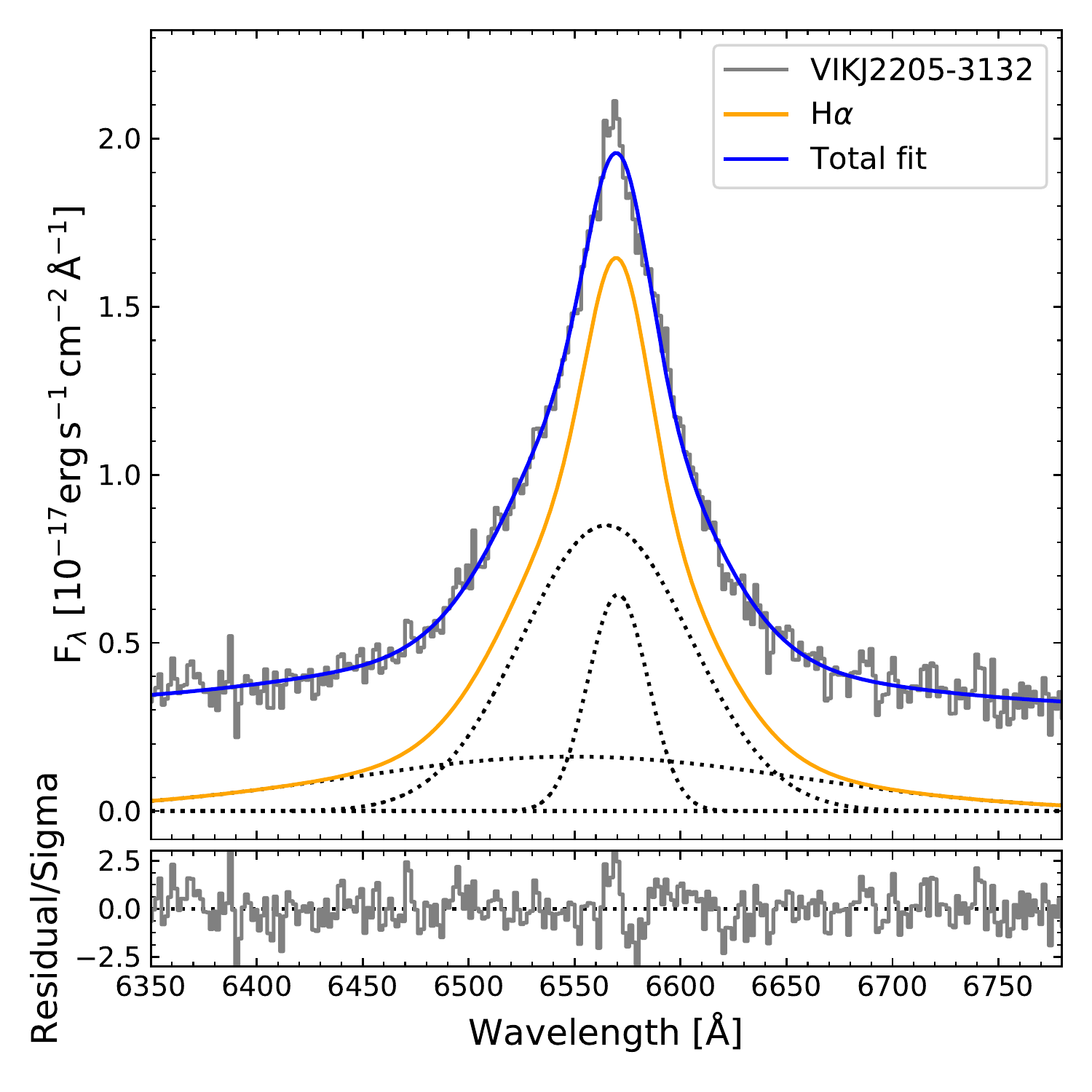}
\includegraphics[width=0.666\columnwidth]{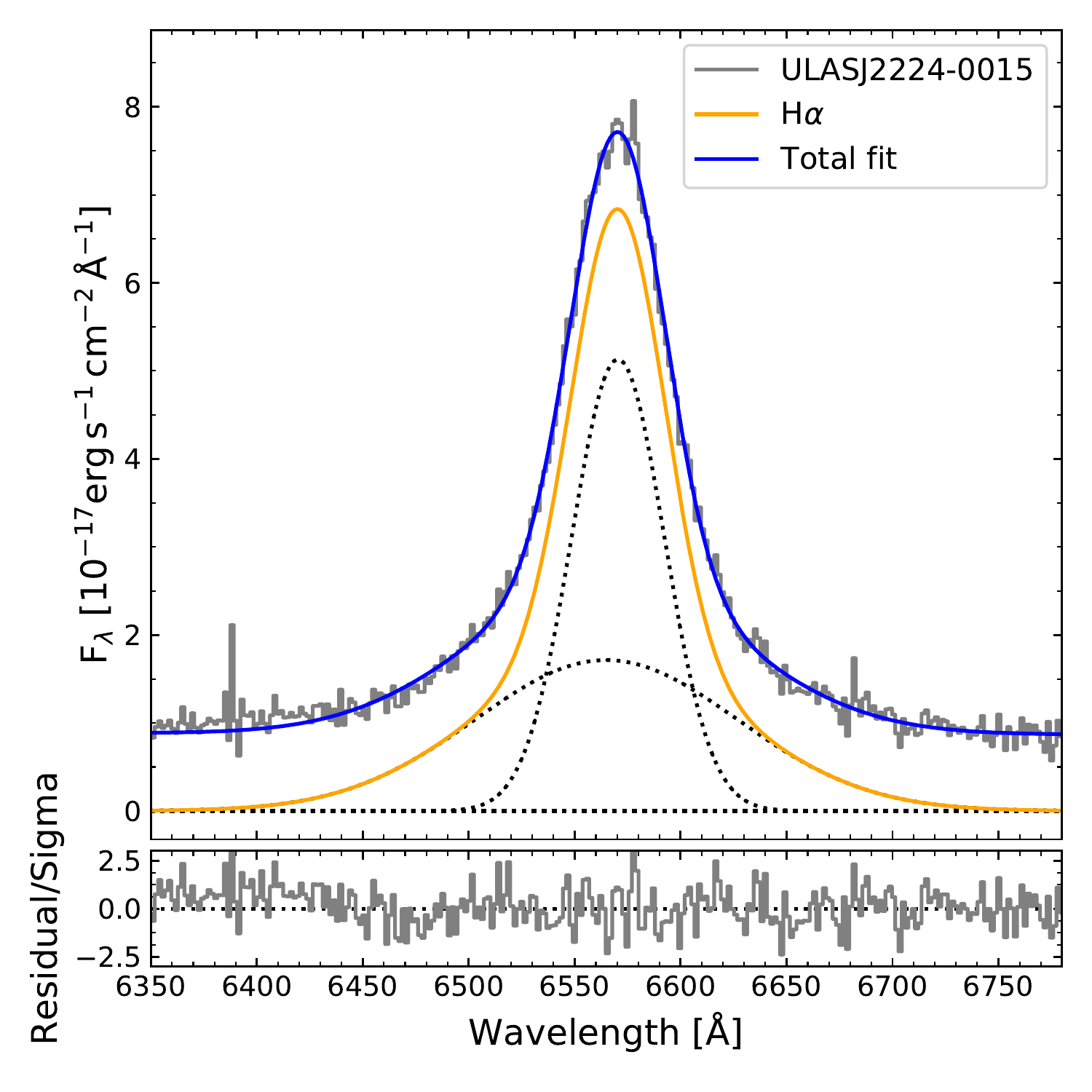}
\includegraphics[width=0.666\columnwidth]{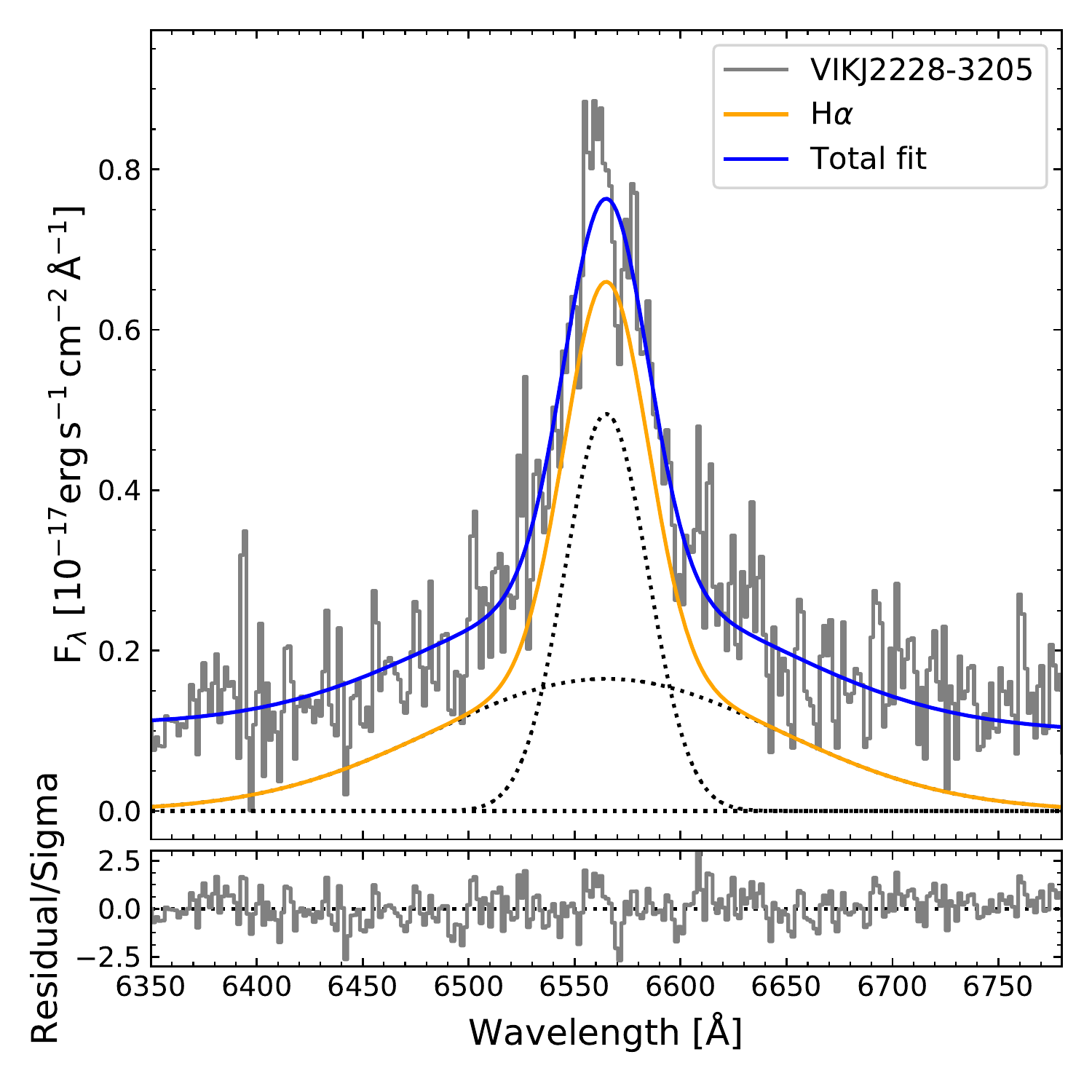}
\includegraphics[width=0.666\columnwidth]{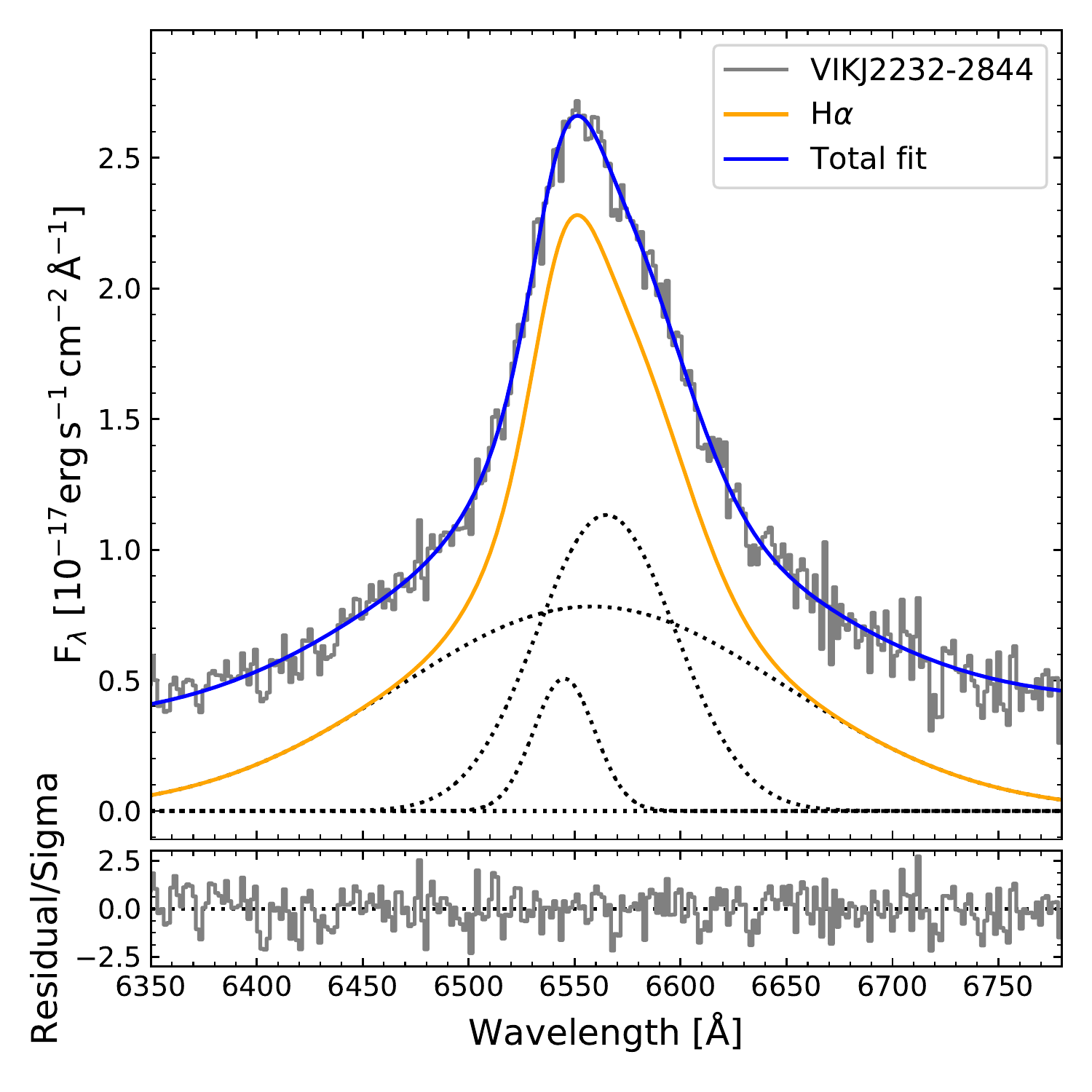}
\includegraphics[width=0.666\columnwidth]{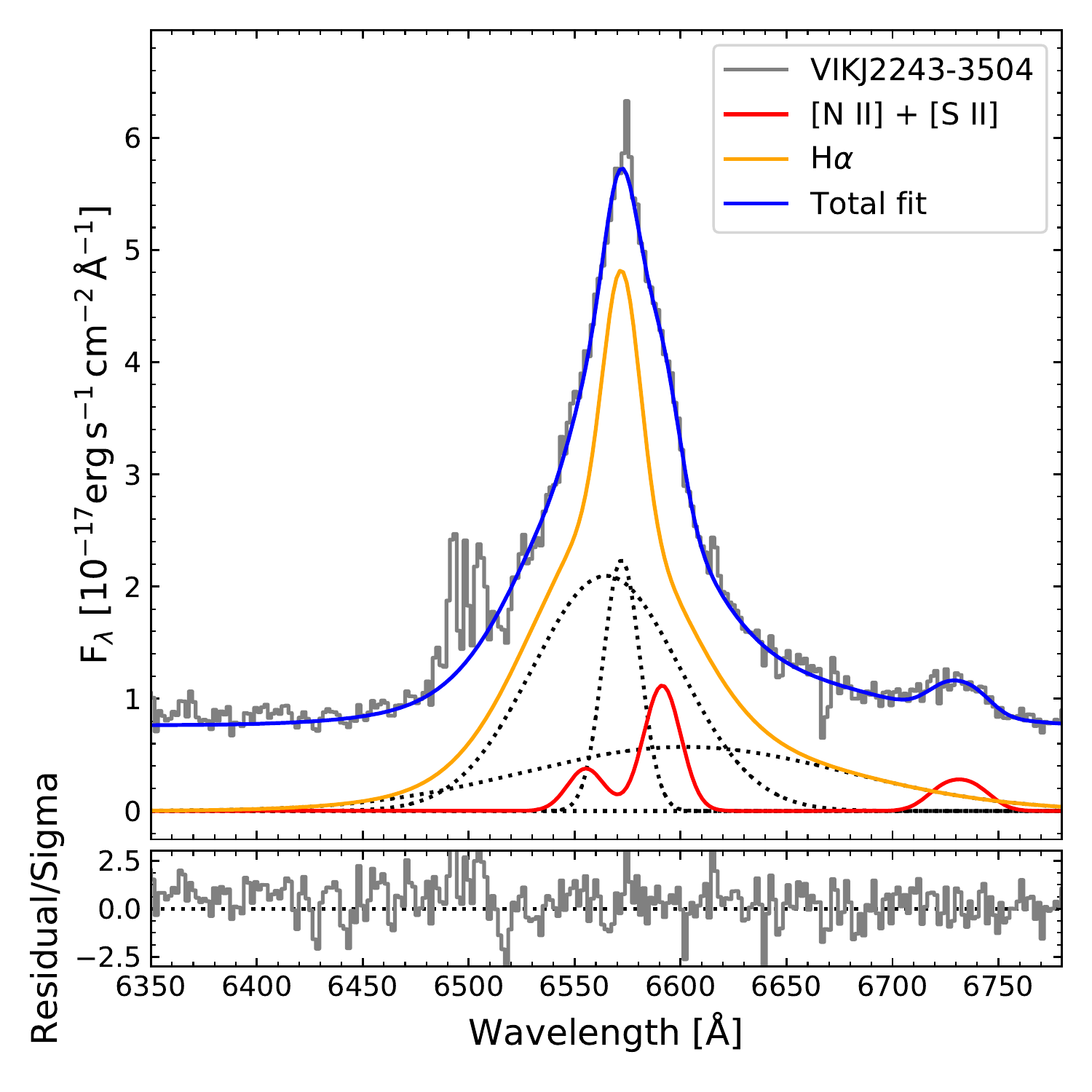}
\includegraphics[width=0.666\columnwidth]{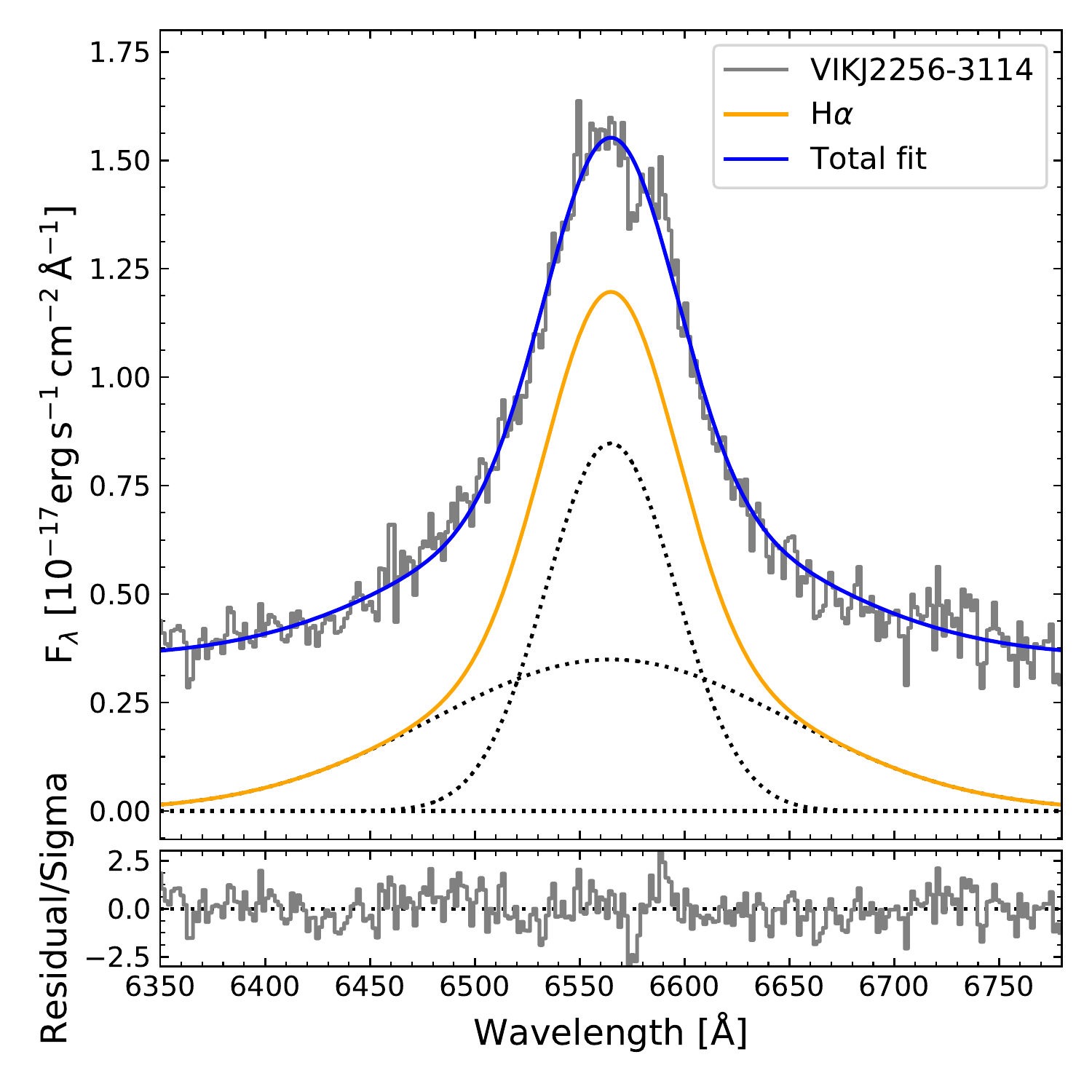}
\includegraphics[width=0.666\columnwidth]{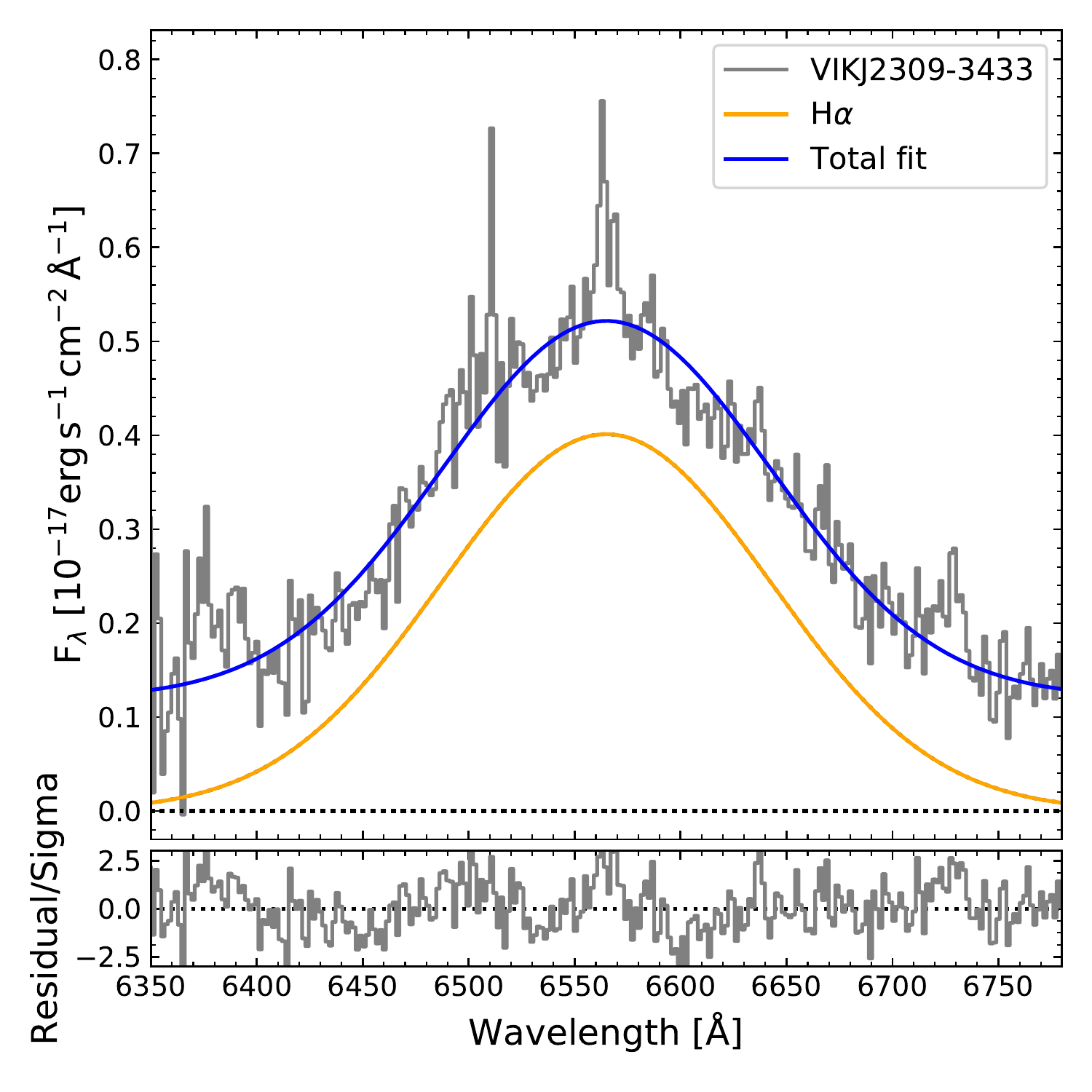}
\includegraphics[width=0.666\columnwidth]{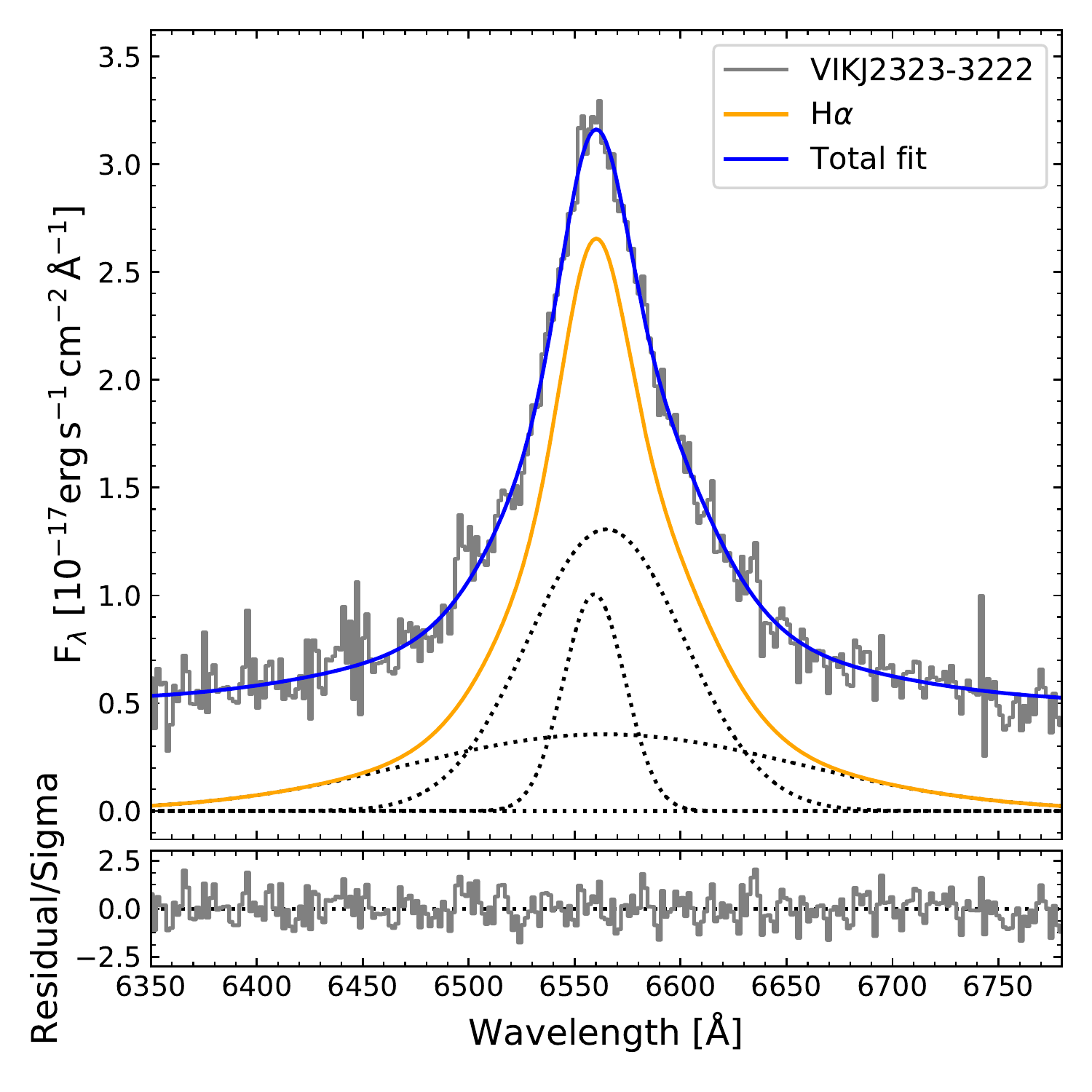}
\includegraphics[width=0.666\columnwidth]{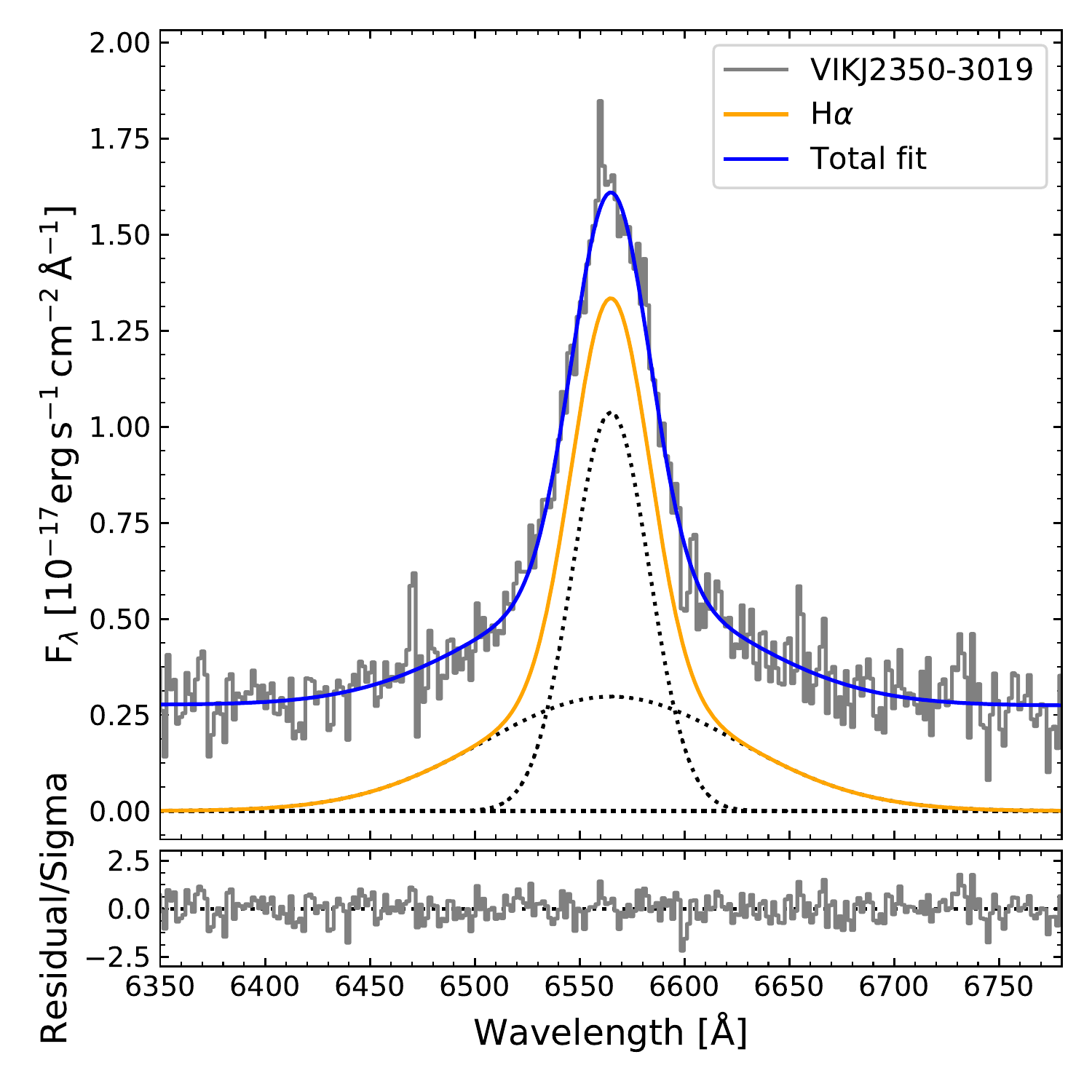}
\end{center}
    \contcaption{}
\end{figure*}

\begin{figure*}
\includegraphics[width=0.666\columnwidth]{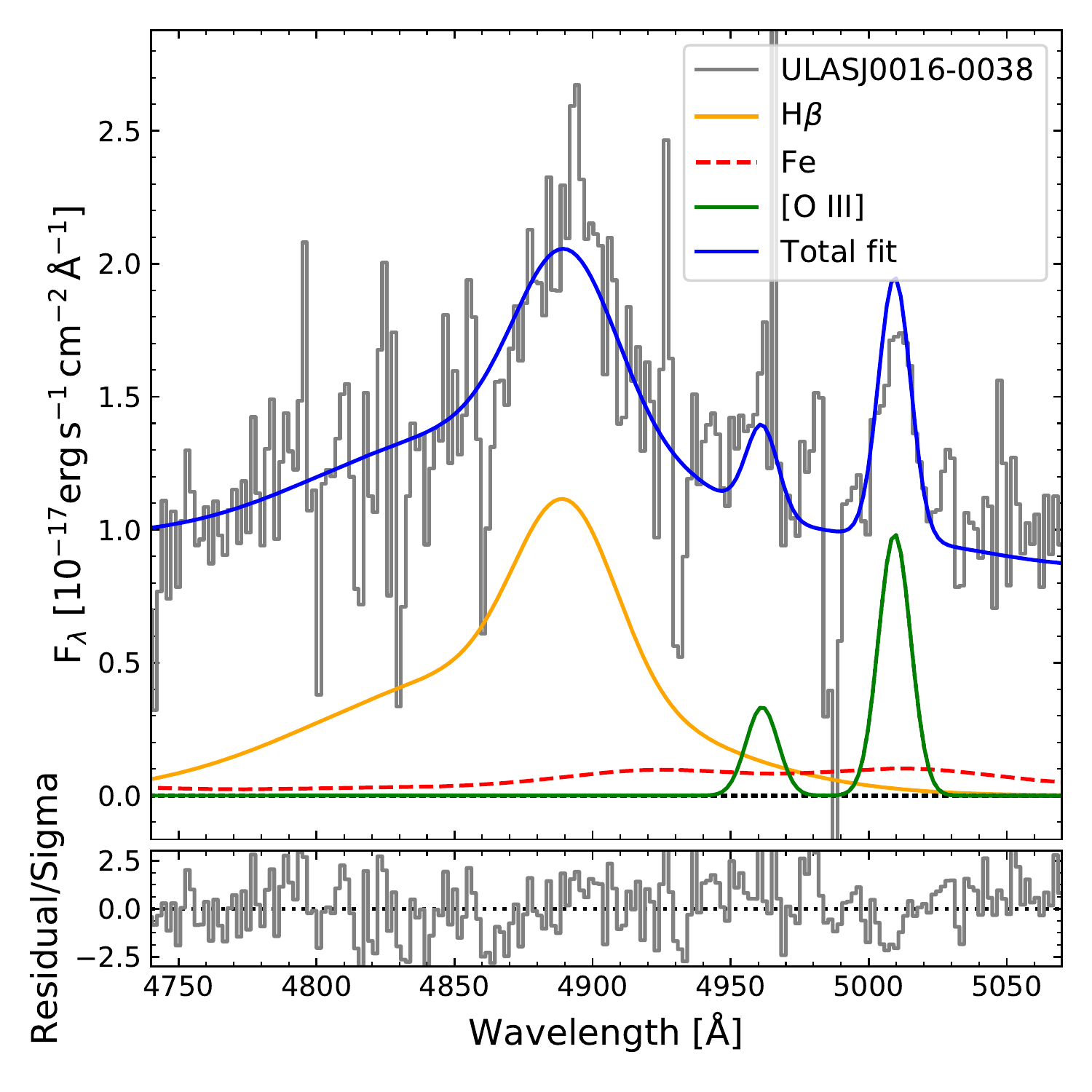}
\includegraphics[width=0.666\columnwidth]{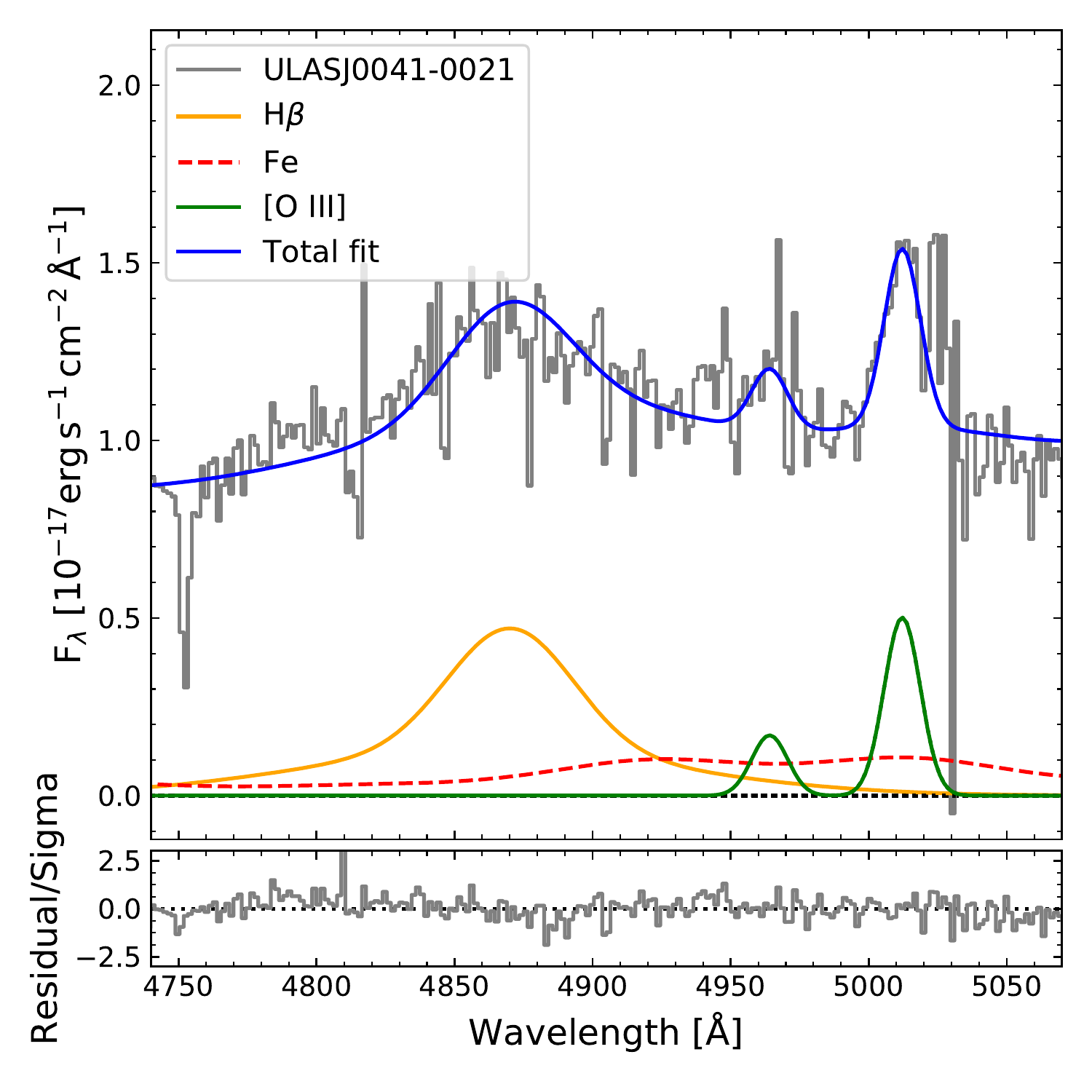}
\includegraphics[width=0.666\columnwidth]{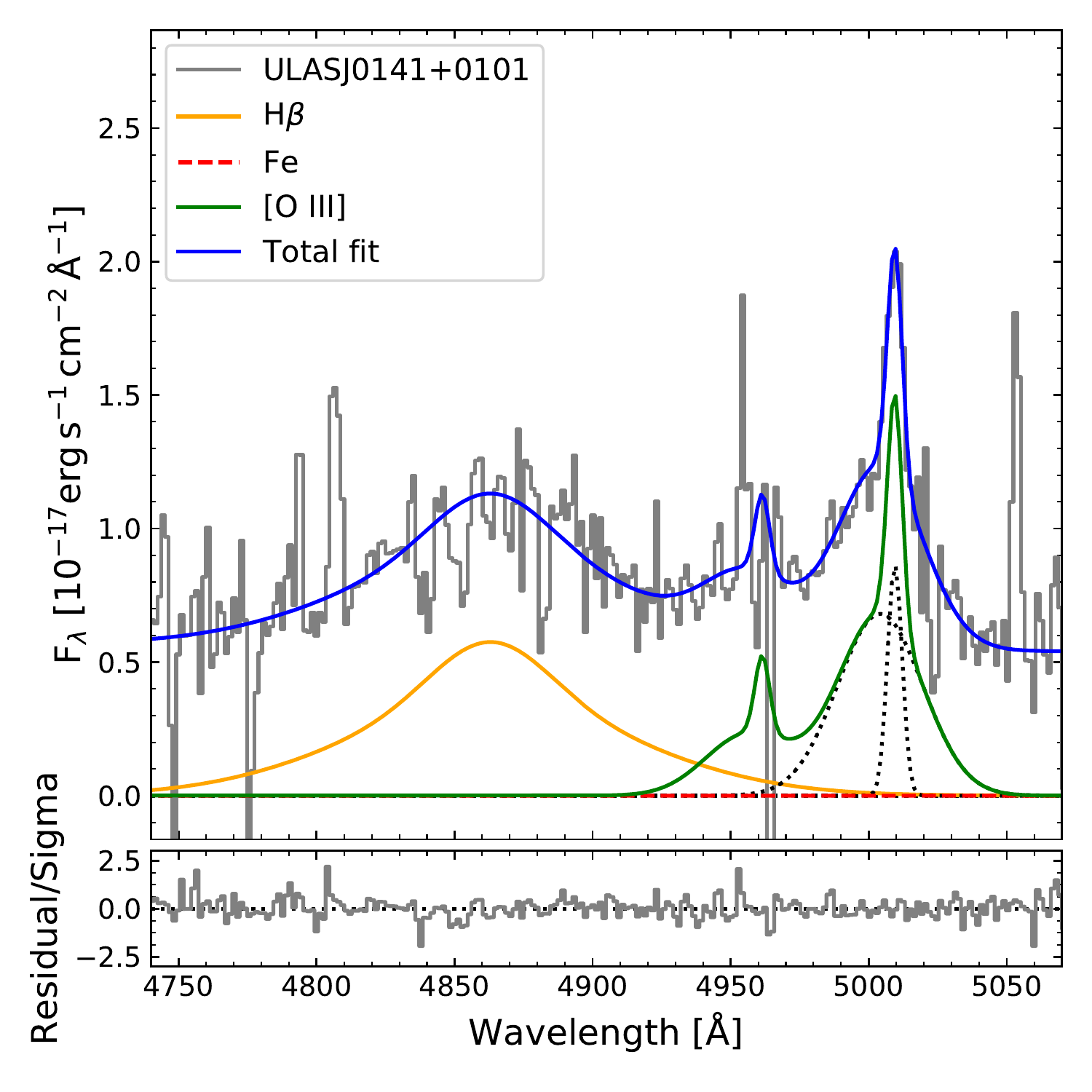}
\includegraphics[width=0.666\columnwidth]{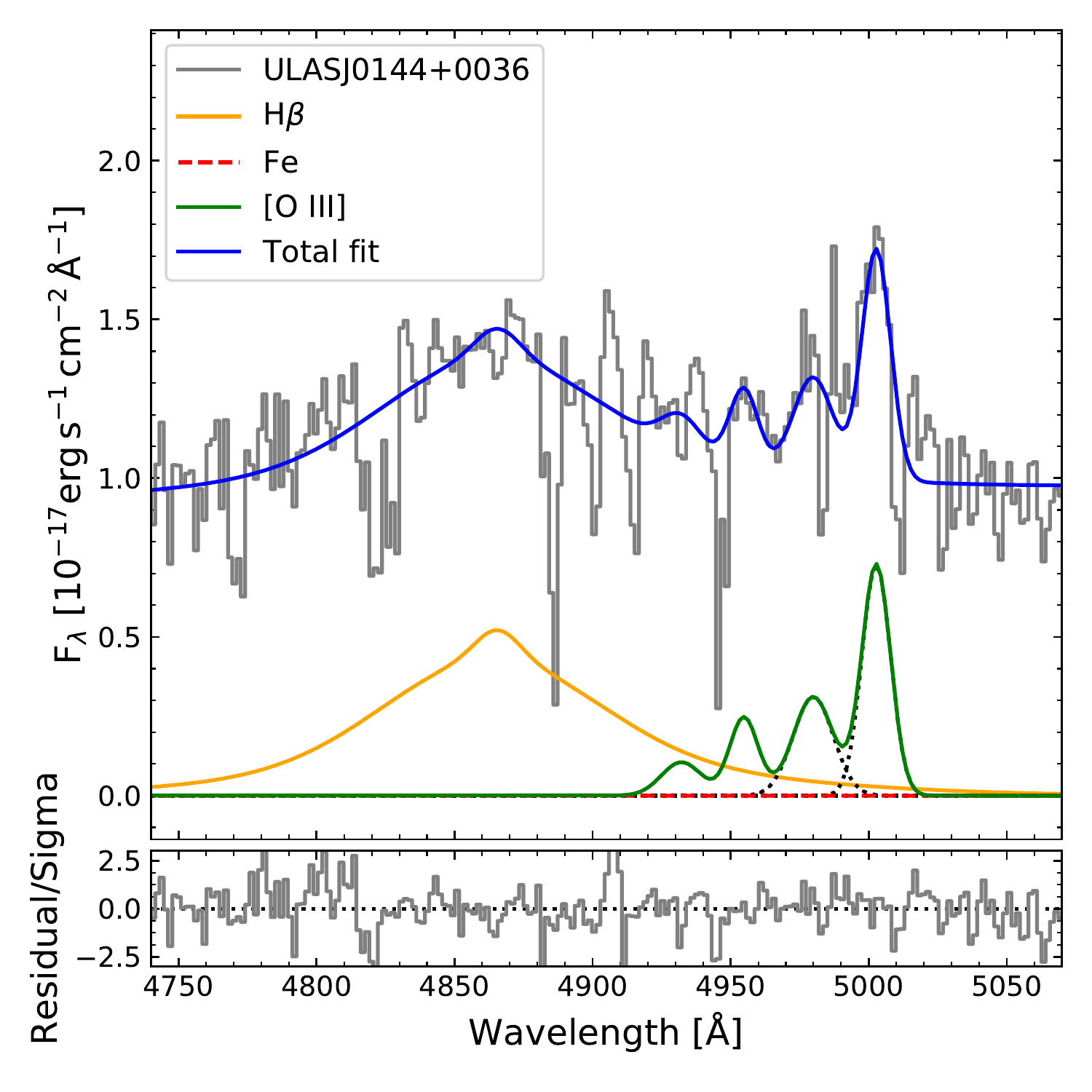}
\includegraphics[width=0.666\columnwidth]{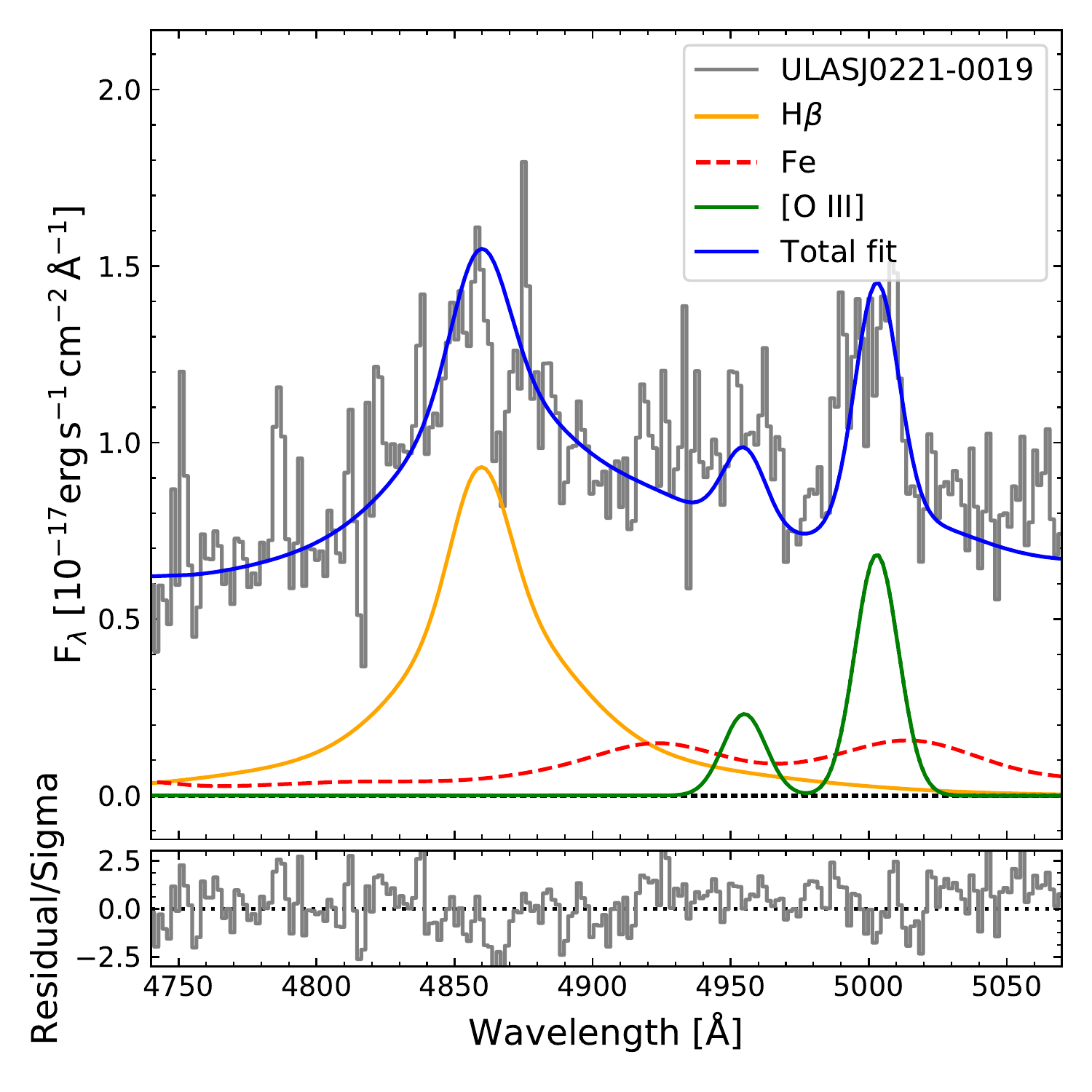}
\includegraphics[width=0.666\columnwidth]{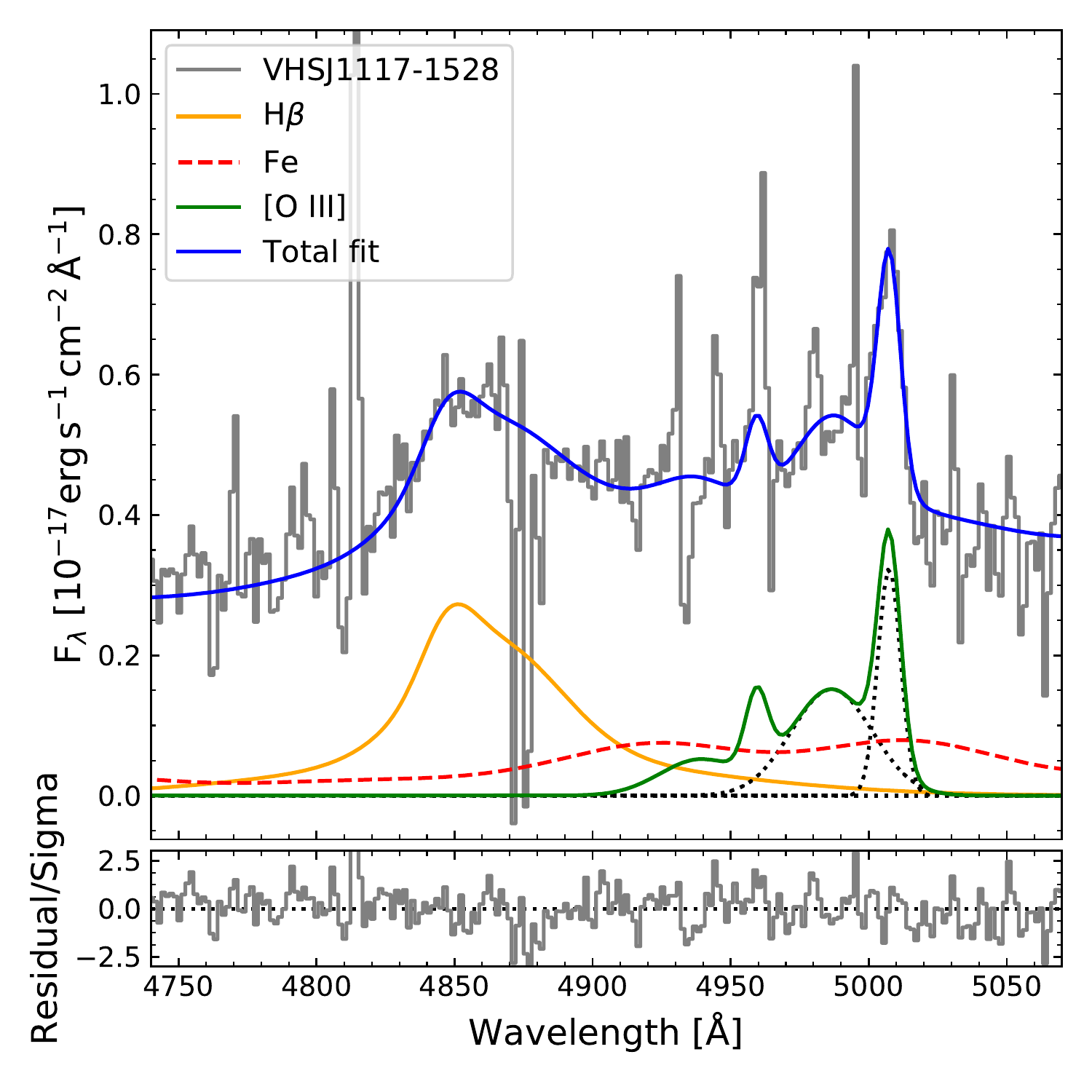}
\includegraphics[width=0.666\columnwidth]{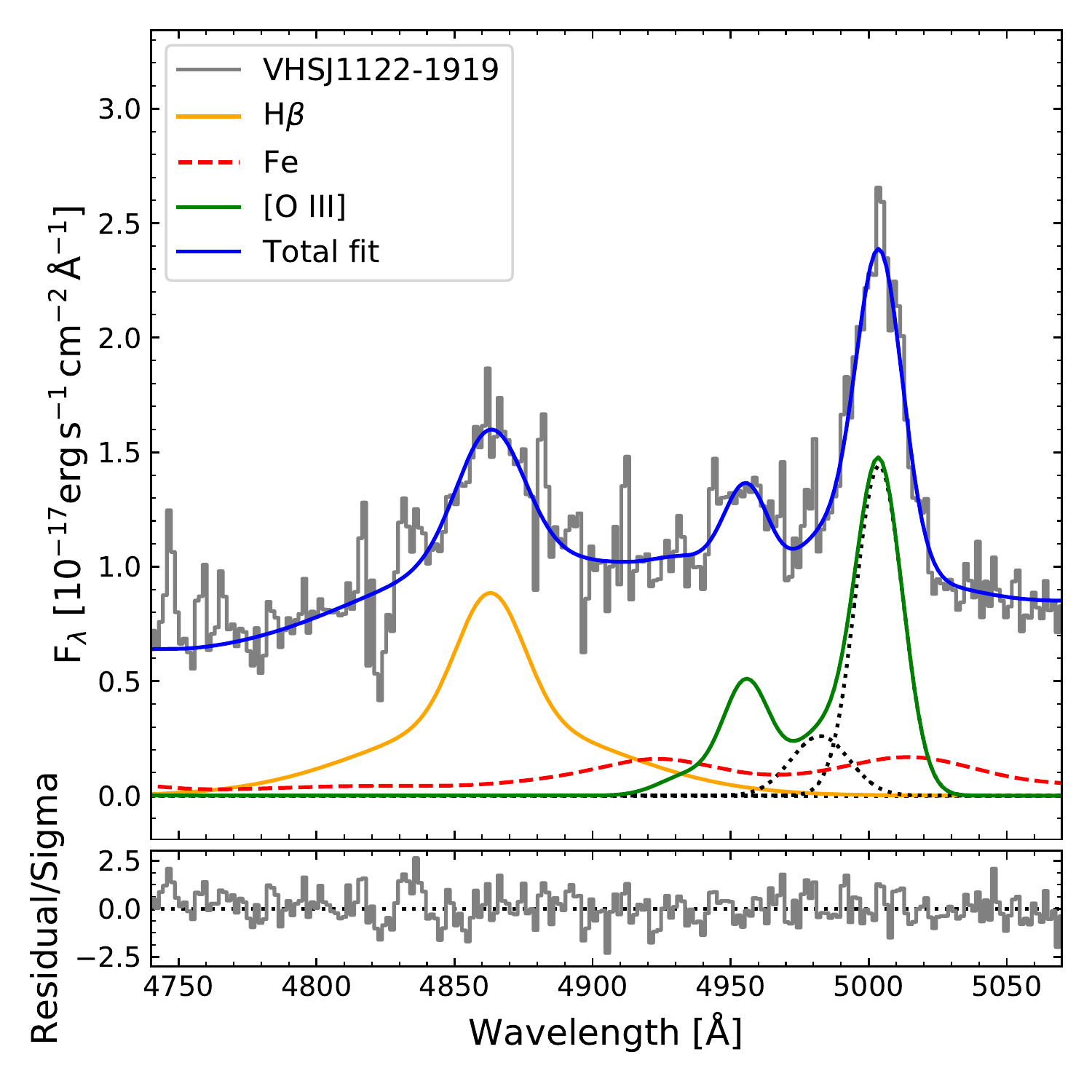}
\includegraphics[width=0.666\columnwidth]{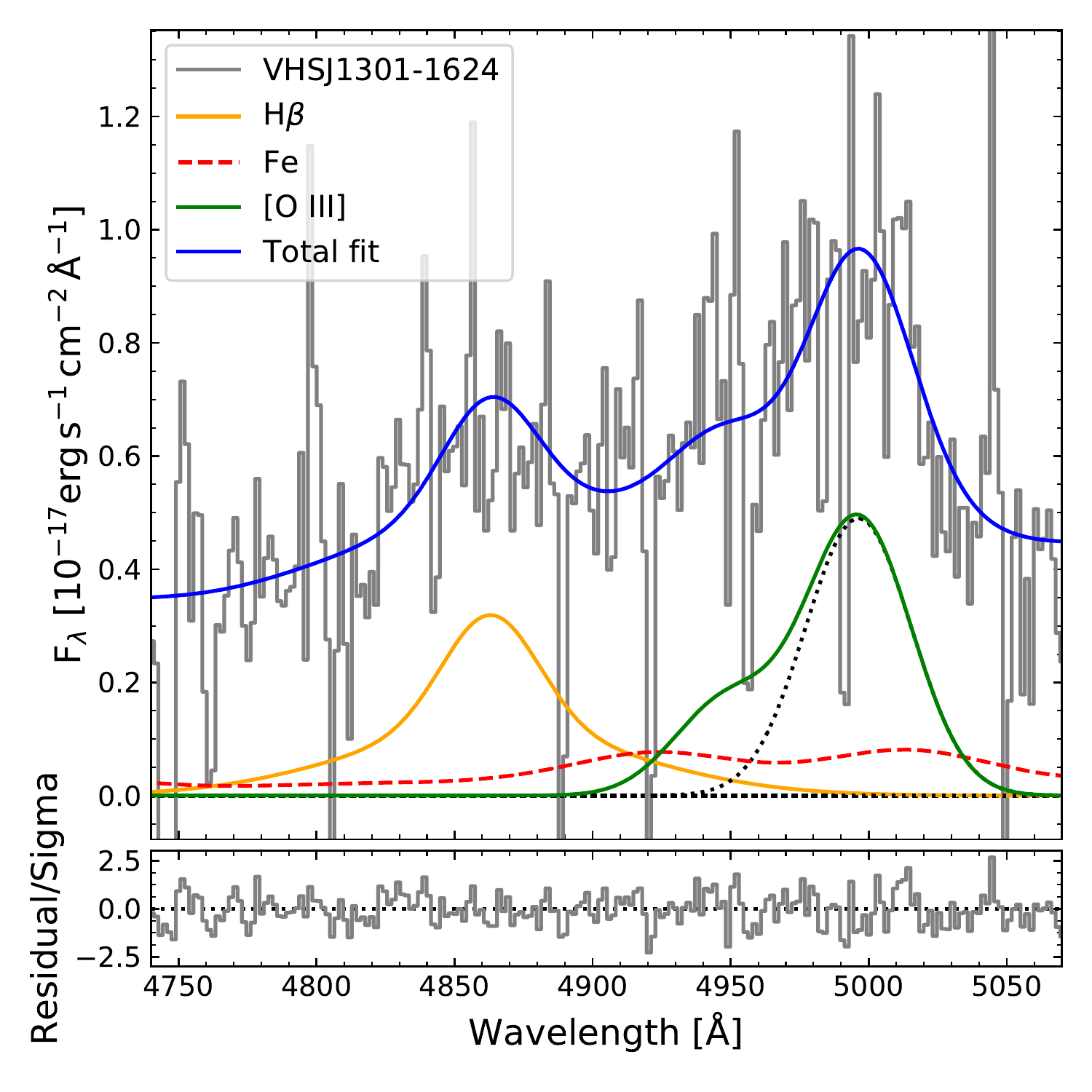}
\includegraphics[width=0.666\columnwidth]{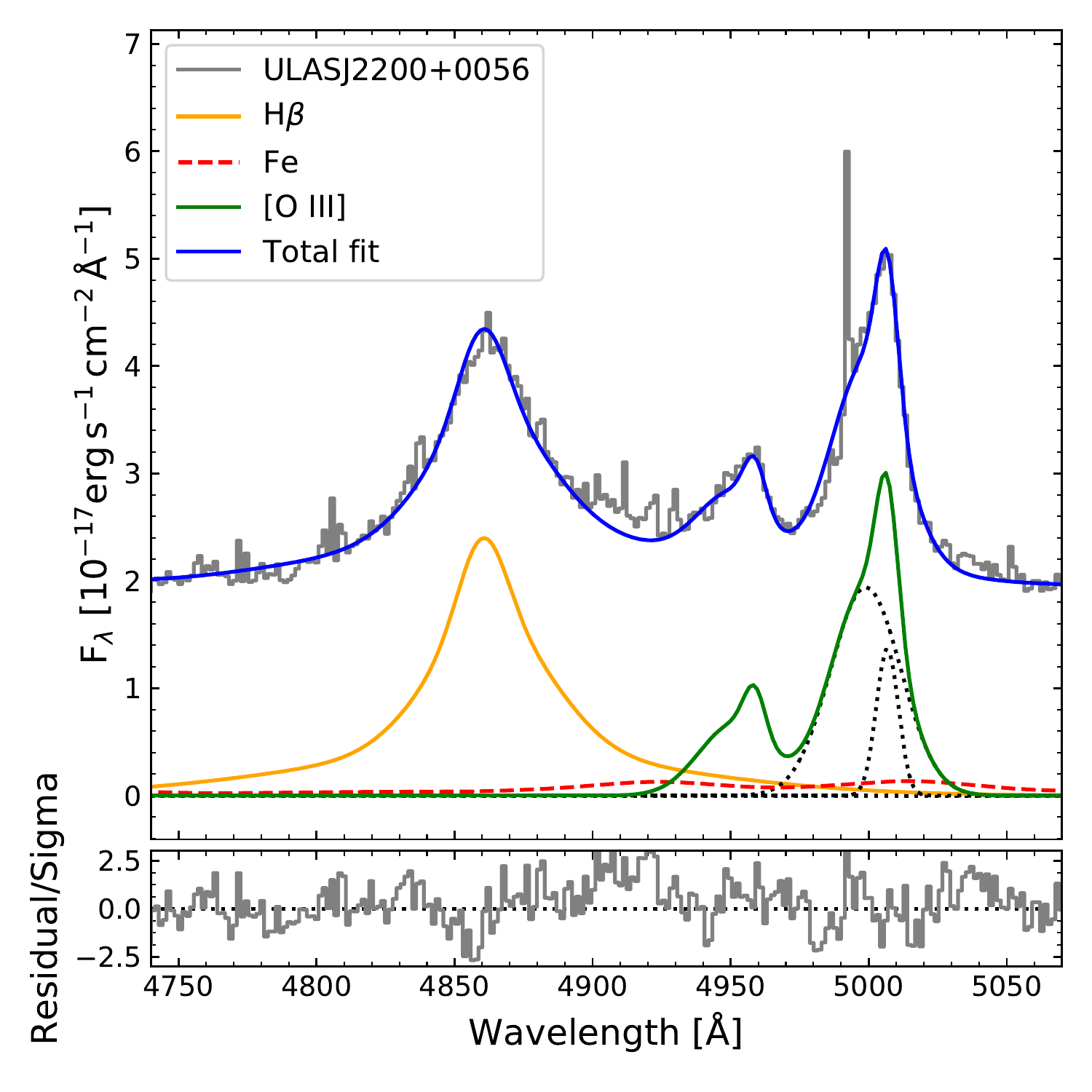}
\includegraphics[width=0.666\columnwidth]{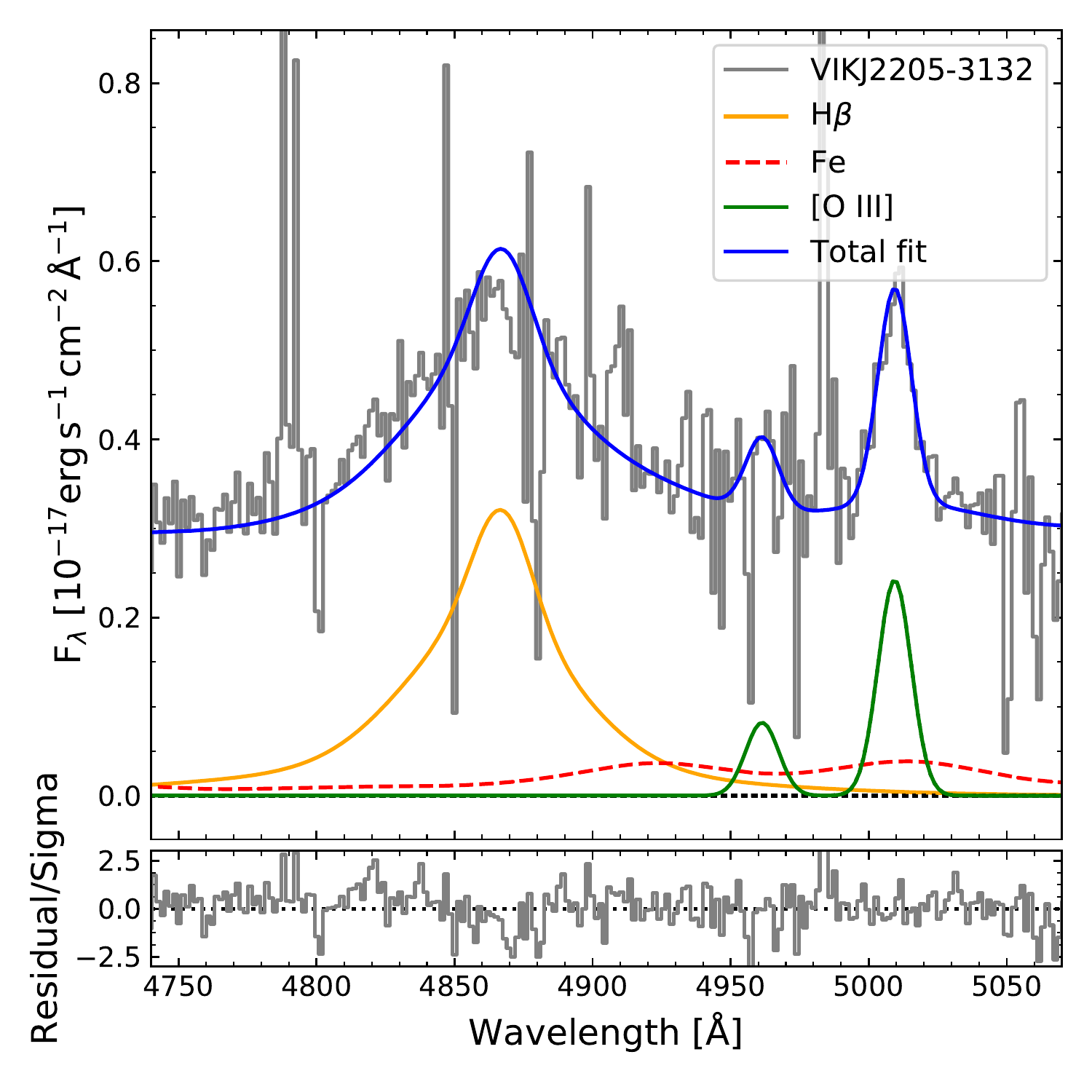}
\includegraphics[width=0.666\columnwidth]{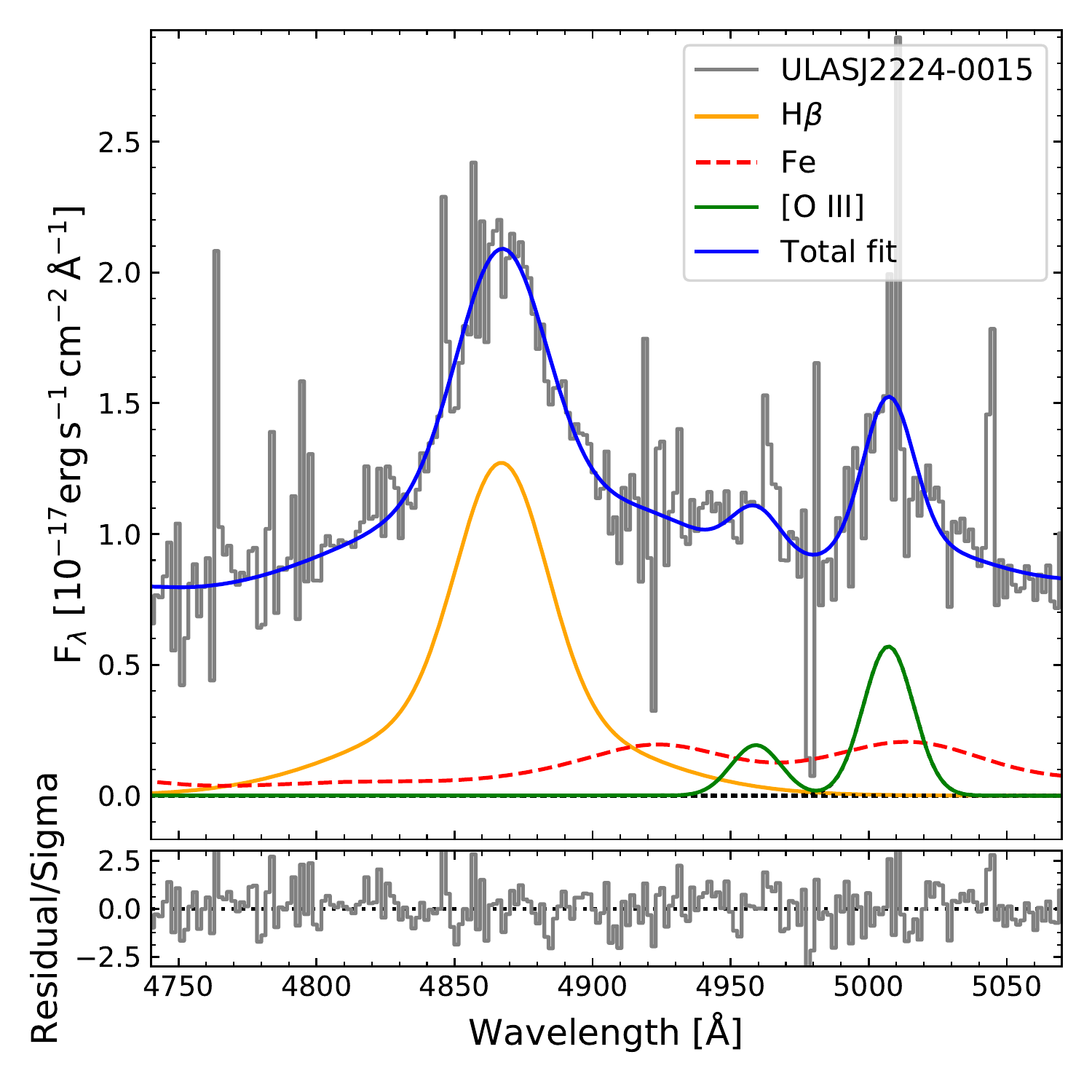}
\includegraphics[width=0.666\columnwidth]{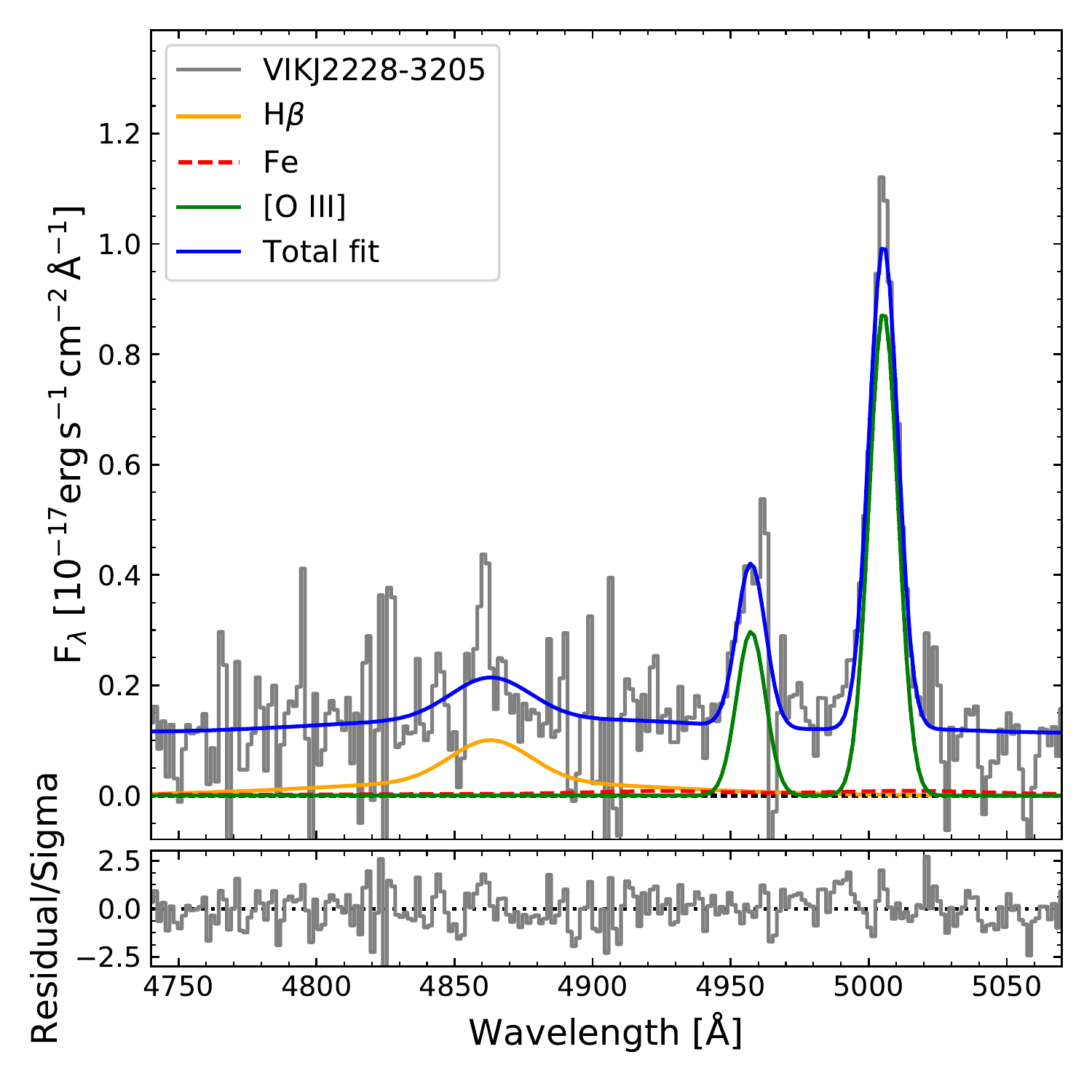}
    \caption{Fits to the region around H\,$\beta$ and [\ion{O}{iii}]. Individual \Oiii$\lambda$5008 Gaussian components are shown with dashed lines. Residuals are shown in the panels below, scaled by the noise.}
    \label{fig:Oiii}
\end{figure*}
\begin{figure*}
\begin{center}
\includegraphics[width=0.666\columnwidth]{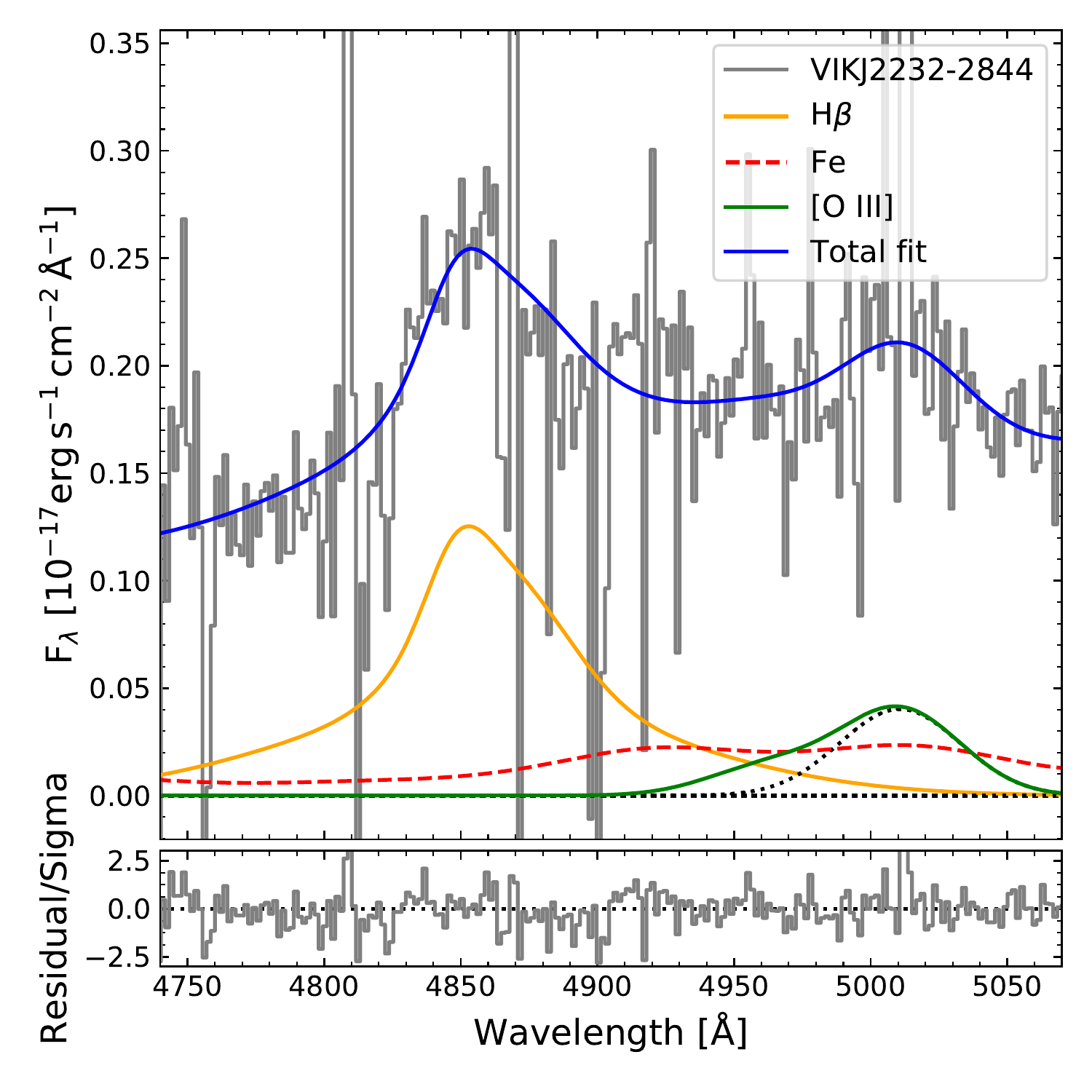}
\includegraphics[width=0.666\columnwidth]{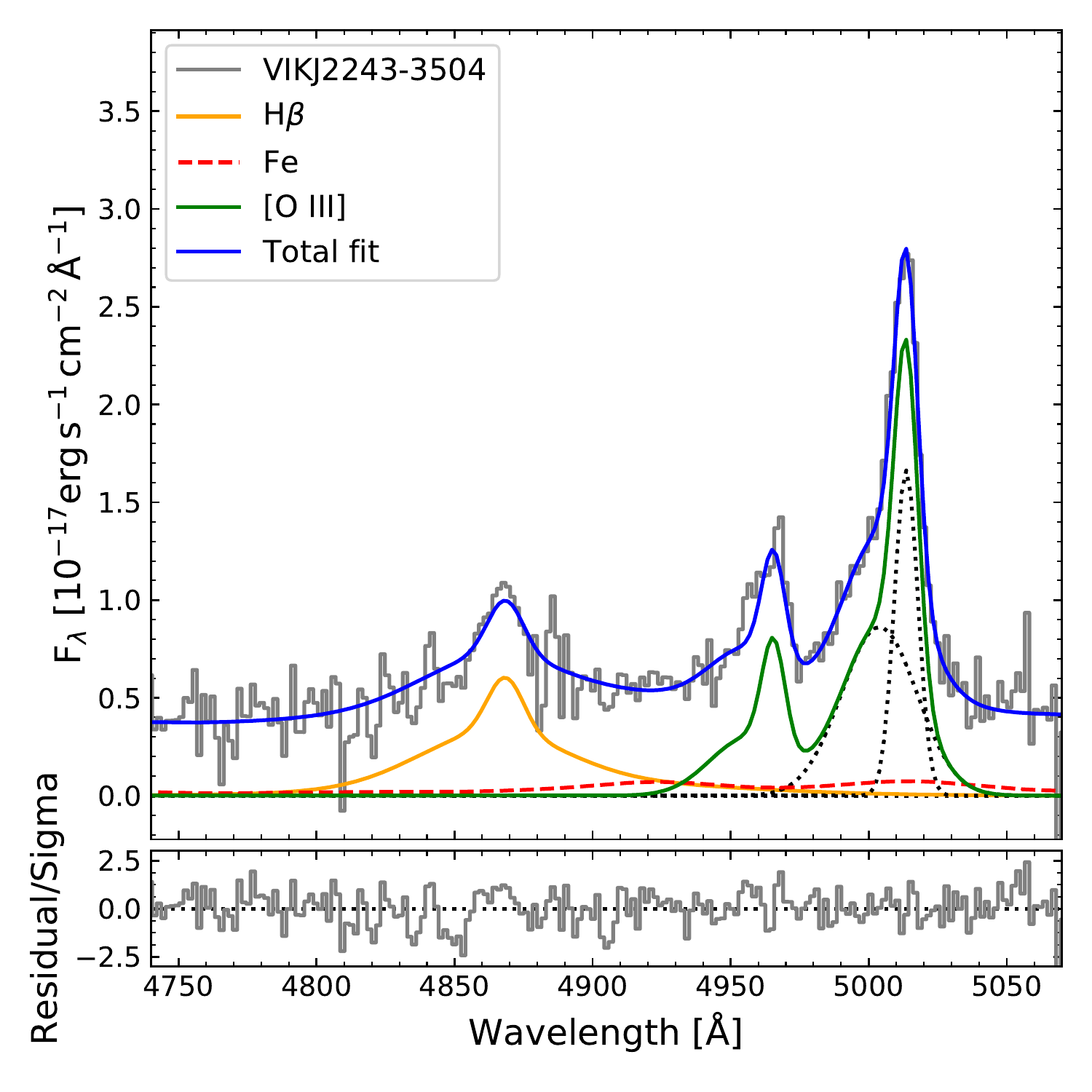}
\includegraphics[width=0.666\columnwidth]{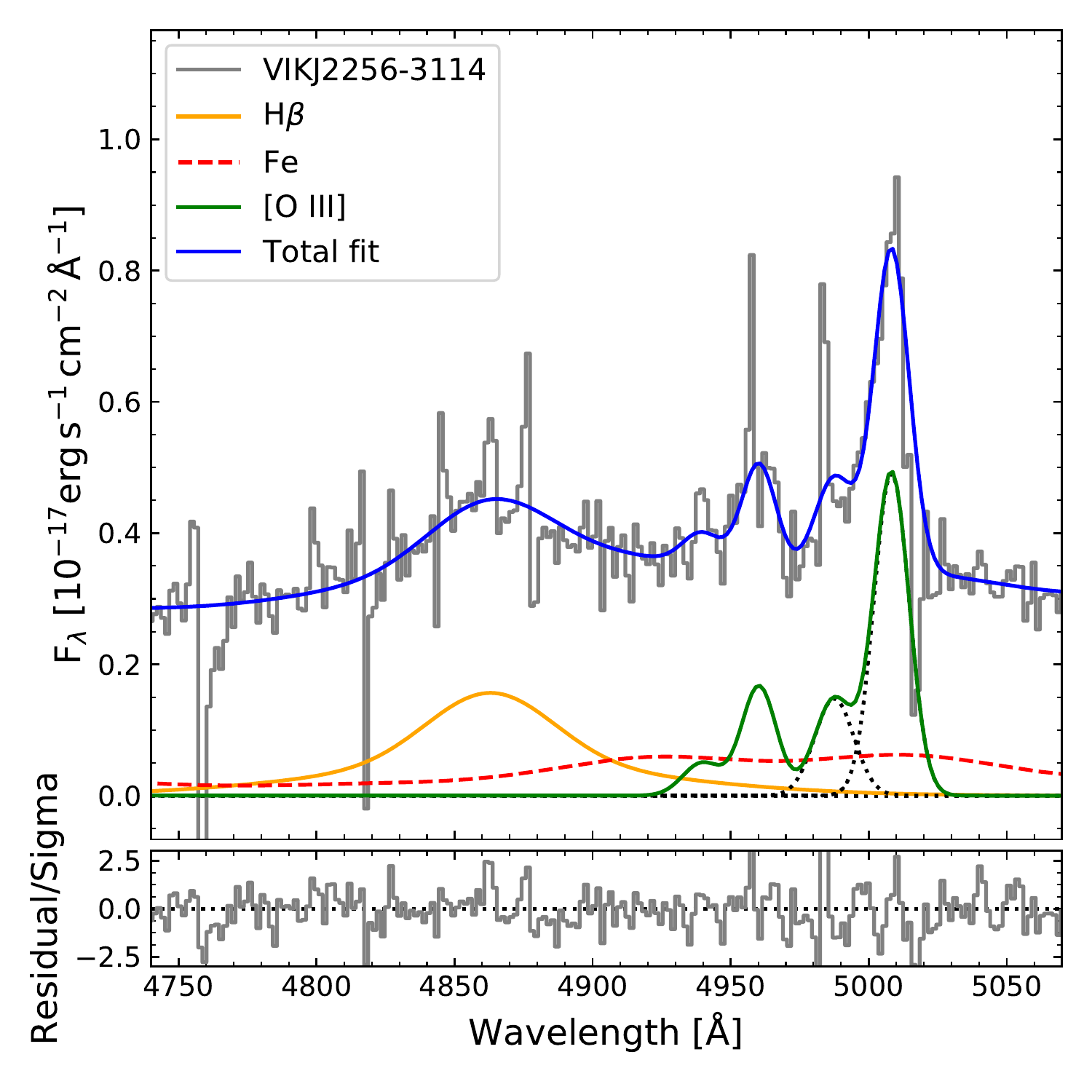}
\includegraphics[width=0.666\columnwidth]{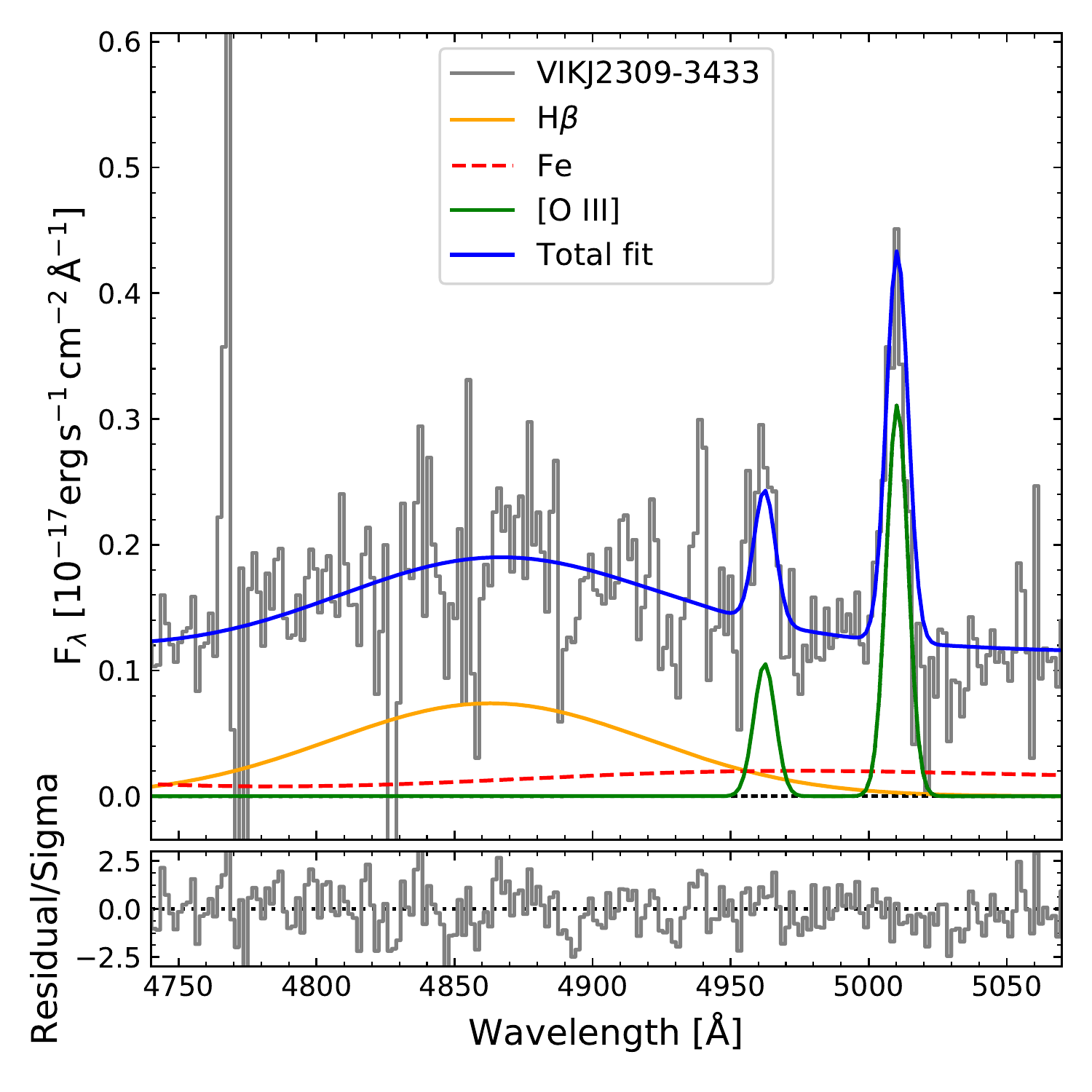}
\includegraphics[width=0.666\columnwidth]{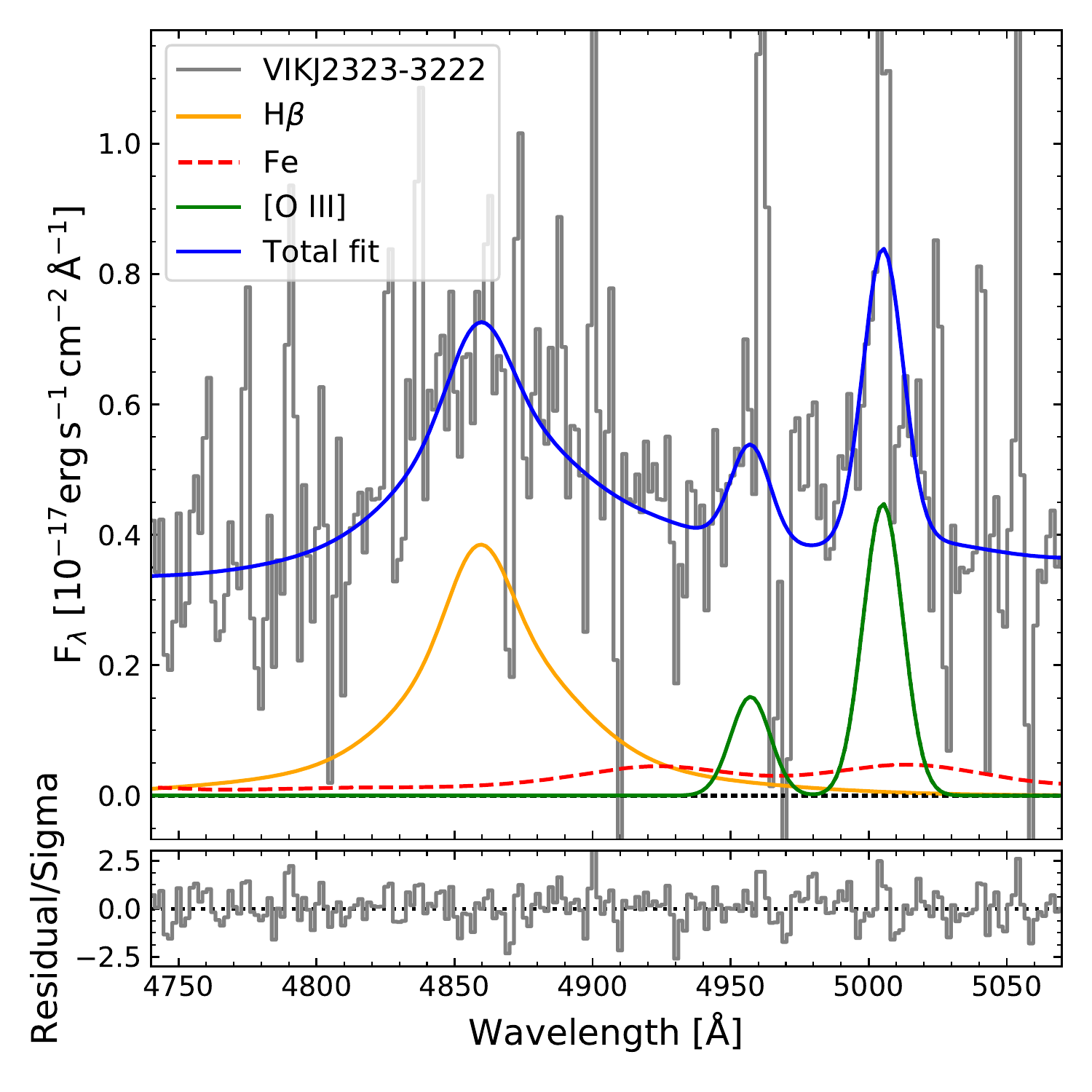}
\includegraphics[width=0.666\columnwidth]{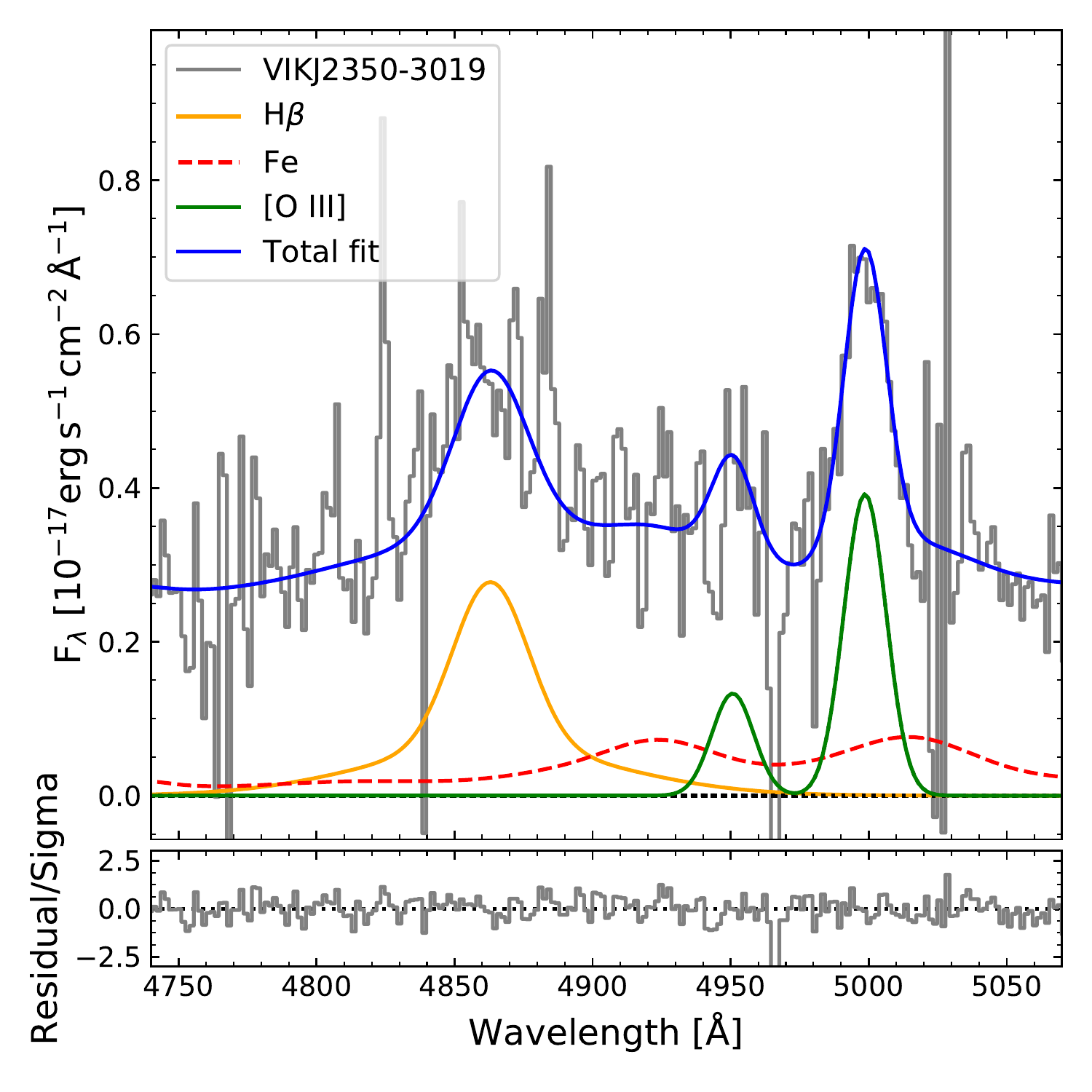}

\end{center}
    \contcaption{}
\end{figure*}

\section{Iron Subtraction}
\label{sec:iron}

We test the sensitivity of our results to the assumed shape of the iron template by replacing the iron template of \citet[][BG92]{BG92} with that of \citet[][KPD10]{Kovacevic10} and re-deriving the \Oiii line properties with an otherwise identical line fitting routine. The iron prescription of KPD10 is much more flexible, in that it allows the ratios of different groups of iron lines to vary, corresponding to an increase in the number of free parameters in the fit and enabling a more accurate description of the iron in objects where iron line ratios are different from I Zw 1, the NLS1 galaxy used to derive the template of BG92.

\begin{figure*}
    \includegraphics[width=\columnwidth]{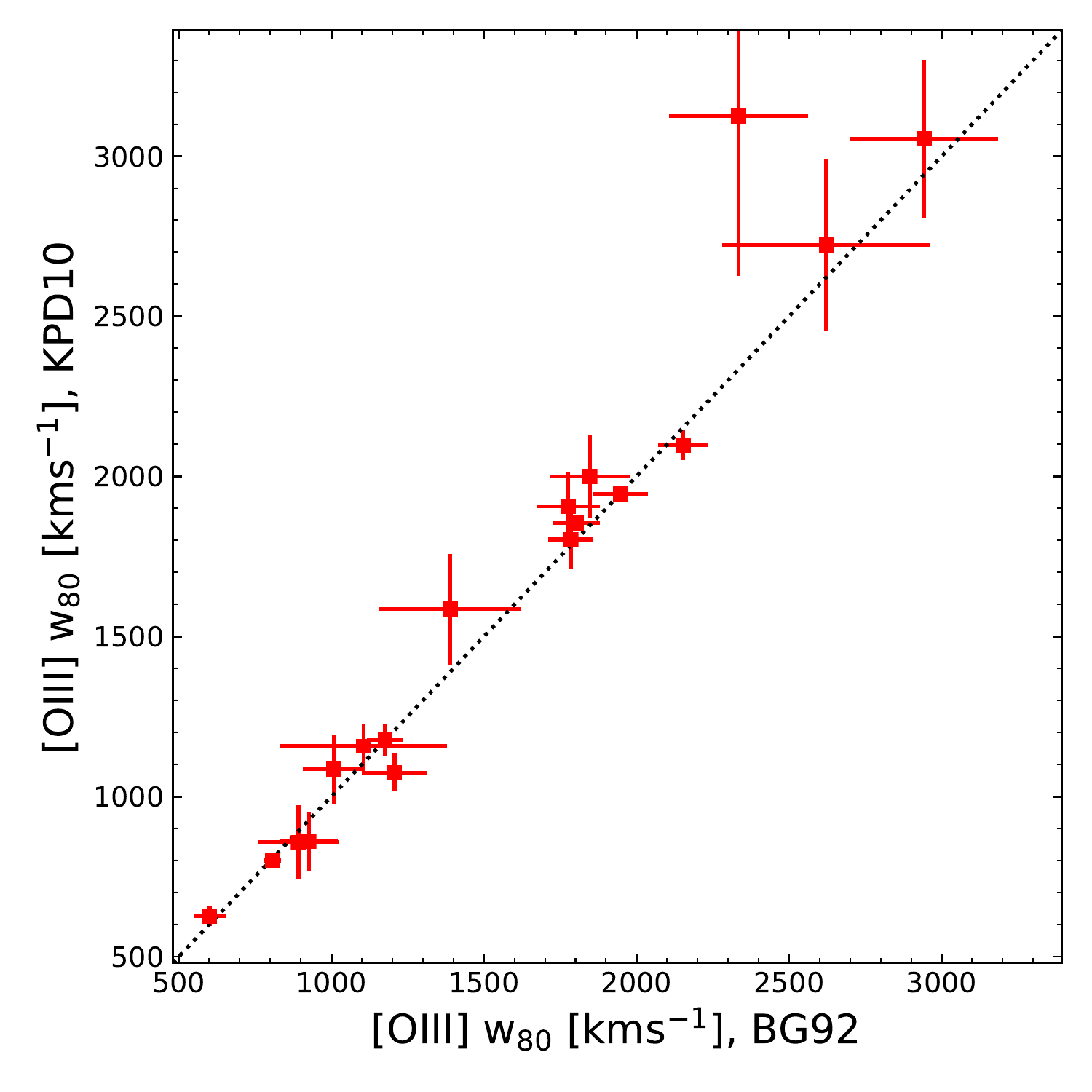}
    \includegraphics[width=\columnwidth]{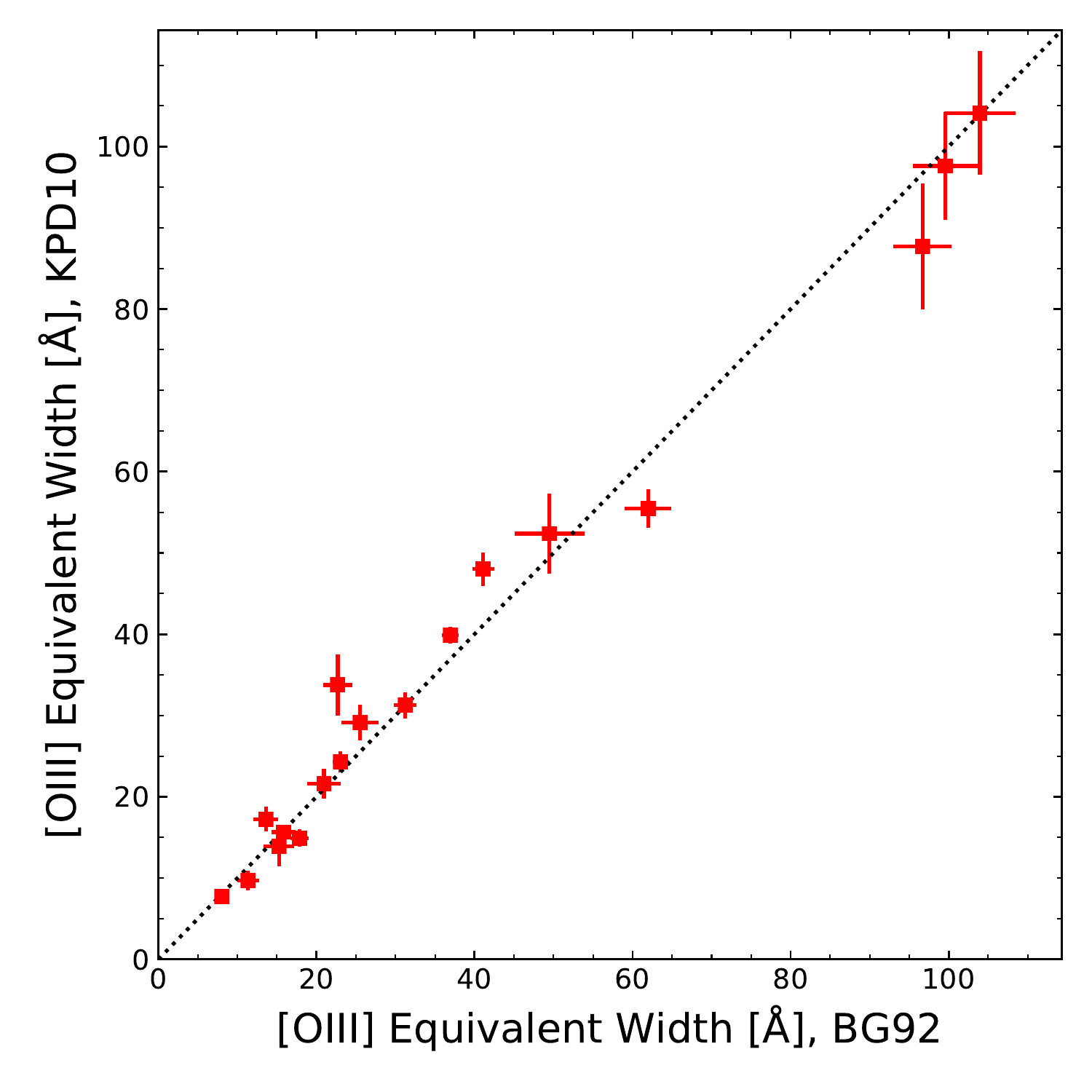}
    \caption{The  \Oiii 80 per cent velocity widths and  equivalent widths for our sample of dust-reddened quasars, derived using the optical iron templates from \citet[][BG92]{BG92} and \citet[][KPD10]{Kovacevic10}. VIK J2232-2844 is not well-fit by the KPD10 iron template and so is excluded. All other aspects of the fitting routine are the same for both templates, and are as described in the main text. The dashed line is the 1:1 relation, and we find the two iron prescriptions give very similar results when considering the distributions of \Oiii line properties across the whole sample.}
    \label{fig:Fe_EW}
\end{figure*}

\begin{figure*}
    \includegraphics[width=\columnwidth]{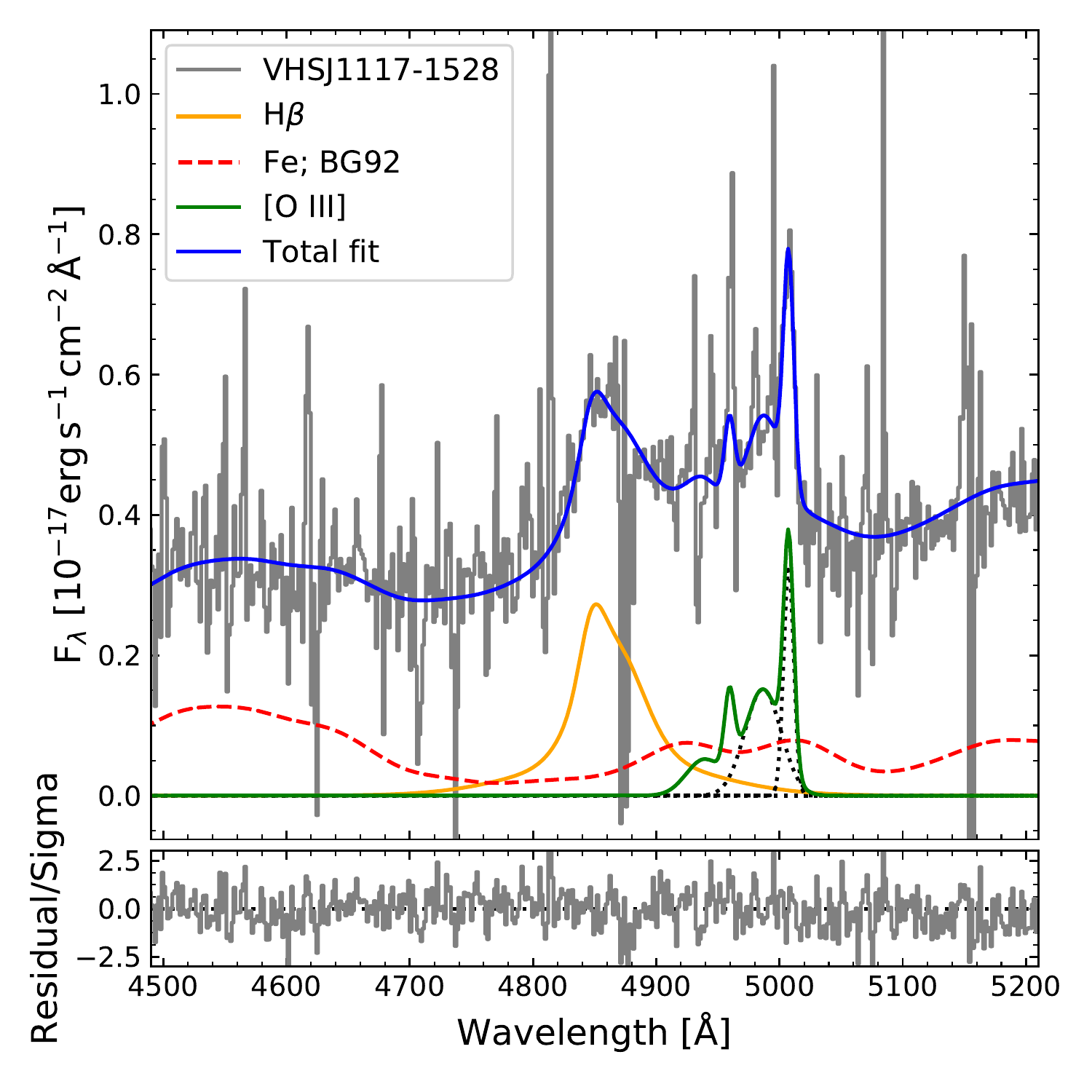}
    \includegraphics[width=\columnwidth]{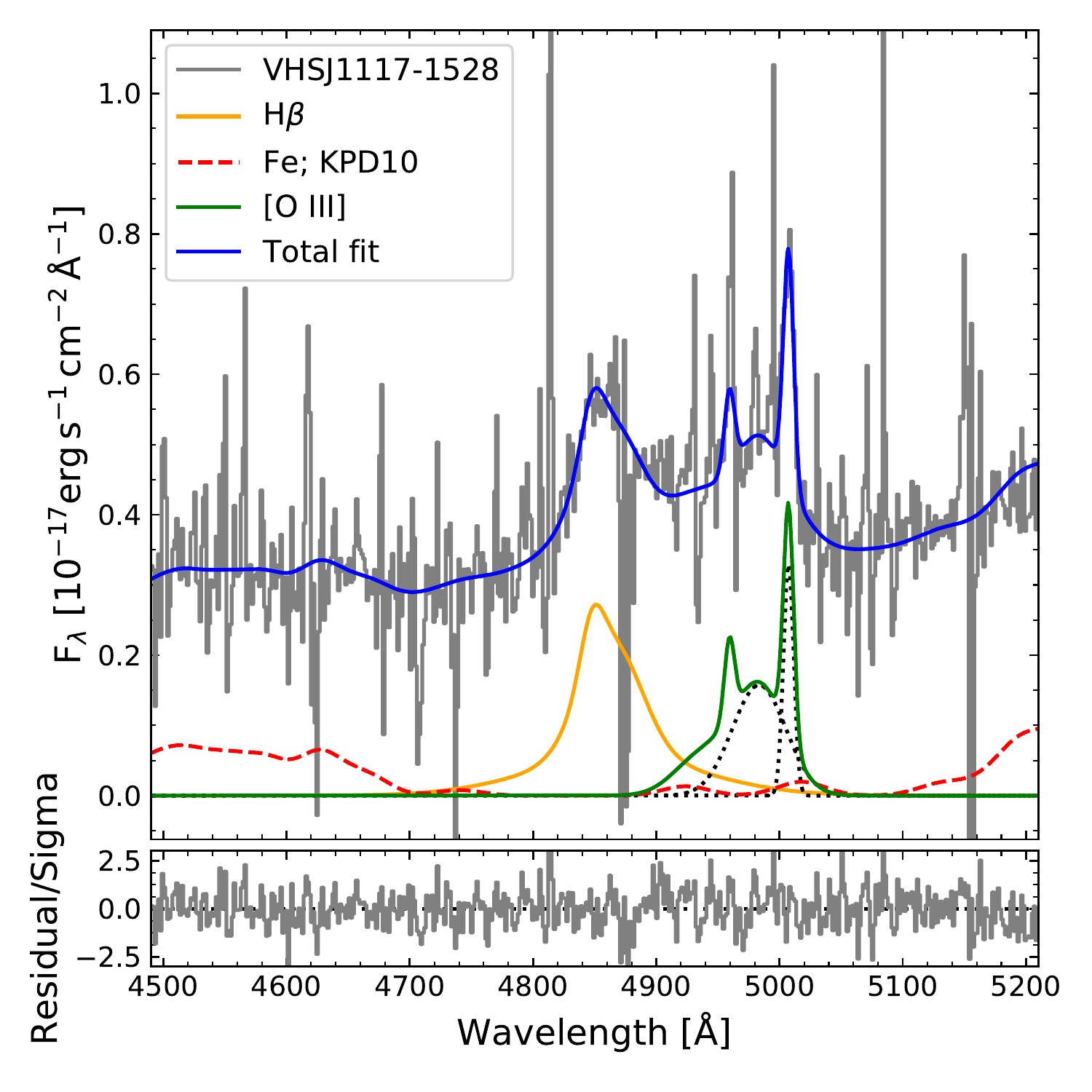}
    \caption{The H\,$\beta$ - \Oiii region in VHS~J1117-1528, fit using the iron prescriptions from \citet[][BG92]{BG92} and \citet[][KPD10]{Kovacevic10}. For individual quasars, the derived line properties are sensitive to the form of the iron prescription used to model the data  when the \Oiii emission line is both broad ($w_{80}>2000$\kmps) and weak ($\rm{EW}<35$\AA). }
    \label{fig:Fe_spec}
\end{figure*}

We compare the main results of our work under the two different iron prescriptions, as shown in Fig.~\ref{fig:Fe_EW}. We visually inspect individual objects, such as 
 VHS~J1117-1528, which has $w_{80}= 2330, 3130$\kmps\ and $\rm{EW}=22.7, 33.8$\,\AA\ when fit with BG92 and KPD10 respectively, making this the object from  our sample with the biggest difference in the derived \Oiii velocity widths. The fit in this quasar (Fig.~\ref{fig:Fe_spec}) is sensitive to the form of the iron prescription due to the fact that the \Oiii emission line is both broad ($w_{80}>2000$\kmps) and weak ($\rm{EW}<35$\,\AA), and we caution other authors to be careful when deriving line properties in similarly extreme objects. 
 
 However, the population statistics in our sample do not change significantly when using a different iron prescription to fit our data; in particular, the results described in Section~\ref{sec:results} are the same when derived with either prescription. 
Therefore, while we are cautious about the iron contribution in individual objects, we are confident that the precise form of the iron template used does not affect the conclusions of this paper.

\bsp    
\label{lastpage}
\end{document}